\documentclass[12pt,preprint]{aastex}
\usepackage{graphicx}
\usepackage[section] {placeins}

\begin{document}
\shorttitle{A New H$\alpha$ Emission-Line Survey in the ONC}
\shortauthors{Szegedi-Elek et al.}
\title{A New H$\alpha$ Emission-Line Survey in the Orion Nebula Cluster}

\author{E. Szegedi-Elek\altaffilmark{1} M. Kun\altaffilmark{1}, B. Reipurth\altaffilmark{2}, A.~P\'al\altaffilmark{1}, L.~G. Bal\'azs\altaffilmark{1}, M. Willman\altaffilmark{2}}
\email{eelza@konkoly.hu}
\altaffiltext{1}{Konkoly Observatory, H-1121 Budapest, Konkoly Thege \'ut 15--17, Hungary}
\altaffiltext{2}{Institute for Astronomy, University of Hawaii at Manoa, 640 N. Aohoku Place, Hilo, HI 96720, USA}

\begin{abstract}
We present results from an H$\alpha$ emission-line survey in a one square degree area centered on the Orion Nebula Cluster, obtained with the {\it Wide Field Grism Spectrograph--2} on the $2.2$-meter telescope of the University of Hawaii. We identified $587$ stars with H$\alpha$ emission, $99$ of which,  located mainly in the outer regions of the observed area, have not appeared in previous H$\alpha$ surveys. We determined the equivalent width (EW) of the line, and based on it classified $372$ stars as classical T Tauri stars (CTTS) and $187$ as weak line T Tauri stars (WTTS).  Simultaneous $r^\prime$, $i^\prime$ photometry indicates a limiting magnitude of $r^\prime \sim 20~mag$, but the sample is incomplete at $r^\prime > 17~mag$. The surface distribution of the H$\alpha$ emission stars reveals a clustered and a dispersed population, the former consisting of younger and more massive young stars than the latter. Comparison of the derived EWs with those found in the literature indicates variability of the H$\alpha$ line. We found that the typical amplitudes of the variability are not greater than a factor 2-3 in the most cases.  We identified a subgroup of low-EW stars with infrared signatures indicative of optically thick accretion disks. We studied the correlations between the equivalent width and other properties of the stars. Based on literature data we examined several properties of our CTTS and WTTS subsamples and found significant differences in mid-infrared color indices, average rotational periods, and spectral energy distribution characteristics of the subsamples.

\end{abstract}
\section{INTRODUCTION}

T Tauri type stars are young, low-mass stars. Their typical spectral types are between F and M \citep{herbig62}. These stars are still accreting material from a circumstellar disk via magnetic channels. In consequence of accretion and outflows, pre-main sequence stars exhibit emission lines, especially strong H$\alpha$ emission.

Some decades ago these stars were discovered because of their strong H$\alpha$ emission via low-dispersion objective prism photographic surveys. Nowadays slitless grism spectroscopy with CCD detectors allows simultaneous detection of numerous stars within the field of view, offering a great opportunity to identify H$\alpha$ emission-line stars \citep[see e.g.][]{nakano2012}. 
A good target for  such an H$\alpha$ survey is the young Orion Nebula Cluster (ONC).

The Orion Nebula is one of the best studied high mass star-forming regions \citep{muench2008, odell2008}. It is situated some $414$ pc from the Sun, based on trigonometric parallax \citep{distance}, in the Orion~A molecular cloud, which covers $29$ deg$\sp{2}$ and contains $1\times 10\sp{5}$ M$\sb{\Sun}$ of molecular gas. It has a very rich \citep[$ > 2000$ members;][]{hillenbrand1997} and young \citep[$\sim1$~Myr;][]{hillenbrand1997} population. At this age many clusters are still embedded, contrary to the ONC, which is observable at optical wavelenghts because the parental cloud along our line-of-sight was removed by the radiation of Trapezium stars.
The stellar population centered on the Trapezium stars is traditionally divided into concentric radial zones. These are as follows: the central $0.3$~pc ($\sim2$~arcmin) is called the {\it Trapezium Cluster}, the inner $3$~pc ($\sim20$~arcmin) is the classical {\it Orion Nebula Cluster}, and the Orion OB1c association extends to more than $25$~pc (approximately $3\degr$) in radius.

\citet{Haro53} found $255$ H$\alpha$ emission line stars in the brightest area of $3.5$ square degrees of the ONC centered on the Trapezium. \citet{Parsam} catalogued $543$ H$\alpha$ emission line stars in a wider $5\degr \times 5\degr$ region also centered on the ONC. \citet{kiso} conducted an objective-prism survey of the ONC. The number of H$\alpha$ emitters was $191$ in the observed area A-0975 ($ 5\degr \times 5\degr$, centered at $\alpha=5^\mathrm{h}20^\mathrm{m}, \delta=-5\degr$) and $415$ in the area A-0976 ($ 5\degr \times 5\degr$, centered at $\alpha =5^\mathrm{h}40^\mathrm{m}, \delta =-5\degr$). In all, the Kiso Survey revealed $606$ stars showing H$\alpha$ emission.

\citet{furesz} carried out a spectroscopic study of $1215$ selected stars covering the ONC and its vicinity. The observations were performed with the Hectoechelle multiobject spectrograph at the $6.5$~m MMT telescope in Arizona in $2004$ and $2005$.

Narrow band filter photometric techniques have also been used to identify new H$\alpha$ emission line stars in the ONC. \citet{dario2009} 
observed the ONC in a field of about $34\arcmin \times 34\arcmin$ with the WFI imager at the ESO/MPI $2.2$~m telescope at La Silla Observatory. They identified $638$ H$\alpha$ sources in the region. 

The goal of our paper is to assess the power of slitless grism spectroscopy by studying the well-known, nearby and rich ONC and compare the results with literature data on the same region, derived from various other observing methods. We describe the observations and data reduction, as well as the ancillary data, used during  statistical analysis in Sect.~\ref{sect_dat}. The results and their comparison with literature data are presented in Sect.~\ref{sect_anal}, and we discuss our results in   Sect.~\ref{sect_disc}. A short summary of the results is given in Sect.~\ref{sect_sum}.


\section{DATA}
\label{sect_dat}

\subsection{Observations and Data Reduction}
\label{sect_obs}

\subsubsection{Slitless Grism Spectroscopy}
\label{sect_ha}

We observed the Orion Nebula Cluster, centered on the Trapezium, with the Wide Field Grism Spectrograph--$2$ (WFGS$2$) installed on the  University of Hawaii $2.2$-meter telescope, on the five nights 2010 December 31, 2011 January 2, 2011 January 25, 2011 February 27, 2011 October 16. We covered an area of almost one square degree centered on RA($2000$)$= 5^\mathrm{h}35^\mathrm{m}16\fs4$ and Dec($2000$)$=-5\degr23\arcmin25\arcsec$ with a mosaic of $25$ overlapping fields. 
We used a $300$ line mm$\sp{-1}$ grism blazed at $ 6500$\,\AA\ and providing a dispersion of $3.8$~\AA~pixel$\sp{-1}$ and a resolving power of $820$. The H$\alpha$ filter had a $500$\,\AA\  passband centered near $6515$\,\AA\ . The detector for WFGS$2$ was a Tektronix $2048\times2048$ CCD, whose pixel size of $24\,\mu$m  corresponded to $0.34$~arcsec on the sky. The field of view was $11.5\arcmin\times11.5\arcmin$. 
For each field, we took a short, $60$~s exposure in order to detect the H$\alpha$ line in bright stars and avoid saturation of the central parts of the nebula. Then for $21$ fields we obtained two frames with $480$~s exposure time, and for the remaining four fields three frames of 300~s exposure.

\begin{figure}
\begin{center}
\includegraphics[scale=0.8]{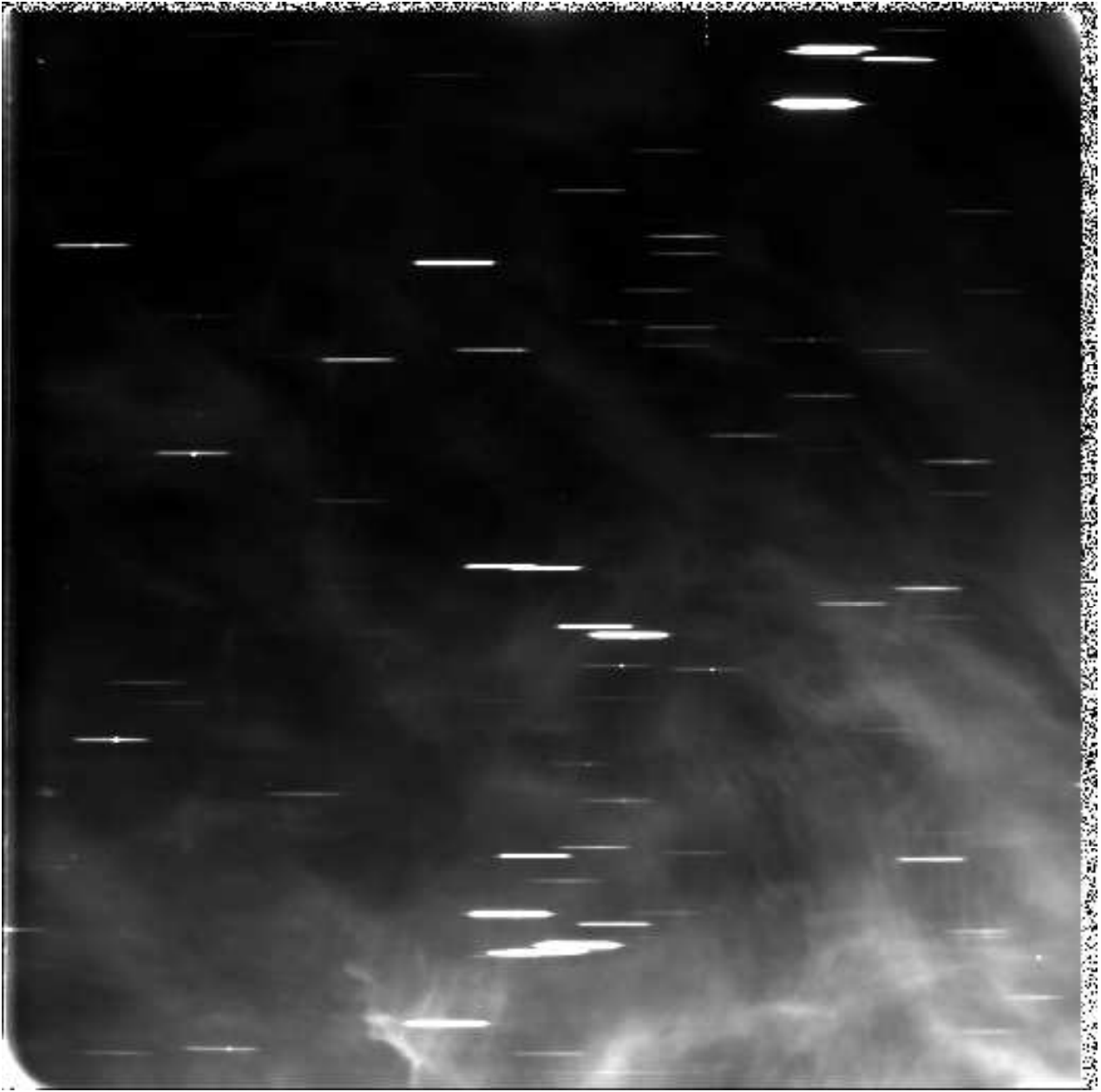}
\figcaption{An example of the reduced and coadded images.
\label{onc16av}}
\end{center}
\end{figure}

Bias substraction and flat-field correction of the images were done in IRAF\footnote{IRAF is distributed by the National Optical Astronomy Observatories, which are operated by the Association of Universities for the Research in Astronomy, Inc., under cooperative agreement with the National Science Foundation. http://iraf.noao.edu/}. Then we used the FITSH, a software package for astronomical image processing\footnote{http://fitsh.szofi.net/}  \citep{pal2012} to remove cosmic rays, coadding the long-exposure images, identify the stars on the images, and transform the pixel coordinates into the equatorial coordinate system. A representative example of the reduced and coadded images can be seen in Fig.~\ref{onc16av}.

We examined all spectra visually for presence of the H$\alpha$ emission line and found $587$ stars with H$\alpha$ emission in the whole observed area. We associated our emission sources with {\it 2MASS\/} \citep{2mass} point-sources. We found matches unambiguously within $2\arcsec$ for all but seven stars. Due to small separation, three pairs of H$\alpha$ stars are associated with single {\it 2MASS\/} sources (J05342073-0530461, J05342207-0501342, and J05355535-0547017). Another H$\alpha$ star, V1504~Ori has no associated {\it 2MASS\/} point source. Since each of these stars can be found in the {\it UKIDSS--DR8 Galactic Clusters Survey\/} \citep{Lawrence}, we label them with their {\it UGCS\/} identifiers. 

The equivalent width ({\it EW}) of the H$\alpha$ emission line was computed as follows:  First, small stamps with size of $30 \times 30$ pixels were cut around the core of the emission line. The size of this stamp was derived from the FWHM of the profile and the characteristic lengths of the structures in the grism frames and found to be stable throughout our observations. In each stamp, the background level (that varied significantly due to the intrinsic background of the Orion Nebula complex) was obtained from the upper $30\times5$ and lower $30\times5$ pixels by either fitting a constant level or a plane, depending on the variation level of the background. We found that in most cases, both fits yielded a smooth background, however, in some cases the two methods led to different results. In such cases we accepted the more ``stable'' solution. The uncertainty of the background level was determined by a simple statistic, i.e. from the fit residuals divided by the square root of the number of involved pixels. Second, these background-subtracted stamps were stacked via the y axis (i.e. almost exactly perpendicular to the grism dispersion), yielding $30$ pixel long spectra. The uncertainty of each spectral point was estimated from the photon noise statistic (taking into account the effective camera gain 1.780~e$\sp{-}$/ADU) and by adding quadratically the scaled background scatter. Third, these spectra were fitted with a Gaussian, parameterized by the four free parameters of zero-point level, center, amplitude and standard deviation. The equivalent width of the emission line was then simply derived from these fitted parameters, i.e. by dividing the integral of the Gaussian (without the zero-point level) by the zero-point level (that is proportional to the instrumental continuum level). The uncertainty of the equivalent widths was obtained from linear error propagation using the fitted results. 

Applying the above described method, we could determine the equivalent widths for $452$ stars. For the remaining stars in which the H$\alpha$ emission was clearly detected, this method could not be applied due to faint continuum or the bright and variable nebulous background. In such cases we estimated the equivalent width using the IRAF `splot' package. However, the equivalent widths could not be determined for $28$ stars. Due to overlapping fields, $50$ of the stars were observed more than once.

\begin{figure}
\begin{center}
\includegraphics[scale=0.8]{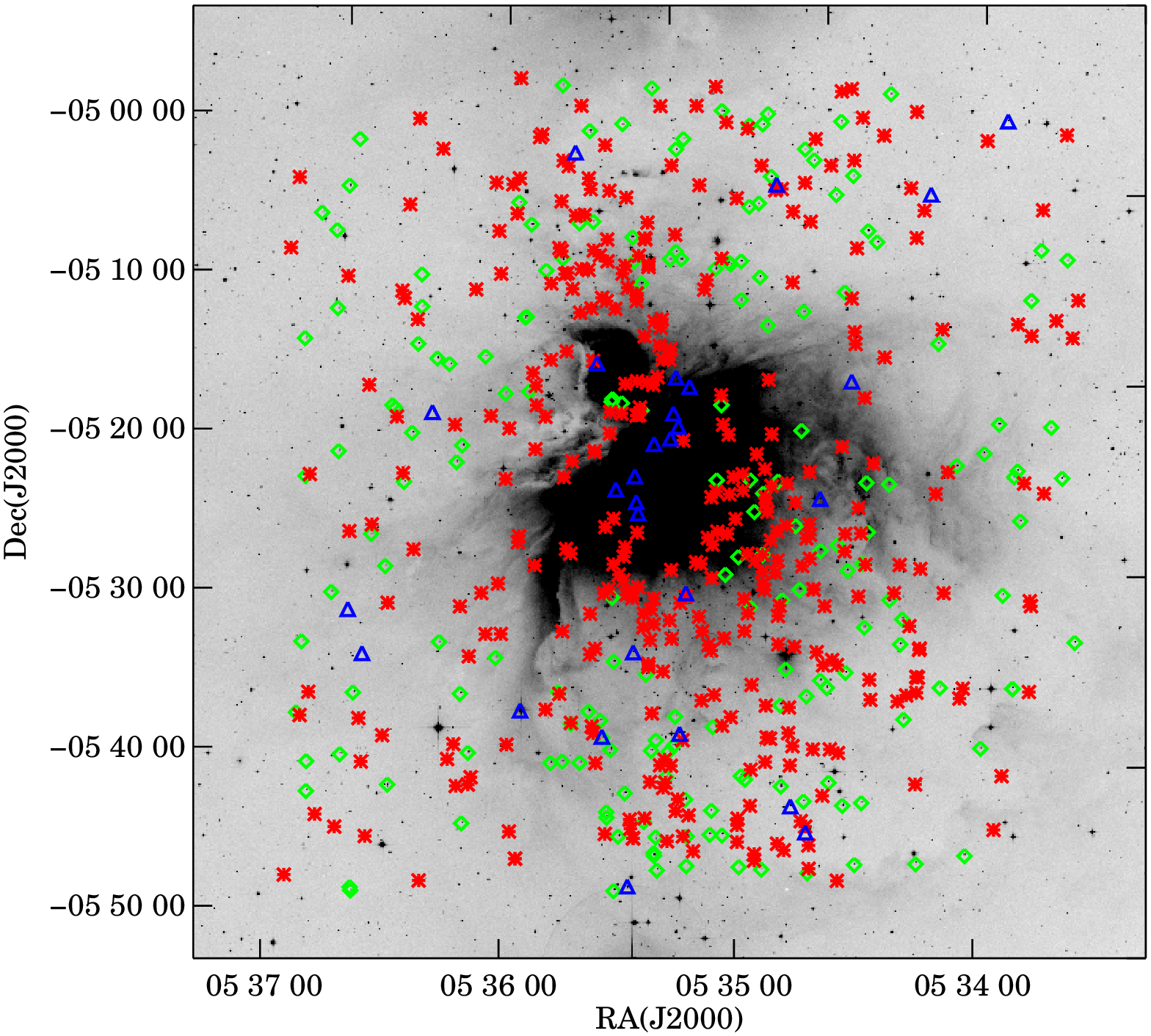}
\figcaption{H$\alpha$ emission stars, identified during the present survey, overplotted on the {\it DSS2 IR} image of the observed area. Asterisks (red in the online version) indicate the $372$ classical T Tauri stars, diamonds (green in the online version) mark the $187$ weak-line T Tauri stars while  triangles (blue in the online version) represent $28$ sources without measured {\it EW\/}(H$\alpha$). 
\label{fig_onc}}
\end{center}
\end{figure}

\begin{figure}
\begin{center}
\includegraphics[scale=0.8]{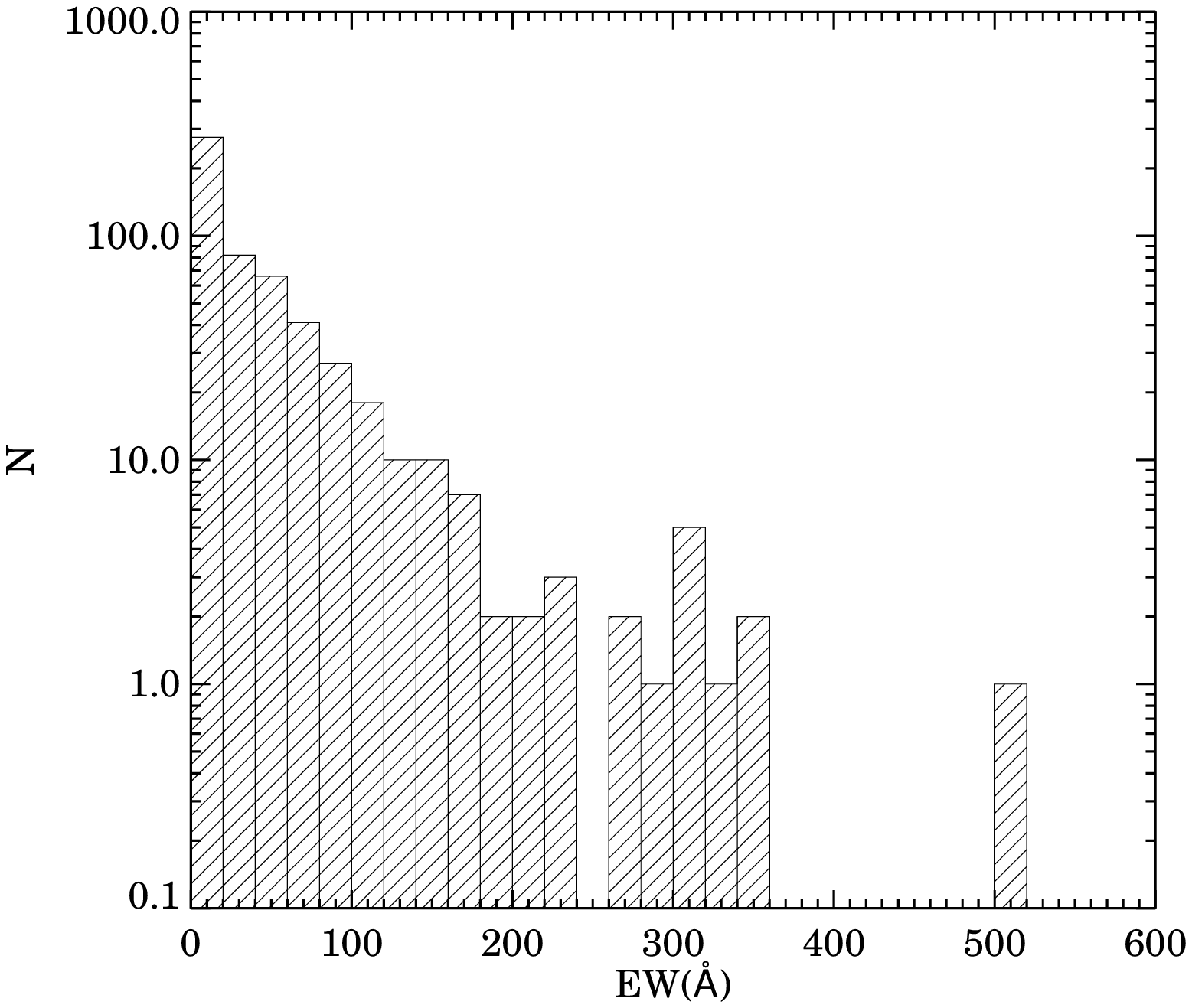}
\figcaption{Distribution of the measured H$\alpha$ equivalent widths.
\label{fig_ewhist}}
\end{center}
\end{figure}

Figure \ref{fig_onc} shows the distribution of our H$\alpha$ emission stars over the observed area centered on the Trapezium. Asterisks indicate the classical T Tauri stars, diamonds mark the weak-line T Tauri stars, while triangles represent sources of uncertain nature (see Sect.~\ref{sect_class}).  Figure~\ref{fig_ewhist} shows the histogram of equivalent widths.

\subsubsection{Photometry}
\label{sect_phot}

Direct images of each field of the mosaic with the same instrument were obtained through  $r^\prime$ and $i^\prime$ filters immediately before the spectroscopic exposures. 
One exposure was taken at each position and in each filter with integration time of $60$ sec. Aperture photometry was carried out with the PHOT task in the DAOPHOT package.
Several stars from {\it The SDSS Photometric Catalog, Release 8} can be found in every frame.
In order to transform inst\-rumental magnitudes into the {\it SDSS} system, we applied those stars and the following transformation equations:  

\begin{displaymath}
r\sb{SDSS}=C\sb{1} \times r\sp{\prime}\sb{instr}+C\sb{2}\times airmass+C\sb{3}
\end{displaymath}
\begin{displaymath}
r\sb{SDSS}-i\sb{SDSS}=C\sb{4}\times (r\sp{\prime}-i\sp{\prime})\sb{instr}+C\sb{5}\times airmass+C\sb{6}
\end{displaymath}

\begin{figure}[!ht]
\begin{center}
\includegraphics{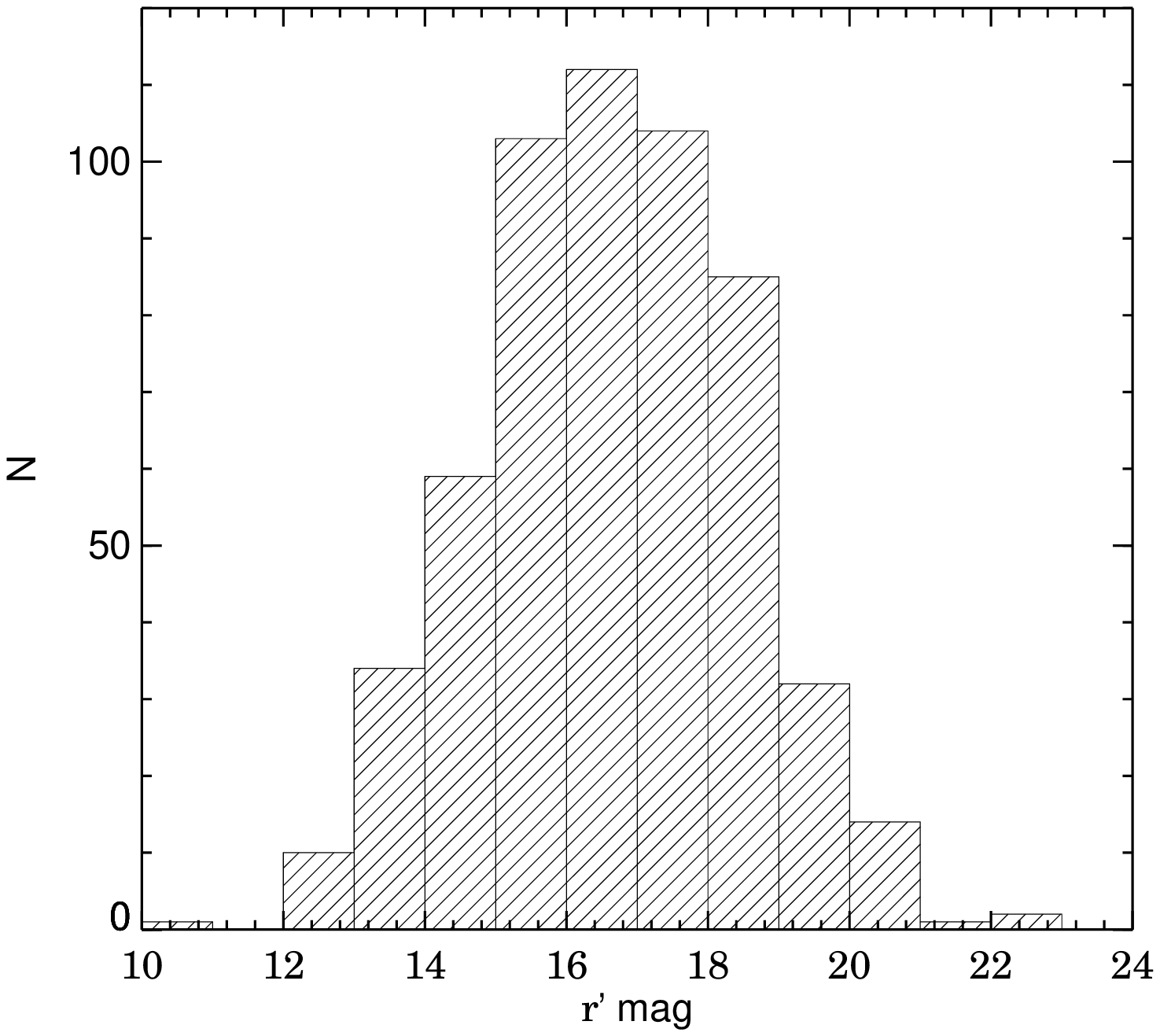}
\figcaption{Distribution of the {\it  $r^\prime$} magnitudes of the H$\alpha$ emission stars.
\label{fig_khist}}
\end{center}
\end{figure}

\begin{figure}[!ht]
\begin{center}
\includegraphics{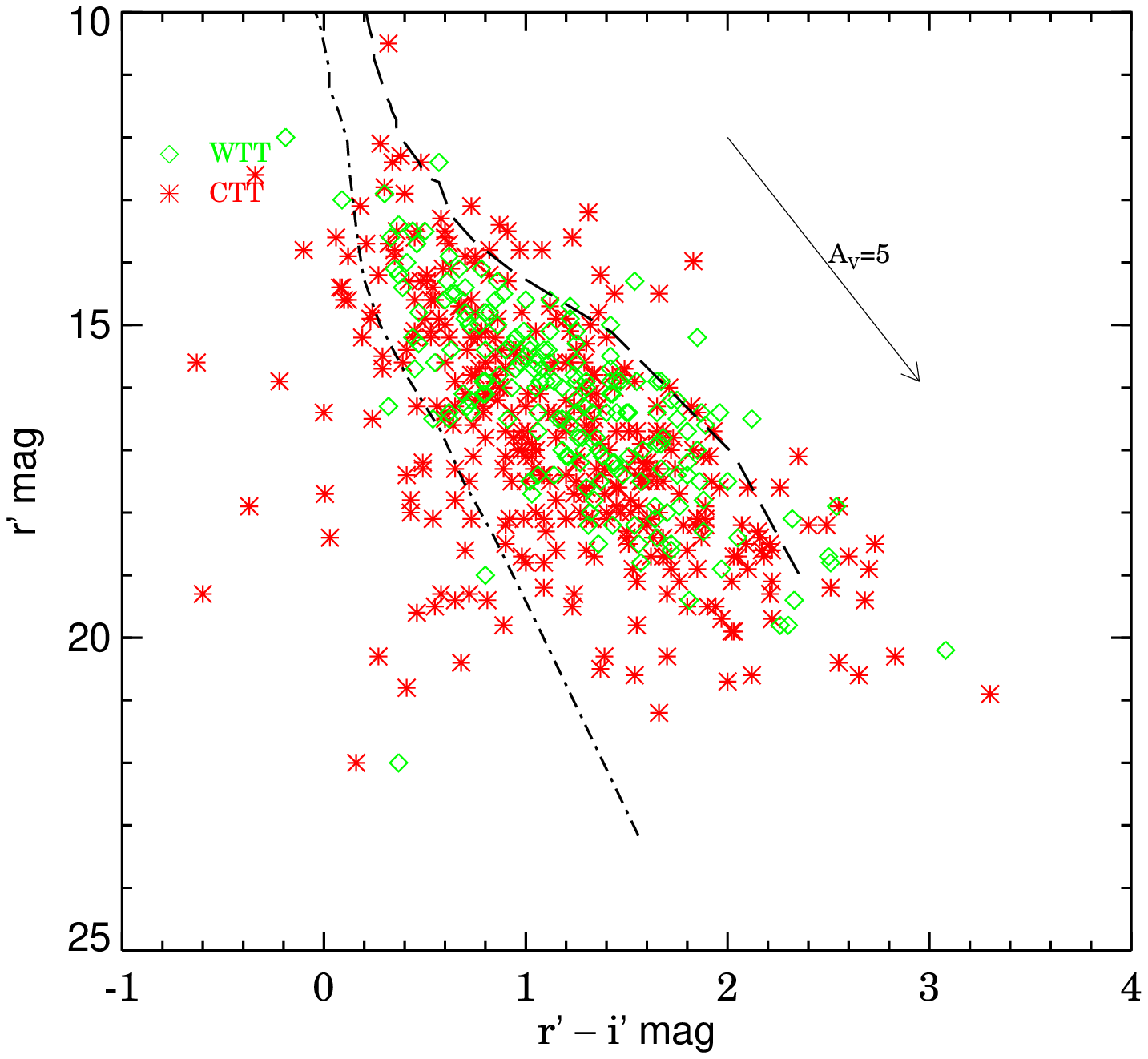}
\figcaption{A $r^\prime$, $r^\prime-i^\prime$ color-magnitude diagram for the H$\alpha$ emission stars. The dash-dotted line represents the zero-age main sequence, while the dashed line indicates the $10^6$-year isochrone \citep*{siess}. The arrow shows the direction of extinction.
\label{fig_cmd}}
\end{center}
\end{figure}

The C$\sb{1}$ $\ldots$ C$\sb{6}$ constants are derived from the fitting procedure. 
Photometric errors of typically $0.1$ mag were derived from quadratic sums of the formal errors of coefficients of the transformation equations.  Fig.~\ref{fig_khist} shows the histogram of the  $r^\prime$   magnitudes of our H$\alpha$ emission stars, and the $r\sp{\prime}$ vs. $r\sp{\prime}-i\sp{\prime}$ color-magnitude diagram is displayed in Fig.~\ref{fig_cmd}.

\subsection{Supplementary Data}
\label{sect_anc}

To characterize the nature of the H$\alpha$ emission stars and investigate possible correlations between the {\it EW} of the  H$\alpha$ line and other stellar properties, we used data available via the {\it VizieR\/} catalogue access tool. Equatorial coordinates were determined using the {\it 2MASS All-Sky Catalog of Point Sources} \citep{2mass} as reference. {\it JHK$_\mathrm{s}$}  magnitudes for all but five stars are found in the same catalog. {\it I} magnitudes can be found in the {\it DENIS} \citep{denis} data release, and mid-infrared ({\it W1}=3.4, {\it W2}=4.5, {\it W3}=12, and {\it W4}=22~$\mu$m) data in  the {\it WISE All-Sky Data Release} \citep{wise2012}. {\it Spitzer IRAC} data can be found for $345$ of our H$\alpha$ emission stars in \citet{megeath12}. Spectral types for $346$ stars can be found in \citet{dario2009},  \citet{rebull2000}, \citet{rebull2001} and \citet{hillenbrand1997}. 

$496$ of our stars are located within the area observed by the {\it XMM-Newton} X-ray observatory, and $257$ of them were detected. Their X-ray data can be found in the {\it XMM-Newton Serendipitous Source Catalogue 2XMM-DR3} \citep{xmm}. Rotation period measurements are available for $214$ stars in \citet{wherbst}, \citet{cieza2007}, and \citet{flaccomio2005}.

\section{ANALYSIS}
\label{sect_anal}

\subsection{Classification}
\label{sect_class}

Classical T Tauri stars (CTTS) are pre-main sequence stars with an accretion disk. Material from the circumstellar disk accretes onto the star via magnetic funnel flows. This accretion process causes broad, often asymmetric H$\alpha$ emission lines. A weak-line T Tauri star (WTTS) can be either a T Tauri star that evolved and no longer has much circumstellar material to accrete from or it is a T Tauri star that for some reason is not accreting much at the moment. Their H$\alpha$ emission lines are much weaker and the origin of the line is mostly chromospheric.
There are several methods to distinguish accreting and non-accreting young stars from each other. The most widely used method  \citep[e.g.][]{martin1998} is based on the equivalent width of H$\alpha$, and classifies as CTTS (accreting) each young star with EW(H$\alpha) > 10$~\AA. \citet{barrado2003} defined an empirical classification scheme  which takes into account the spectral type as well, while \citet{WB2003} distinguished accreting and non-accreting T Tauri stars based on the width of the H$\alpha$ emission lines. 	 

To distinguish accreting and non-accreting stars, we applied the criteria by \citet{barrado2003} for the $346$ stars having spectral types in the literature \citep{dario2009,rebull2000,rebull2001,hillenbrand1997}, and adopted {\it EW }= 13\,\AA\ as a boundary between accreting and non-accreting stars for the unclassified population, which is fainter on average than the classified subsample ($\langle J \rangle = 12.39$, and  $\langle J \rangle = 13.06$  for the classified and unclassified group, respectively). This EW value corresponds to the saturation limit of chromospheric activity at M3--M4 type. 

We list the $587$ sources in Tables~\ref{tabshort_ctts}, \ref{tabshort_wtts}, and \ref{tab_unc}. 
Table~\ref{tabshort_ctts} contains the $372$ stars classified as classical T~Tauri stars, Table~\ref{tabshort_wtts} lists the $187$ weak-line T~Tauri stars, and Table~\ref{tab_unc} contains the $28$ stars without measured equivalent width. Column~1 of each Table gives their {\it 2MASS\/} designation of the star. The newly identified H$\alpha$ emission stars are marked by an asterisk following their {\it 2MASS\/} identifier. Equivalent widths, measured on the dates given in the column head (yyyymmdd), are listed in Cols.~2--6, column~7 gives the {\it $r^\prime$} magnitudes, column~8 contains the {\it $r^\prime -$  $i^\prime$} color index, while the last column contains cross-identification from the {\it General Catalogue of Variable Stars} \citep{GCVS}, \citet{kiso}, \citet{hillenbrand1998}, and \citet{furesz}.
\newpage

\begin{deluxetable}{llllllllll}

\tabletypesize{\scriptsize}
\tablecolumns{8}
\tablewidth{0pt}
\rotate
\tablecaption{H$\alpha$ emission stars classified as classical T Tauri stars\label{tabshort_ctts}}
\tablehead{
\colhead{2MASS} & \multicolumn{5}{c}{EW (\AA)} &\colhead{$r^\prime$}&\colhead{$r^\prime-i^\prime$} &\colhead{Cross-identification}\\
\cline{2-6}
\colhead{}&\colhead{20101231}&\colhead{20110102}&\colhead  {20110125}& \colhead{20110227}&\colhead{20111016}&\colhead{}&\colhead{}&\colhead{}\\

}
\startdata 
 	J05333304$-$0511555	 								&								 	 								&								 	49.6	(4.9)	 								&								 	 								&								 	 								&																	&				16.3				&				0.46				&						  RZ Ori,  Kiso A-0976 45, \\&&&&&&&& [FHM2008] S1-ap75 \\	 
 	J05333443$-$0514177	 								&								 	 								&								 	22	(1.5)	 								&								 	 								&								 	 								&																	&				16.7				&			0.90			&						  [FHM2008] S1-ap78 \\	 
 	J05333588$-$0501324	 								&								 	 								&								 	148.5	(38.9)	 								&								 	 								&								 	 								&																	&				15.8				&				0.85				&						  VY Ori,  Kiso A-0976 50, \\&&&&&&&&   [FHM2008] S2-ap44 \\		 
 	J05333855$-$0513125	 								&								 	 								&								 	60.7	(4.3)	 								&								 	 								&								 	 								&																	&				17.0				&				0.96				&						  Kiso A-0976 51,  [FHM2008] F21-ap90 \\	 
 	J05334167$-$0524042	 								&								 	 								&								 	57.9	(3.1)	 								&								 	 								&								 	 								&																	&				16.4				&			0.60			&						 Kiso A-0976 52,  [FHM2008] F21-ap63 \\
 	J05334192$-$0506148	 								&								 	 								&								 	293.0	(177.9)	 								&								 	 								&								 	 								&																	&				20.3				&				1.39				&						  [FHM2008] S2-ap37 \\		 
 	J05334479$-$0514098	 								&								 	 								&								 	14.7	(1.1)	 								&								 	 								&								 	 								&																	&				13.8				&				0.35				&						  V729 Ori,  Kiso A-0976 53 \\	 
 	J05334493$-$0531085	 								&								 	 								&								 	44.7	(1.4)	 								&								 	 								&								 	 								&																	&				15.2				&				0.73				&						 V386 Ori \\
 	J05334525$-$0530498	 								&								 	 								&								 	7.7	(2.4)	 								&								 	 								&								 	 								&																	&				14.1				&				0.59				&						  VZ Ori,  Kiso A-0976 54 \\	 
 	J05334545$-$0536323	 								&								 	 								&								 	119 	(8.8)	 								&								 	 								&								 	 								&																	&				17.9				&				1.89				&						 V726 Ori, Kiso A-0976 55,  \\&&&&&&&&  [FHM2008] S1-ap42 \\
 
\enddata
\tablecomments{Table \ref{tabshort_ctts} is published in its entirety in the
electronic edition of the {\it Astrophysical Journal}.  A portion is
shown here for guidance regarding its form and content.}

\end{deluxetable}

\clearpage


\newpage

\begin{deluxetable}{lllllllll}

\tabletypesize{\scriptsize}
\tablecolumns{10}
\tablewidth{0pt}
\rotate
\tablecaption{H$\alpha$ emission stars classified as weak-line T Tauri stars \label{tabshort_wtts}}
\tablehead{
\colhead{2MASS} & \multicolumn{5}{c}{EW (\AA)} &\colhead{$r^\prime$}&\colhead{$r^\prime-i^\prime$}&\colhead{Cross-identification}\\
\cline{2-6}
\colhead{}&\colhead{20101231}&\colhead{20110102}&\colhead{20110125}& \colhead{20110227}&\colhead{20111016}&\colhead{}&\colhead{}&\colhead{}\\

}
\startdata
J05333705$-$0523069*									&									10.1 (4.3)									&																		&																		&																		&																			&				16.4				&				0.73				&								V725 Ori \\					
	J05334954$-$0536208											&																						&													4.0 (0.3)													&																						&																						&																					&				16.5				&				0.62				&									V354 Ori,  Kiso A-0976 63,\\&&&&&&&& [FHM2008] S3-ap60 \\					
	J05335210$-$0530284											&																						&													7.2 (1.0)												&																						&																						&																					&				16.5				&				1.16				&									[FHM2008] F21-ap53\\						
	J05340797$-$0536170											&																						&													7.8 (0.2)													&																						&																						&																					&				18.4				&				1.65				&									V396 Ori, Kiso A-0976 74, \\&&&&&&&&	[FHM2008] S1-ap33 	\\					
	J05340835$-$0514387*											&																						&										$<$ 2.7												&																						&																						&																					&				18.2				&				1.31				&									\\							
	J05341714$-$0538168											&											5.4 (1.2)											&											3.0 (2.2) 											&																						&																						&																					&				16.0				&			1.18				&										Kiso A-0976 85, [FHM2008] F22-ap240,\\&&&&&&&& [H97b] 10075  \\								
	J05342616$-$0526304											&											$>$ 5.7										&																						&																						&																							&																					&				14.2				&				0.37				&									V1956 Ori, [FHM2008] S2-ap239, \\&&&&&&&&	 [H97b] 3109  \\							
	J05342650$-$0523239											&										11.4 (1.1)										&																				&																						&																							&																					&				16.4				&				1.25				&									[FHM2008] F22-ap24, [H97b] 3080    \\							
	J05342751$-$0528284											&										5.3 (0.6)										&																				&																						&																							&																					&				14.9				&				0.70				&									 [FHM2008] F21-ap54, [H97b] 3120 \\							
	J05342960$-$0547247											&																						&												9.4 (0.7)												&																						&																						&																					&				16.1				&				0.69				&									V935 Ori,  [FHM2008] F31-ap81\\							

\enddata
\tablenotetext{*}{indicates sources, not detected by previous  H$\alpha$ surveys.}
\tablecomments{Table \ref{tabshort_wtts} is published in its entirety in the
electronic edition of the {\it Astrophysical Journal}.  A portion is
shown here for guidance regarding its form and content.}

\end{deluxetable}

\clearpage

\newpage

\begin{deluxetable}{ll}
\tabletypesize{\scriptsize}
\tablecolumns{7}
\tablewidth{0pt}
\tablecaption{H$\alpha$ stars without classification\label{tab_unc}}

\tablehead{
\colhead{2MASS}   &\colhead{Cross-identification}\\
\colhead{}   &\colhead{}\\

}
\startdata

				J05335074$-$0500394	 			&				  V1902 Ori,  [FHM2008] F11$-$ap39 \\	
				J05341019$-$0505155*						&		 \\	
				J05343025$-$0517012			&			 [H97b] 12\\
				J05343822$-$0524236		&			 V1997 Ori, 	[FHM2008] F21$-$ap62, [H97b] 40 \\
				J05344197$-$0545224*								&		V2009 Ori\\	
				J05344576$-$0543439*							&		V770 Ori  \\
				J05344922$-$0504380*	 				&			 	 			\\
				J05351121$-$0517209							&		V2214 Ori, [FHM2008] S1$-$ap86,  [H97b] 358\\	
				J05351216$-$0530201							&		V1276 Ori,	[FHM2008] S2$-$ap218,  [H97b] 379 \\	
				J05351376$-$0539100									&		LY Ori, [H97b] 433 \\
				J05351405$-$0519520							&		V2252 Ori,	[FHM2008] S1$-$ap57, [H97b] 429\\	
				J05351463$-$0516461						&		[FHM2008] S1$-$ap93, [H97b] 444,  [H97b] 20475  \\				
				J05351534$-$0519021							&		[FHM2008] F21$-$ap148,	[H97b] 469, [H97b] 20341\\						
				J05351606$-$0520363								&		V2280 Ori, [H97b] 504b,  [H97b] 9125  \\
				J05352021$-$0520569										&		TU Ori, [H97b] 640 \\	
				J05352425$-$0525186				&			V2421 Ori, 	[FHM2008] S1$-$ap231, [H97b] 750 \\							
				J05352469$-$0524357							&		 V1400 Ori, [H97b] 762 \\									
				J05352505$-$0522585									&		 [H97b] 766a \\	
				J05352547$-$0534028							&			[FHM2008] S3$-$ap197,  [H97b] 781 \\
				J05352699$-$0548460*							&		V2458 Ori \\					
				J05352986$-$0523484									&		 [H97b] 845 \\					
				J05353340$-$0539212					&	 [H97b] 10702 \\
				J05353461$-$0515528							&		[FHM2008] S2$-$ap181, 	 [H97b] 903 \\					
				J05354002$-$0502370						&		\\		
				J05355408$-$0537423		&			 Kiso A$-$0976 219 , [H97b] 2271 \\	
				J05361618$-$0518560*										&		[H97b] 5069 \\		
				J05363761$-$0531189*							&			\\			
				J05363404$-$0534054									&		 [FHM2008] F21$-$ap217 \\

\enddata

\tablenotetext{*}{* indicates sources, not detected by previous  H$\alpha$ surveys. }
\end{deluxetable}

\clearpage
\subsection{Discriminant Analysis}
\label{sect_stat1}

To test whether our WTTS and CTTS samples indeed represent physically different objects, we applied a discriminant analysis. It is a common ingredient of professional statistical packages. We used {\it SPSS\/}\footnote{SPSS is a registered trademark. Website: www-01.ibm.com/software/analytics/spss/}  in the computations. Discriminant analysis aims to find differences between groups in the multivariate parameter space, orders membership probabilities and one may use this scheme for classifying additional cases not having assigned group memberships. During this process we used those variables  in which a significant number of cases remain after removing missing data. To fit with this requirement we used the {\it I}$-${\it J}, {\it J}$-${\it H}, {\it H}$-${\it K}$_\mathrm{s}$, {\it K}$_\mathrm{s}-$[W1], [W1]$-$[W2], [W2]$-$[W3], and [W3]$-$[W4] color indices, available for 241 of the 372 CTTS and 134 of the 187 WTTS. Using the analysis we look for the linear combination of the selected color indices, the {\em discriminant variable\/}, which gives maximal separation between the groups of the cases. 

The program calculated the means and standard deviations of quantities included in the analysis, along with the number of cases having valid measurements in all of the variables used in the computation. The results are shown in  Table~\ref{disc_anal}. A quick look at the Table gives an impression that both the means and standard deviations are greater in the CTT group. The statistical tests show that, except for [W2]$-$[W3]  and [W3]$-$[W4],  the difference between the WTT and CTT groups is highly significant. 

\begin{deluxetable}{lllll}

\tabletypesize{\scriptsize}
\tablecolumns{4}
\tablewidth{0pt}
\tablecaption{The data used for discriminant analysis\label{disc_anal}}
\tablehead{
\colhead{Type} & \colhead{Quantity}  &  \colhead{Mean}  & \colhead{Std. Dev.}& \colhead{Number of stars} \\

}
\startdata

WTT   &  I$-$J  &   1.67  &  0.31& 134 \\
      &  J$-$H & 0.74 & 0.13 & 134\\
      &  H$-$K$_\mathrm{s}$ & 0.30 & 0.12&134  \\
      &  K$_\mathrm{s}$$-$[W1] & 0.29 & 0.24&134  \\
      &  [W1]$-$[W2] & 0.26 & 0.20 & 134\\
      &  [W2]$-$[W3] & 2.66 & 1.20 &134 \\
      &  [W3]$-$[W4] & 3.15 & 1.11  &134\\
\cline{1-5}
CTT  & I$-$J & 1.89 & 0.52&  241\\
     & J$-$H & 0.89 & 0.24 & 241\\
     & H$-$K$_\mathrm{s}$ & 0.49 & 0.21 & 241\\
     & K$_\mathrm{s}-$[W1] & 0.66 & 0.39 &241 \\
     & [W1]$-$[W2] & 0.54 & 0.23 & 241\\
     & [W2]$-$[W3] & 2.80 & 1.01 & 241\\
     & [W3]$-$[W4] & 3.17 & 1.42 & 241\\
\enddata
\end{deluxetable}

We found that the strongest contribution to the discriminant variable is given by [W1]$-$[W2], while those of [W2]$-$[W3]  and [W3]$-$[W4]  are negligible. From an astrophysical point of view this result can be interpreted in the following way. The spectral energy distributions (SEDs) of both the WTT and CTT classes are built up as a superposition of the photospheric and circumstellar thermal emission over the wavelengths covered by the near and mid infrared colors. At shorter wavelengths the contribution of the photospheric emission is the strongest and the opposite is true at the other end of the wavelengths domain of the input variables. The wavelengths of {\it W1} and {\it W2} are between these extremes. The larger average {\it I}$-${\it J} and {\it J}$-${\it H} color indices of the CTTS group possibly  indicate greater extinction suffered by these stars, since infrared excesses are less significant in these colours. 

Based on the canonical discriminant function (the discriminant variable) resulting from the analysis one can control the reliability of the original (a-priori) classification based on the equivalent width of the H$\alpha$ emission line. The histograms of the discriminant function for the WTT and CTT groups demonstrate that these groups are not disjoint, but rather overlapping with respect to zero, the a-posteriori boundary between the two types of T~Tauri stars. We found that $25$ of our $187$ WTTS have IR colors similar to CTTS, and $90$ of the $372$ CTTS exhibit infrared properties similar to WTTS.

The discriminant function allowed us to estimate the probable types of the $28$ unclassified stars. Although, the {\it 2MASS\/} ({\it J,H,K$_\mathrm{s}$}) colors of the unclassified sources are complete, a significant fraction of the {\it WISE\/} measurements are missing. The {\em Expectation and Maximization} (EM) algorithm of the {\em Missing Value Analysis} procedure, implemented in the {\it SPSS} statistical package, was used to generate a maximum likelihood estimation for the missing values. The discriminant analysis suggests that $13$ stars from the $28$ unknown sources are weak-line T Tauri stars, while the remaining $15$ sources can be classified as classical T Tauri stars.

\subsection{Comparison with Previous H$\alpha$ Emission Surveys}
\label{sect_comp}

\begin{figure}[!ht]
\begin{center}
\includegraphics[scale=0.8]{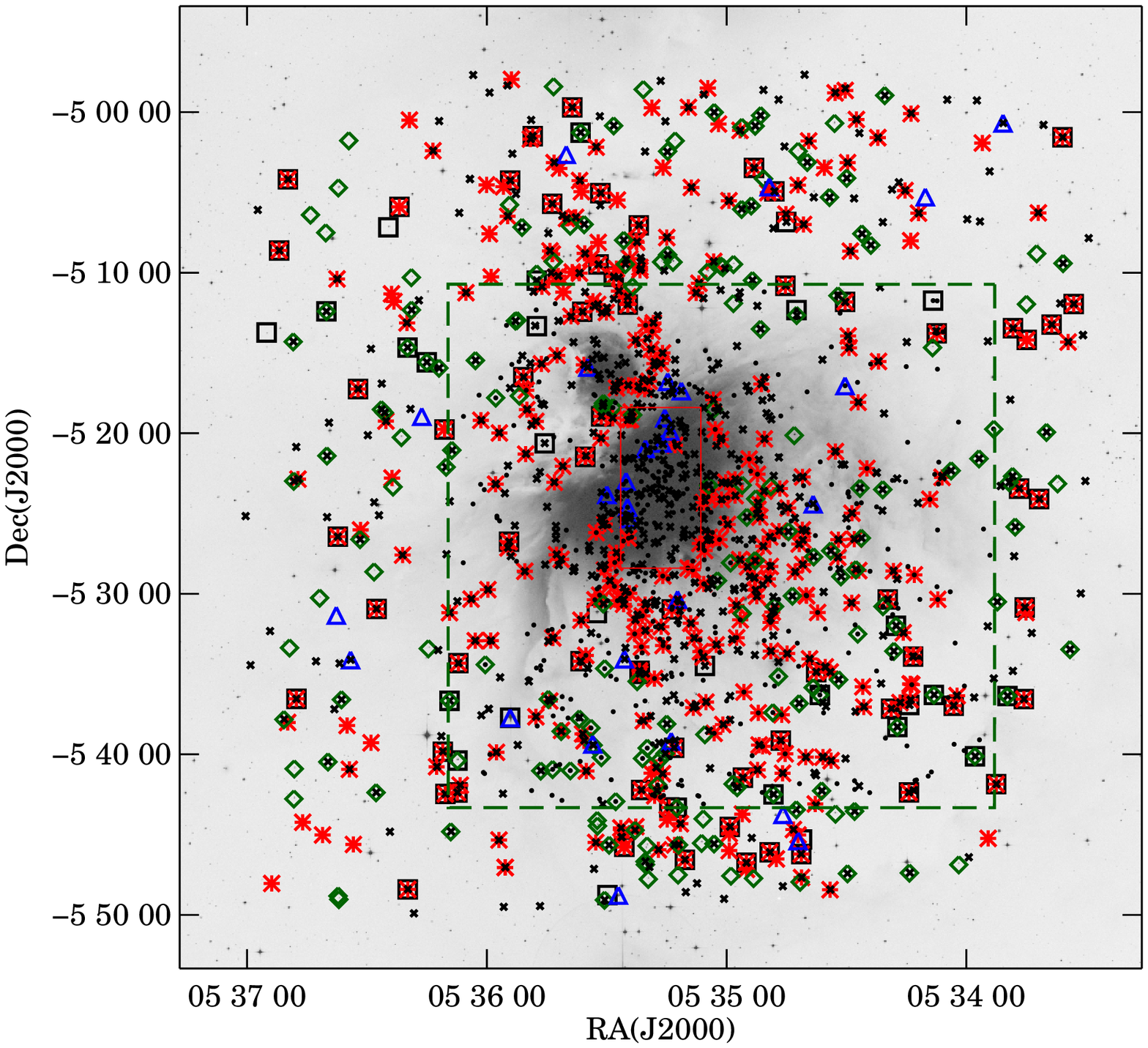}\vspace{1cm}
\figcaption{Distribution of the H$\alpha$ emission stars detected by the present survey (symbols are same as in Fig.~\ref{fig_onc}), by \citet{kiso} (squares), \citet{furesz} (crosses), and \citet{dario2009} (dots), overplotted on a $1$ square degree {\it DSS2 infrared} image centered on the Trapezium. The area bordered by the dashed square (green in the online version) was the target of \citet{dario2009}. The size of the small rectangle at the center of the map (red in the electronic version) is $5\arcmin \times 10\arcmin$, and indicates the nebulosity-dominated region in our {\it WFGS2\/} images.
\label{fig_comp}}
\end{center}
\end{figure}

We compare our sample of H$\alpha$ emission stars with those detected by the Kiso survey \citep{kiso}, \citet{furesz}, and \citet{dario2009} in Fig.~\ref{fig_comp}. The Kiso survey and the survey by \citet{furesz}
covered the whole area of our survey, while the photometric H$\alpha$ survey of \citet{dario2009} covered just the central $34\arcmin \times34\arcmin$. Due to the bright nebulosity we could barely detect sources in the densest central region of the ONC, and for the few detected sources EW(H$\alpha$) could not be determined, therefore comparison with previous H$\alpha$ surveys is justified only outside the central area of $\sim$~$5\arcmin$ in right ascension and $\sim$~$10\arcmin$ in declination. 

Based on the {\it VizieR} database, 92 H$\alpha$ stars identified by the Kiso survey are located in the area covered by our observations. Out of these 92 stars 79 show emission in our images. Comparison of these two surveys suggests that most of the stars detected by \citet{kiso} were strong emitters. 763 stars observed by \citet{furesz} are located in the area of our survey. Among those we detected H$\alpha$ emission in 337 stars. A detailed comparison shows that, outside the central region, we detected most stars for which \citet{furesz} measured EW(H$\alpha$) $> 10$\,\AA, and their low-EW stars, when detected, appeared as WTTS in our survey as well. 

274 of the 638 H$\alpha$ emission stars identified by \citet{dario2009} coincide with our emission stars. Figure~\ref{fig_comp} shows that most of our H$\alpha$ emission stars, located within the area observed by \citet{dario2009}, are found in their data base. Those of their H$\alpha$ stars that were missed by our survey are either located within the central $5\arcmin \times 10\arcmin$ area, or are fainter than $V \sim 19$~mag.

\begin{figure}[!ht]
\begin{center}
\includegraphics[scale=0.8]{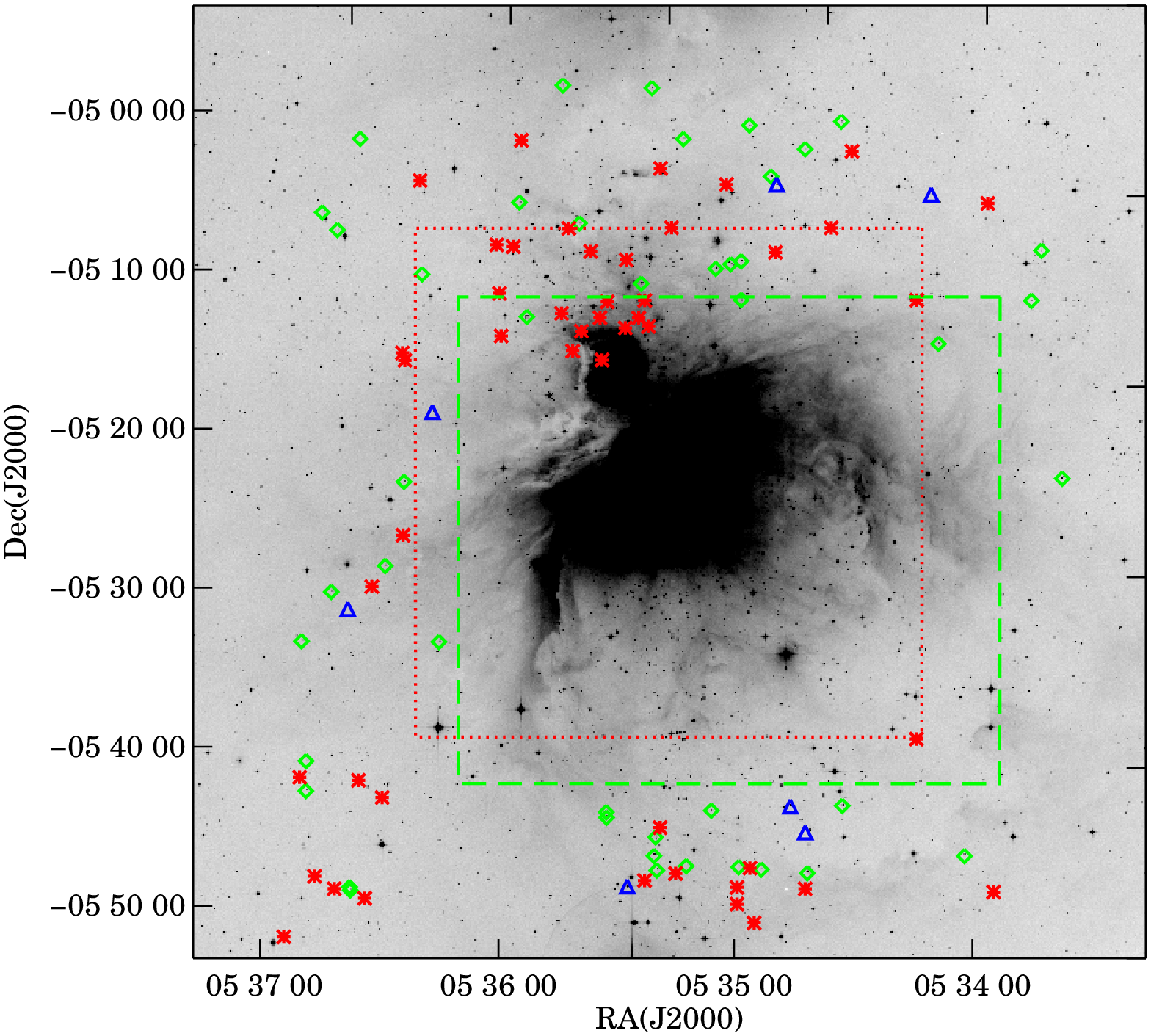}\vspace{1cm}
\figcaption{The newly identified H$\alpha$ emission stars plotted on the {\it DSS2 infrared} image of the ONC. Symbols are same as in Fig.~\ref{fig_onc}. The dotted square (red in the online version) indicates the boundary of the region studied by \citet{hillenbrand1997}, and the area bordered by the dashed square (green in the online version) was the target of \citet{dario2009}.
\label{fig_kimaradt}}
\end{center}
\end{figure}

\begin{figure}[!ht]
\begin{center}
\includegraphics[scale=0.8]{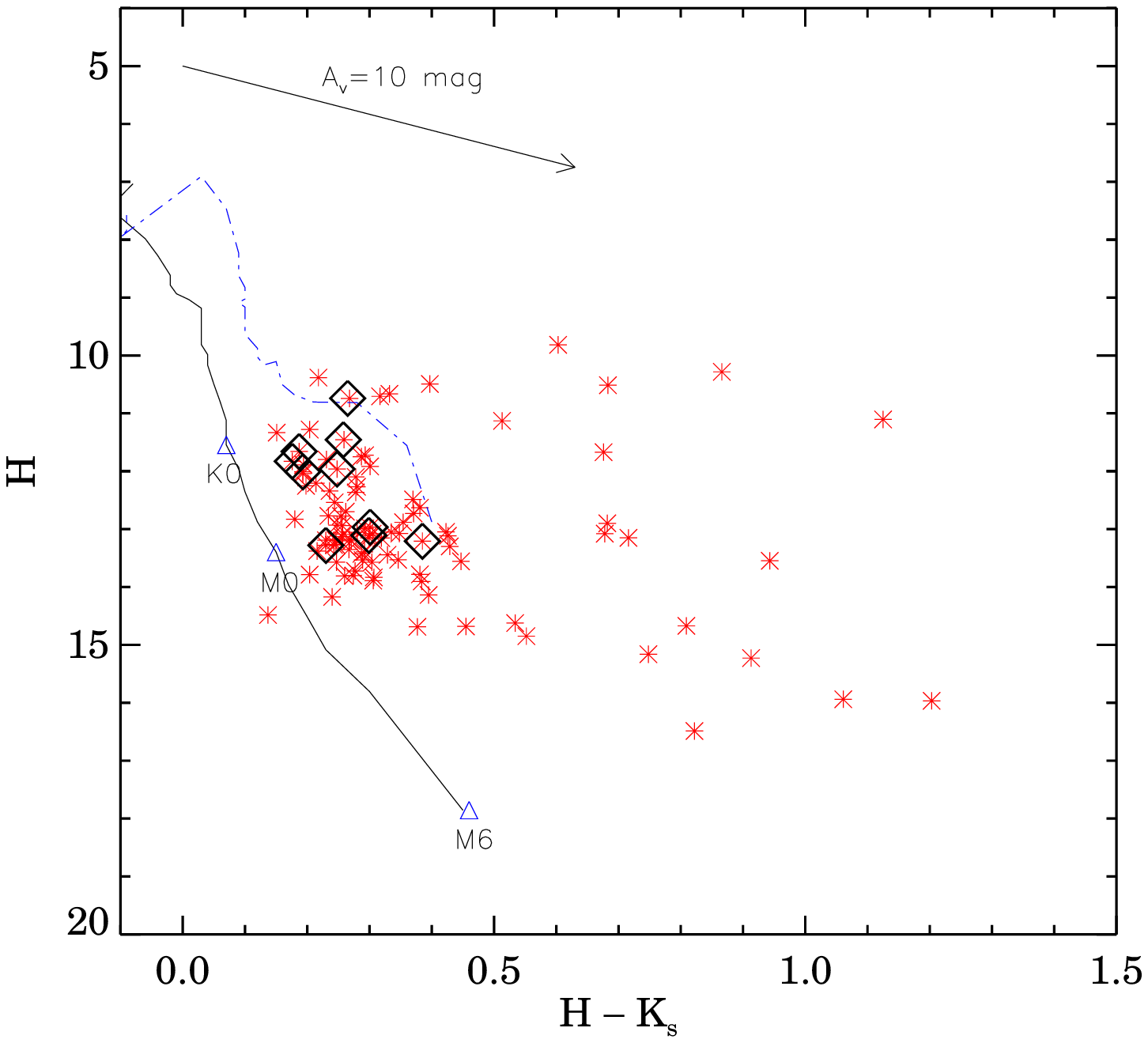}\vspace{1cm}
\figcaption{The {\it H\/} vs. $H-K_\mathrm{s}$ color-magnitude diagram of the newly identified H$\alpha$ emission stars. Dash-dotted line represents the $10^6$-year isochrone, and the solid line indicates the zero-age main sequence of the pre-main sequence evolutionary model of \citet{siess}. The arrow shows the direction of the interstellar extinction. Diamonds indicate the positions of the known ONC members \citep{hillenbrand1998} of the sample.
\label{fig_color_mag}}
\end{center}
\end{figure}

 $99$ of our H$\alpha$ emission stars were detected by neither of the above surveys nor by earlier surveys. The distribution of these stars can be seen in Fig.~\ref{fig_kimaradt}. Most of them are located outside the central, thoroughly studied region of the ONC. Forty-eight stars belong to the CTTS, $44$ to the WTTS, while the remaining $7$ stars to the third group of uncertain nature. Several of them are known variable stars or cluster members. In particular, 15 of the near-infrared variables studied by \citet*{Carpenter} appear in this sample. Fig.~\ref{fig_color_mag} shows the {\it H\/} vs. $H-K_\mathrm{s}$ color-magnitude diagram of the newly identified H$\alpha$ emission line stars, based on the {\it 2MASS\/} data. Diamonds indicate M type cluster members identified by \citet{hillenbrand1998}. The diagram suggests that the new emission line stars may be pre-main sequence stars located at the
distance of the ONC.

\subsection{Comparison with Other ONC Surveys}
\label{sample_onc}

The most complete census of the disk-bearing population of the ONC can be found in the {\it Spitzer\/} data base \citep{megeath12}, whereas the diskless young stars  down to 0.1-0.2 M$_\sun$ were identified by the {\it Chandra Orion Ultradeep Project} X-ray survey \citep{preibischa,preibischb}. To test whether our H$\alpha$ emission stars represent a typical subsample of the cluster population or not, we examined the number and ratio of the infrared and X-ray sources in the $17\arcmin \times 17\arcmin$ field of view of {\it Chandra}, excluding the central $5\arcmin \times 10\arcmin$ region where the nebulosity prevented us from detecting emission line stars. There are 758 X-ray-selected young stars in this area, 369 (48.7\%)  of which are IR-selected disk-bearing stars. We detected 101 H$\alpha$ emission stars, some 13 \% of the whole known young stellar population in the same area. This sample contains 86 CTTSs and 15 WTTSs. Our data show an apparent CTTS fraction of 85\%, indicating that our survey, as expected, is biased toward disk-bearing stars.

\begin{figure}[!ht]
\begin{center}
\includegraphics[scale=0.8]{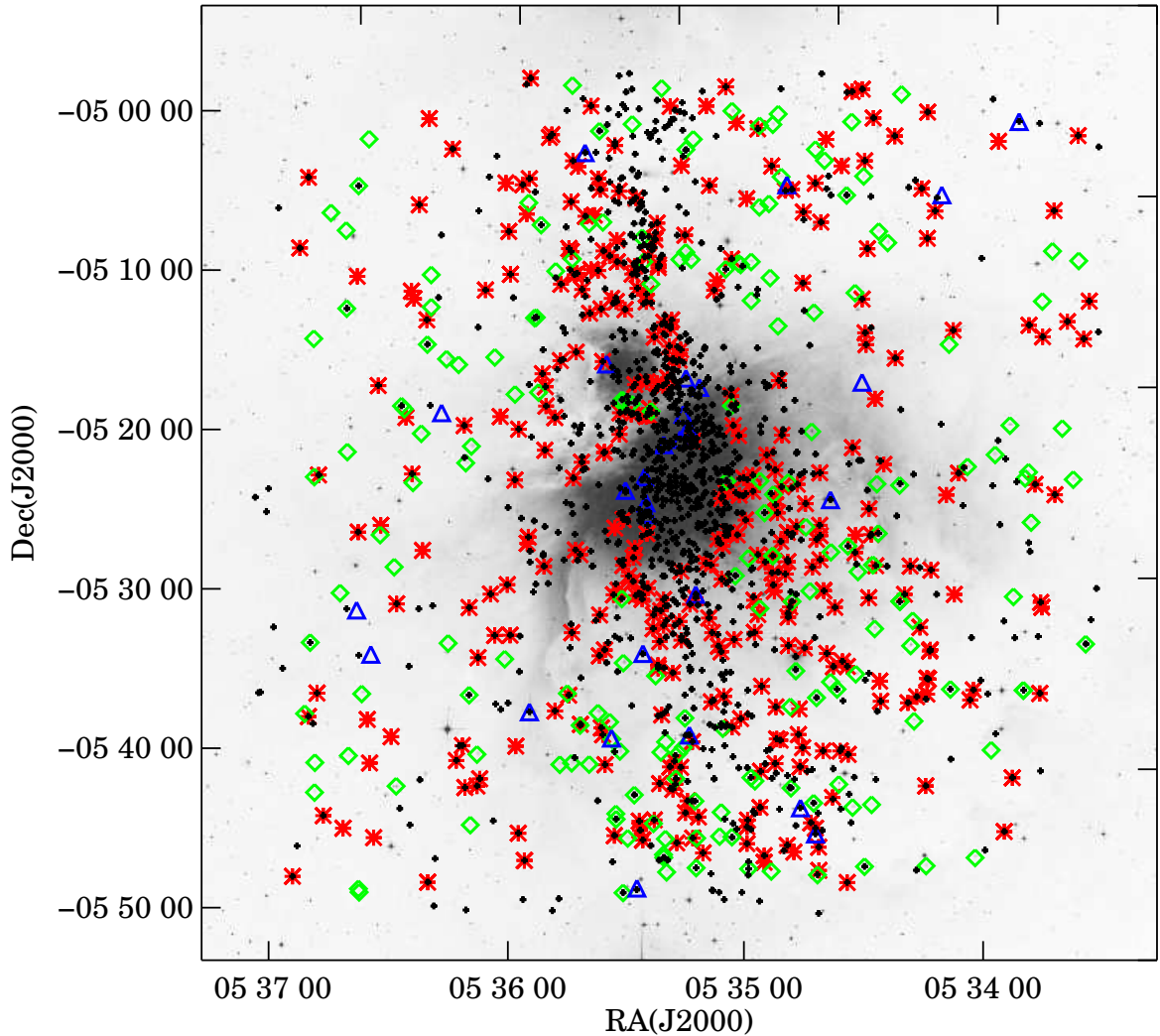}\vspace{1cm}
\figcaption{Surface distribution of our H$\alpha$ emission stars (symbols are same as in Fig.~\ref{fig_onc}) and the {\it Spitzer\/} sources identified as young stars by \citet{megeath12} (small plusses). 
\label{fig_megeath}}
\end{center}
\end{figure}

Since both infrared excess and H$\alpha$ emission in young stellar objects are disk indicators, overlapping of the samples of H$\alpha$ emission star and the infrared-excess stars of the same region is expected. In Fig.~\ref{fig_megeath} the surface distribution of our stars is displayed together with the mid-infrared excess stars identified by \citet{megeath12}, based on {\it Spitzer\/} observations. It can be seen that numerous sources are in common, but of course there are sources which were detected in only one of the surveys.  
Both Figs.~\ref{fig_comp} and \ref{fig_megeath} show that outside the cluster core our results overlap well with those of other surveys, and reveal new cluster / association members in the outer, lower density regions of ONC. Although the $r^\prime$ vs. $r^\prime-i^\prime$ color-magnitude diagram (Fig.~\ref{fig_cmd}) indicates a limiting magnitude of $r^\prime \sim 20$~mag, Fig.~\ref{fig_khist} shows that our sampling starts to be fairly incomplete beyond $r^\prime \sim$ 16--17~mag.

\subsection{Surface Distribution of the H$\alpha$ Emission Stars}
\label{sect_surf.dist}

A look at the surface distribution of the H$\alpha$ emission stars in Figure~\ref{fig_onc} gives the impression of clustering of the stars. The surface distribution of the merged YSO samples (Figs.~\ref{fig_comp} and \ref{fig_megeath}) also suggests the presence of small apparent stellar groups in the outer regions of the field, sampled by our survey. We examined the reality of the clusterings in our sample by an appropriate statistical procedure. To test our impression we generated a fully random case  with the same numbers of CTTS and WTTS as in the real case, and used the {\it Fast Nearest Neighbor Search Algorithms and Applications} (FNN) package of the {\it R} project\footnote{www.r-project.org} to prove the reality of the clustering in the surface distribution. We compared the cumulative distributions of the nearest neighbor distances in the real and random samples applying a Kolmogorov--Smirnov test. The results of KS statistics indicate that the distribution of the H$\alpha$ emission stars in the observed area indeed shows some clustering with respect to the fully random surface distribution.

To study the nature of the clustered and distributed population and reveal the possible difference between their measured properties we first defined clusters in the surface distribution. Based on the maximum departure of the real and random distribution we assigned an object to a cluster if its 4th nearest neighbor had a distance of $d < 130\arcsec$. Comparison of the cluster members and outliers have shown that the mean brightness of the cluster stars is greater at all wavelengths. The difference is increasing toward the longer wavelengths. It is about $0.5$~mag in the NIR and reaches about 1~mag at {\it W4}. For testing the significance of the difference between the corresponding groups we applied a statistical T-test. According to the applied T-test these differences are highly significant. In contrast, however, there is no significant difference in EW(H$\alpha$).  

Regarding the cluster members, a T-test demonstrates that the difference between the CTTS and WTTS in the cluster environment is smaller than the standard error of the mean in the {\it K}$_\mathrm{s}$ brightness. In the {\it J\/} band, however, the CTT class is fainter. This may be accounted for by the greater extinction of the circumstellar matter, or alternatively, may indicate that the weak H$\alpha$ emission of the fainter WTTS population remained undetected.

\subsection{Variability of The Equivalent Width of H$\alpha$ Line}
\label{sect_var}

\begin{figure*}
\centerline{
\includegraphics[width=7cm]{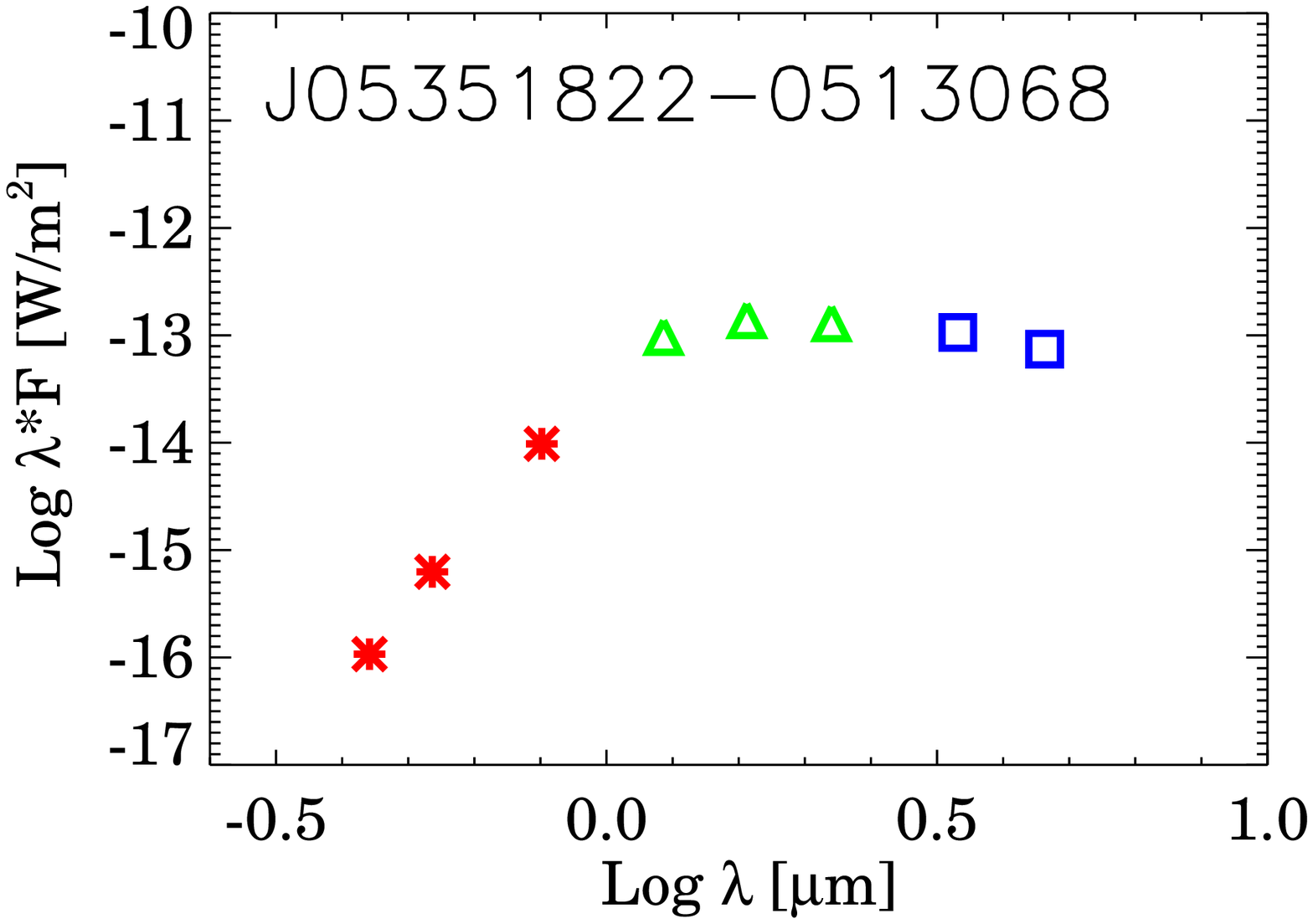}\hspace{1cm}\includegraphics[width=7cm]{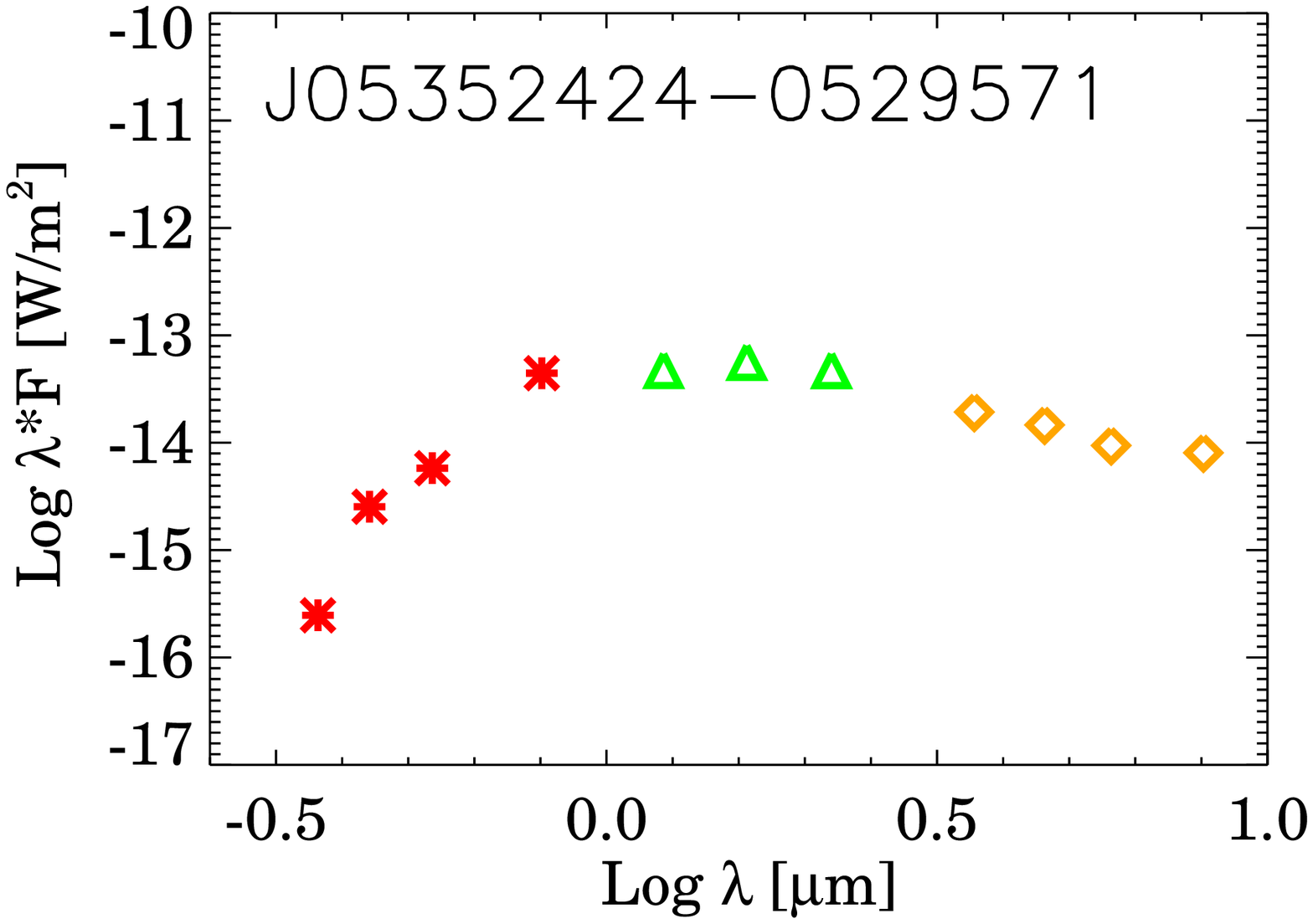}}
\centerline{
\includegraphics[width=7cm]{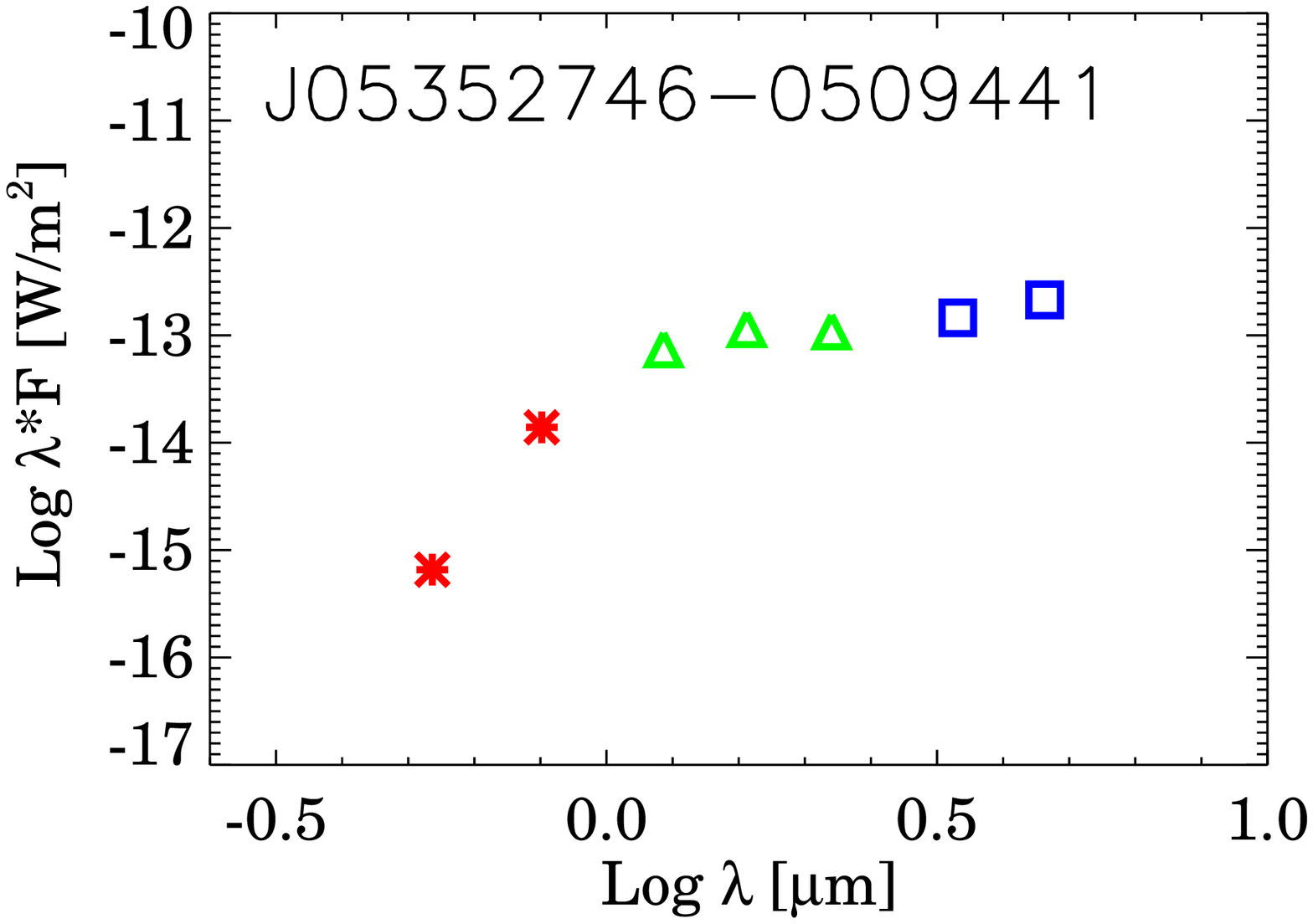}\hspace{1cm}\includegraphics[width=7cm]{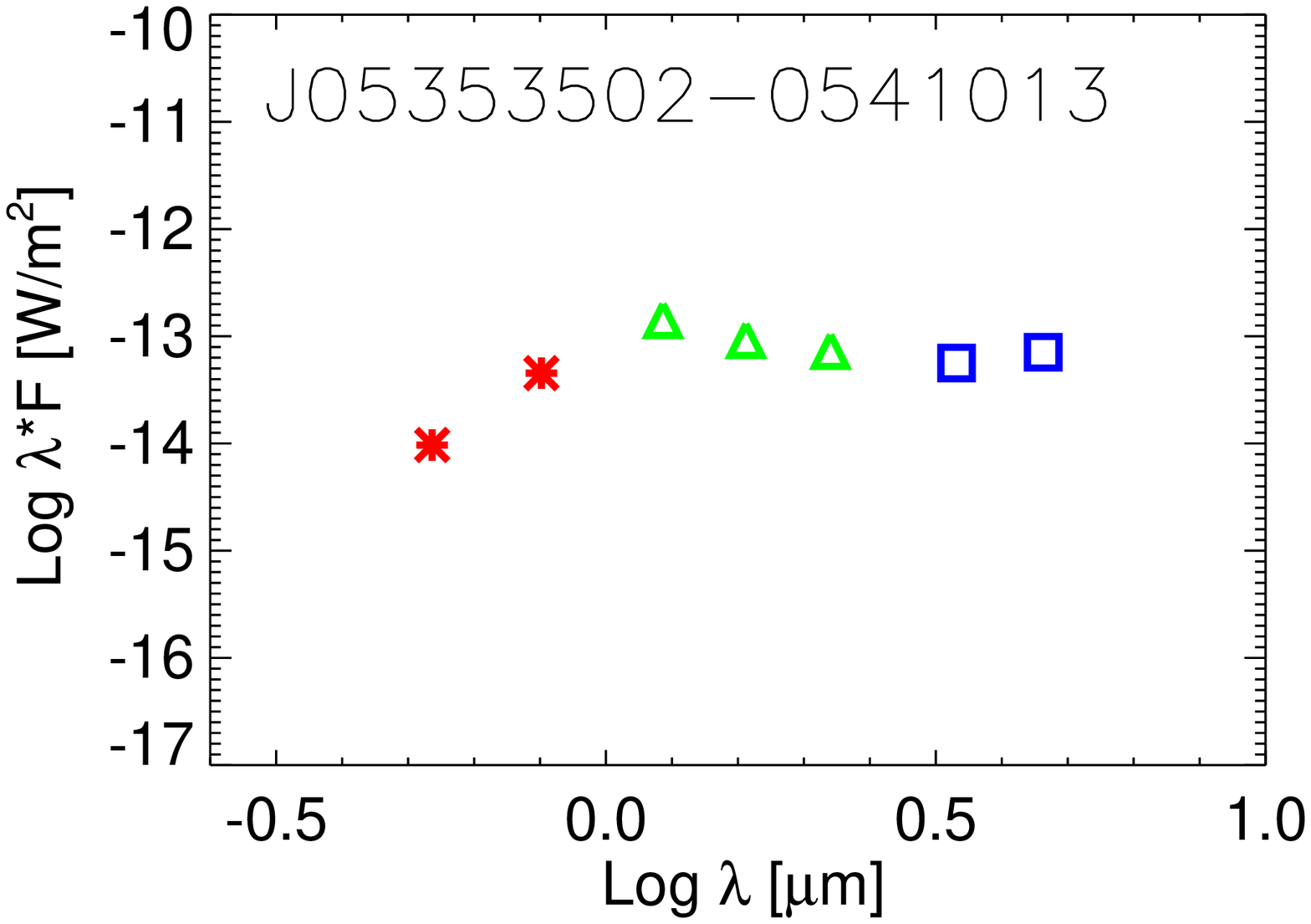}}
\centerline{
\includegraphics[width=7cm]{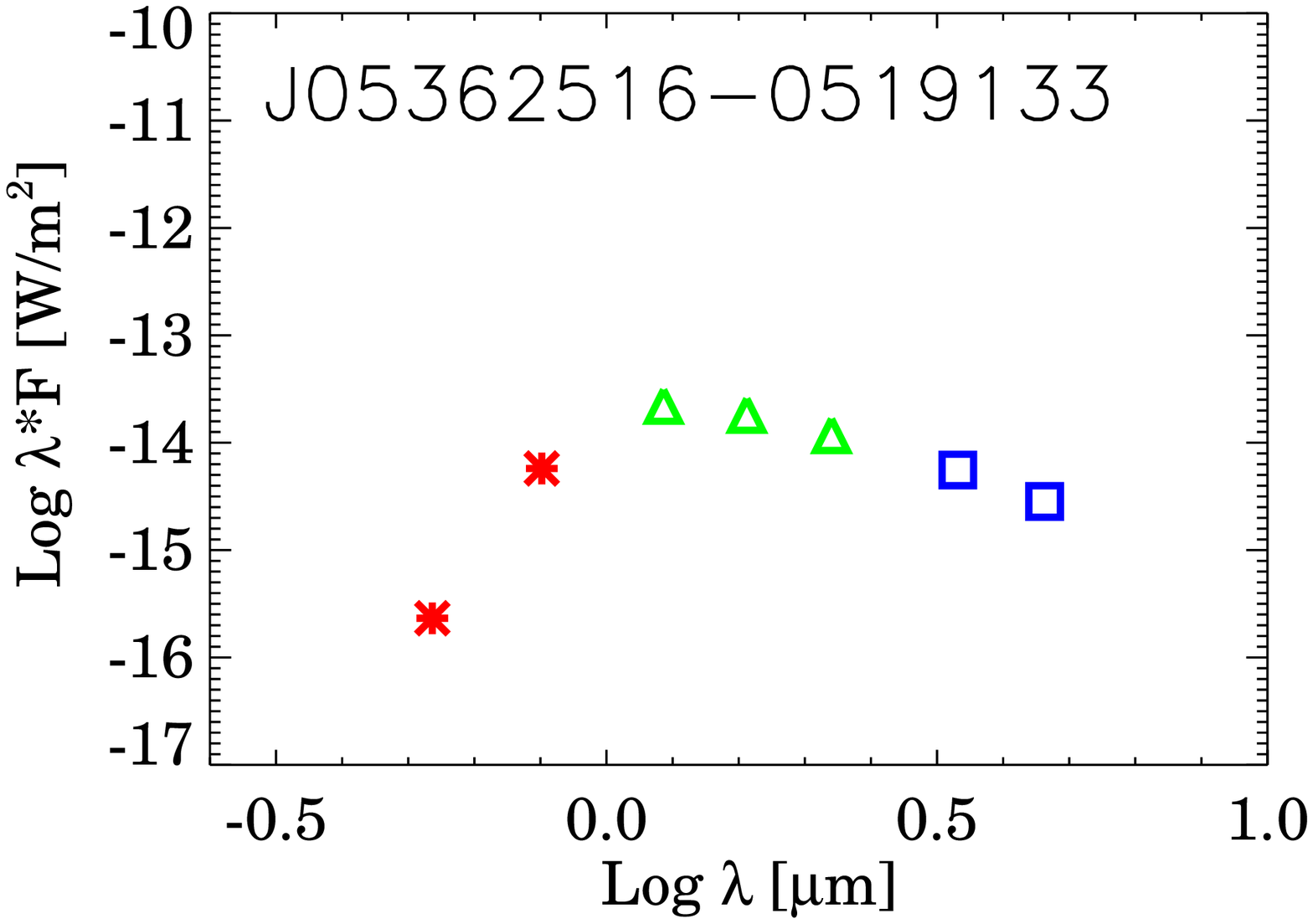}\hspace{1cm}\includegraphics[width=7cm]{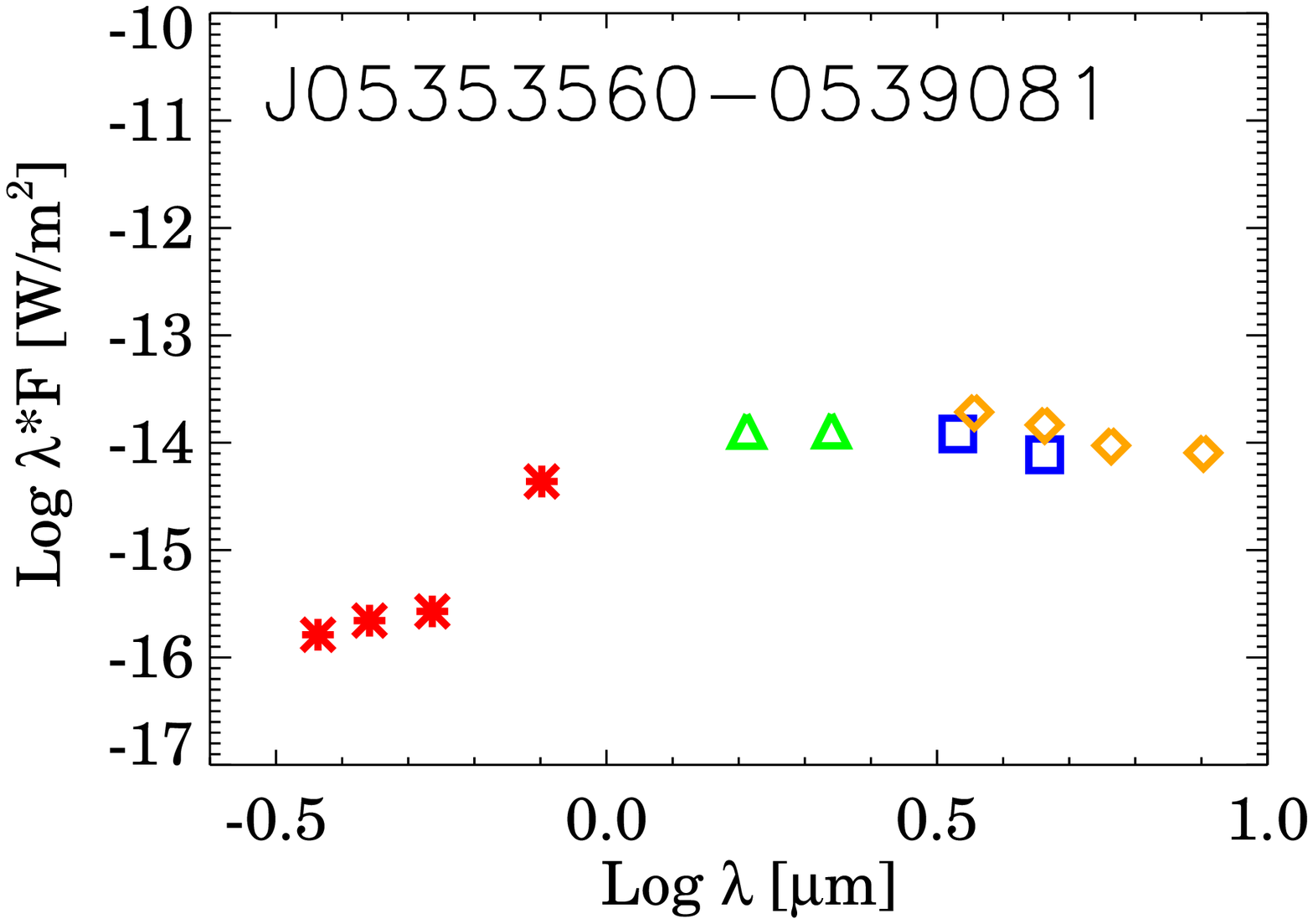}}
\caption{Spectral energy distribution for stars with very short timescale changes in EW. Optical data, asterisks (red in the online version) are from \citet{dario2009}, {\it 2MASS} data, triangle (green in the online version) from \citet{2mass} and {\it WISE All-Sky Data Release}, square (blue in the online version) are from \citet{wise2012}. {\it Spitzer } data, diamonds (yellow in the online version) can be found in \citet{megeath12}.  
\label{fig_varsed}}
\end{figure*}

The strength and shape of the H$\alpha$ emission line can show temporal variability due to variable accretion rate and wind. EW(H$\alpha$) can change even in the absence of flux variations, due to the photometric variations of the stellar continuum. The subsample of H$\alpha$ sources observed several times due to overlapping of the image fields allowed us to examine the short time scale (hours to months) variability. Seven stars were observed twice within a few hours. They showed significant short time scale variations in the H$\alpha$ equivalent width. Of these seven stars, six possess flat or rising (Class~I) spectral energy distribution (Fig.~\ref{fig_varsed}). This suggests that these stars are very young and may still be surrounded by an envelope, and this can lead to significant variability of the accretion rate and accretion-related wind. 
Nine of the 11 stars which were observed twice or more within a few days, showed no significant variation, and the remaining two ones exhibited small (some $30$ percent) variations in EW(H$\alpha$).
Stars observed on longer timescales (some weeks or months) show a wide range of variations in EW(H$\alpha$)  (see Table~\ref{tabshort_ctts}).

\begin{figure}
\centerline{
\includegraphics[width=6cm]{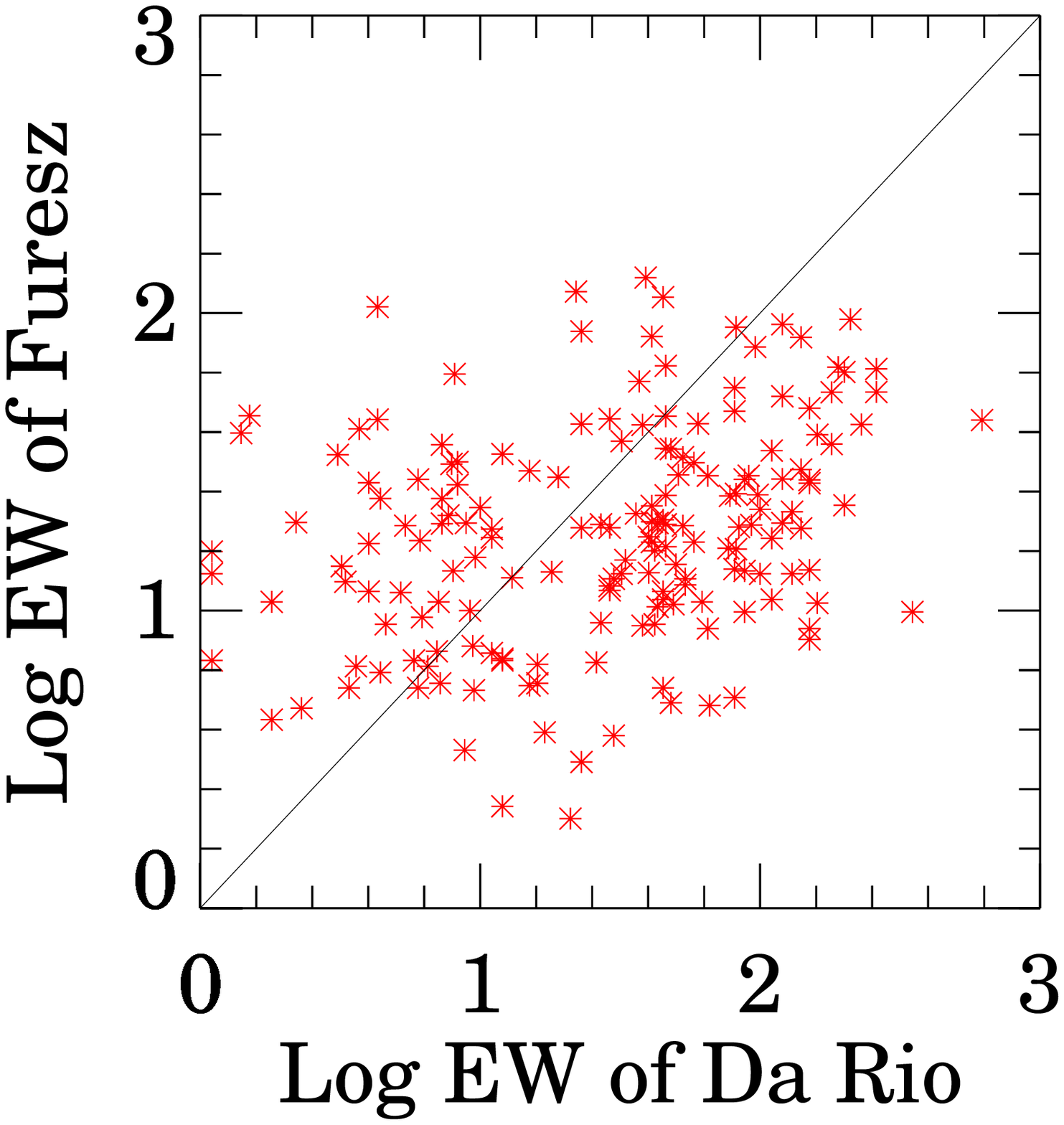}\hspace{0.5cm}\includegraphics[width=6cm]{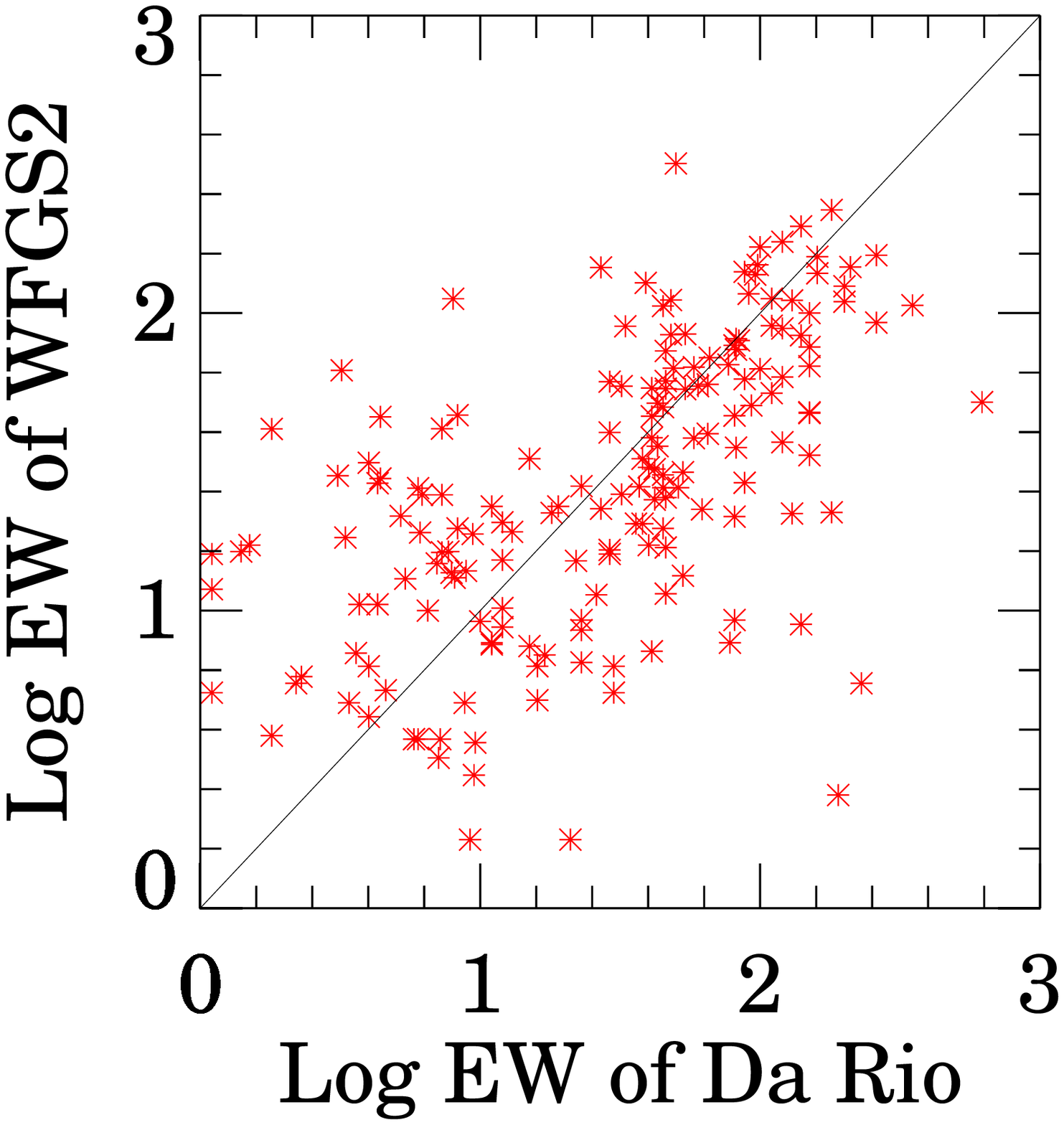}\hspace{0.5cm}\includegraphics[width=6cm]{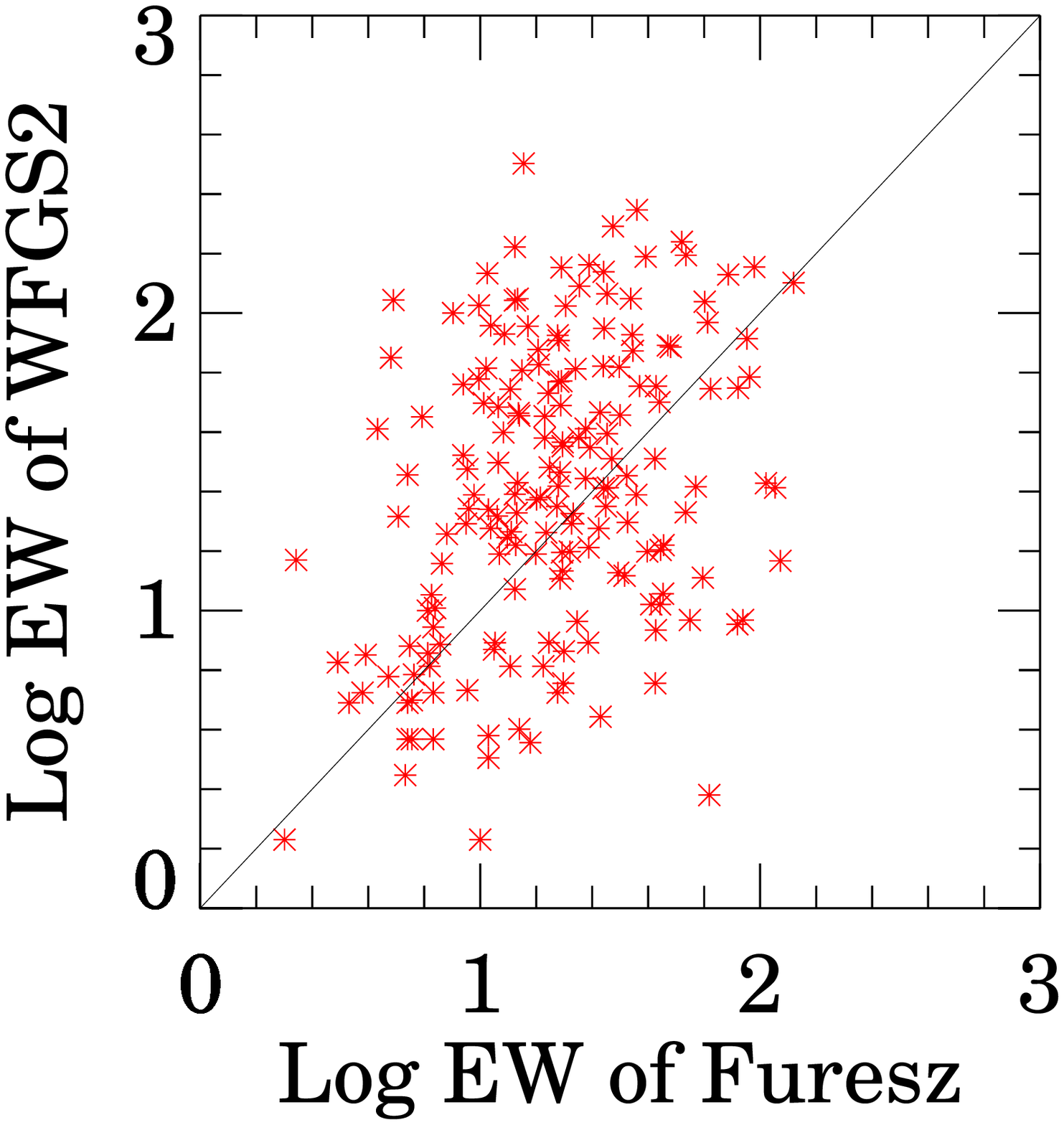}}
\vspace{0.5cm}
\caption{Comparison of the log EW(H$\alpha$) values measured with different methods, for the stars common in the H$\alpha$ surveys
of \citet{furesz}, \citet{dario2009} and the present survey.  \label{fig_method}}
\end{figure}

We compared the derived equivalent widths with those published  by \citet{furesz} and \citet{dario2009}. This comparison allowed us to study the changes of the equivalent widths on time scales of a few years. Part of the observed differences may originate from the fact that the equivalent widths in the compared works were determined with different methods. To check this point, we selected all of the sources common among these three surveys (Fig.~\ref{fig_method}) and calculated the significance of correlations using Kendall's tau statistics. Based on the statistics we find that the surveys of \citet{furesz} and \citet{dario2009} are less correlated, while our survey and the survey of \citet{dario2009} are well correlated. The comparison between the measurements of \citet{furesz} (high resolution) and our survey (low resolution) suggest a quite  good correlation at lower EW values.

The results are presented in Figs. 12 and 13. The histogram with upward diagonal hatching (red in the online version) represents the CTTS, while the histogram with horizontal hatching symbols the WTTS. It can be seen that the amplitudes of variations are  moderate in the greater part of the sample:  most of the observed changes in EW are within a factor of 2--3,  whereas changes as high as 10--20-fold appear in a few cases. The greater EW ratios in Fig. 13 reflect the lower level of correlation between these data sets. 

\begin{figure}[!ht]
\begin{center}
\includegraphics{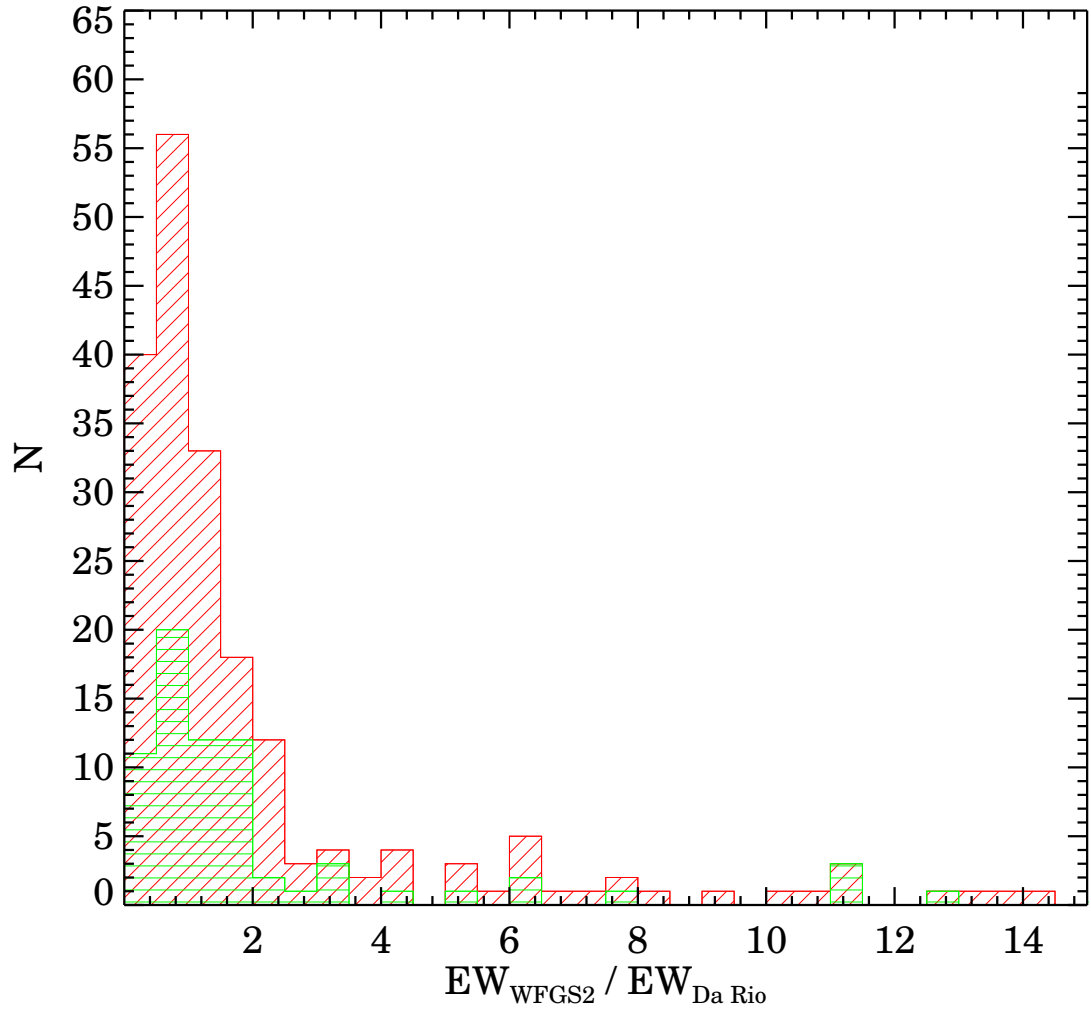}
\figcaption{Distribution of the changes in EW(H$\alpha$) compared with EW data from \citet{dario2009}. Histogram with upward diagonal hatching (red in the online version) indicates the CTTS sample, while the histogram with horizontal hatching (green in the online version) symbols the WTTS.
\label{fig_dario}}
\end{center}
\end{figure}

\newpage
\begin{figure}[!ht]
\begin{center}
\includegraphics{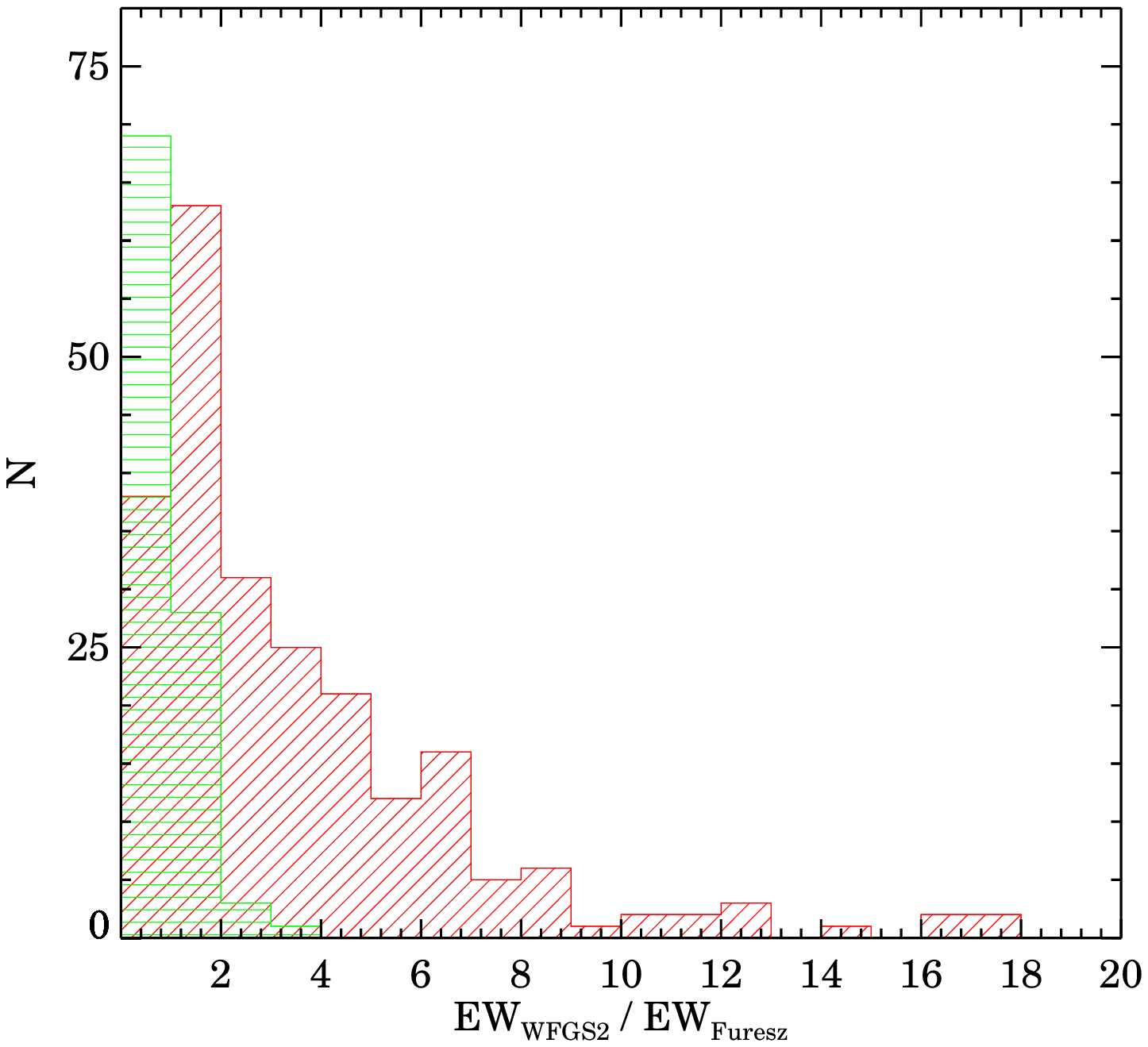}
\figcaption{Same as Fig.~\ref{fig_dario} but compared with data from \citet{furesz}. 
\label{fig_furesz}}
\end{center}
\end{figure}

\section{DISCUSSION}
\label{sect_disc}

\subsection{Accreting and Non-Accreting Stars}
\label{sect_disc1}

\begin{figure}[!ht]
\begin{center}
\includegraphics{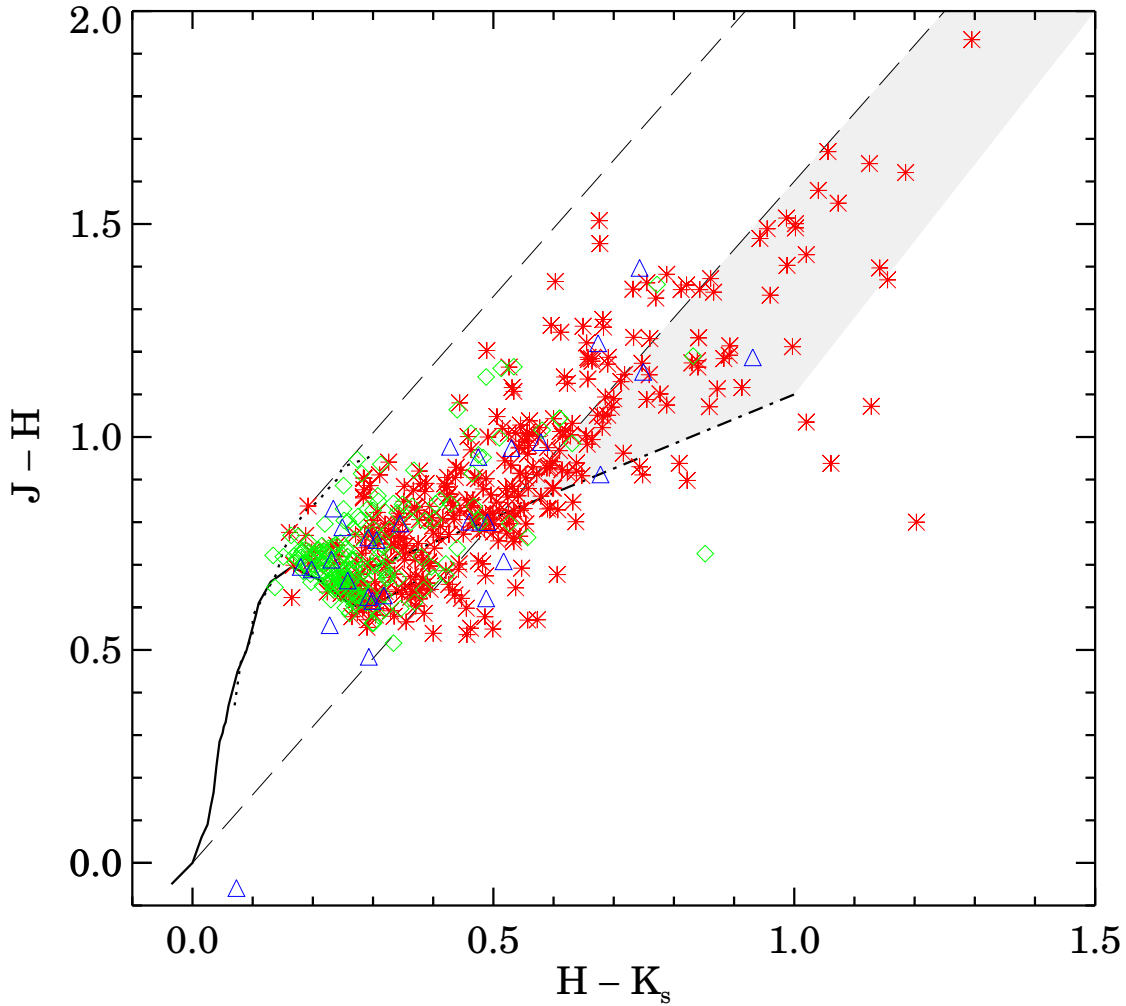}
\figcaption{H$\alpha$ emitters on the {\it 2MASS} {\it J}$-${\it H} vs. {\it H}$-${\it K}$_\mathrm{s}$ diagram. The solid curve shows the colors of the zero-age main sequence, and dotted line is giant branch. The long-dashed lines delimit the area occupied by the reddened  normal stars \citep{cardelli1989}. The dash-dotted line is the locus of unreddened T Tauri stars \citep*{Meyer97}, and the gray shaded band indicates the area of the reddened K$\sb{s}$-excess stars. Asterisks (red in the online version) indicate the classical T Tauri stars, diamonds (green in the online version) mark WTTS and triangles (blue in the online version) are unclassified H$\alpha$ emission stars.
\label{fig_tcd}}
\end{center}
\end{figure}

The classical and weak-line subclasses of T~Tauri stars are generally correlated with the infrared signature of the accretion disks. Figure~\ref{fig_tcd} is the {\it J}$-${\it H}  vs. {\it H}$-${\it K}$_\mathrm{s}$ two-color diagram for the stars listed in Tables \ref{tabshort_ctts}, \ref{tabshort_wtts}, and \ref{tab_unc} with the colors from {\it 2MASS\/} Point Source Catalog. 
This figure shows that only a few of the low-EW stars (diamonds) exhibit near-infrared excess, while a significant part of the high-EW population (asterisks) and a few stars without EW data (triangles) are distributed over the region occupied by reddened stars associated with accretion disks. Stars which are lying below the CTTS locus and right from the zero-age main sequence can be explained by the circumstellar disk models of \citet{lada1992}. The models take into account the surface temperature, disk inclinations, and radial temperature profiles. The modeled systems with stellar surface temperature from $3000$-$12000$ K can explain this part of the two-color diagram. Our sample mainly consists of stars of K and M spectral type, but we detected H$\alpha$ emission in four known Herbig Ae/Be stars too. 	

\begin{figure}
\begin{center}
\includegraphics[scale=0.8]{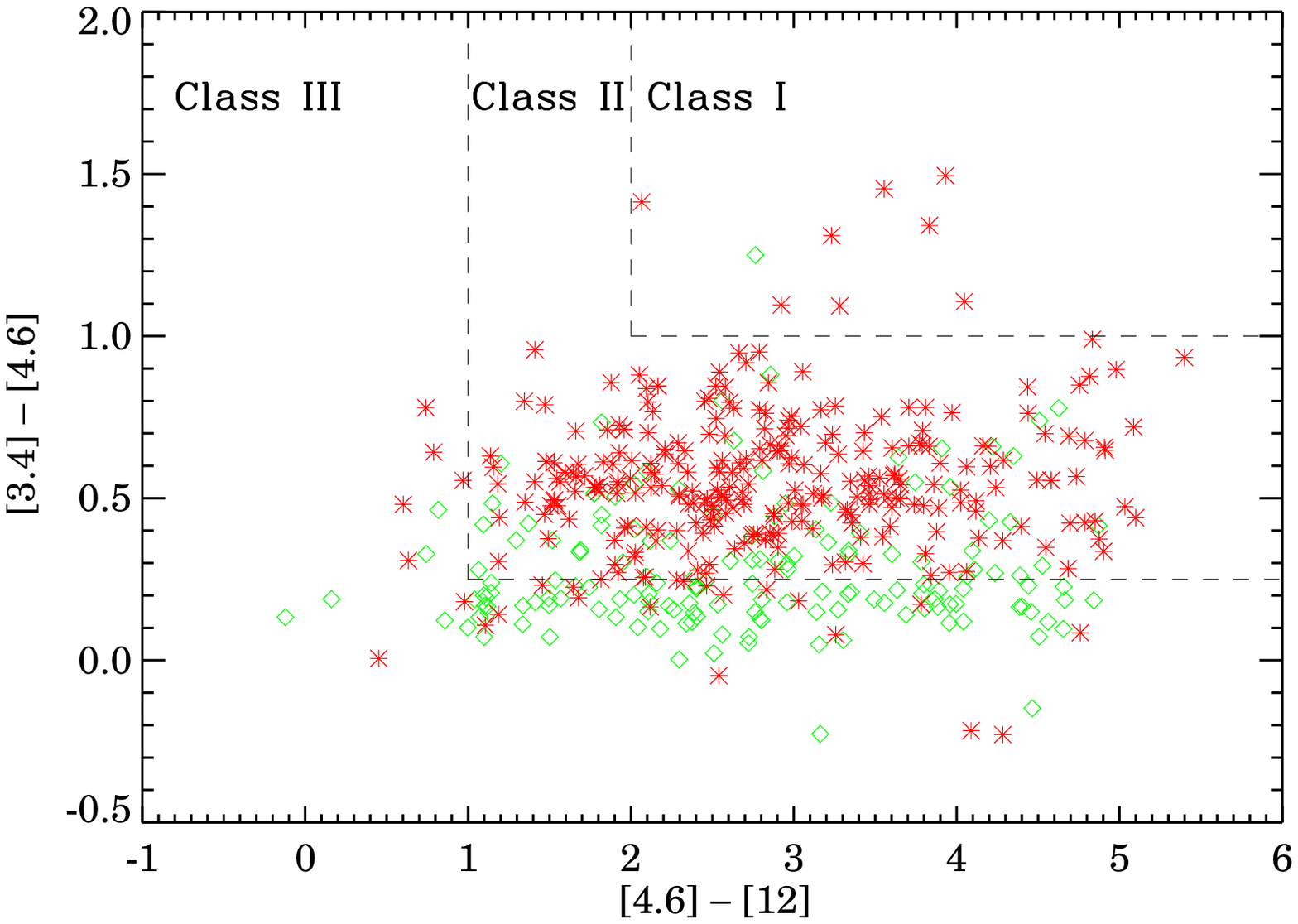}
\figcaption{ {\it WISE} $[3.4]-[4.6]$ vs. $[4.6]-[12]$ color-color diagram. Dashed lines indicate the boundaries between Class I, Class II and Class III sources adopted from \citet{koenig2012}. Asterisks (red in the online version) indicate the CTTS, while diamonds (green in the online version) represent the WTTS.
\label{wisekat}}
\end{center}
\end{figure}

Figure~\ref{wisekat} shows the {\it WISE\/} $[3.4]-[4.6]$ vs. $[4.6]-[12]$ color-color diagram of our CTTS and WTTS sample. The boundaries dividing Class~I, Class~II, and Class~III sources \citep{koenig2012} are indicated. Comparing this plot with the similar diagram of the Taurus star forming region  \citep{koenig2012} shows that the $[4.6]-[12]$ color indices of our stars are unusually high. The reason for this may be that due to the wide point spread function of the {\it WISE\/} and the strong background of the Orion Nebula at 12\,$\mu$m, fluxes measured in the {\it W3} band are seriously contaminated by emission from small nebular clumps or faint neighboring stars \citep{koenig2012}.  To test this hypothesis we examined the positions in the {\it WISE} $[3.4]-[4.6]$ vs. $[4.6]-[12]$  color-color diagram of all {\it Chandra\/} sources \citep{getman} of the ONC without {\it Spitzer\/} detection. The strong X-ray emission and lack of mid-infrared excess ensure their diskless young star nature. We found that these X-ray selected cluster members also exhibited high $[4.6]$--$[12]$ color indices, supporting that the 12~$\mu$m flux probably originates from the environment.  The $[3.4]-[4.6]$ colors indicate that  most of our CTTSs are Class~II infrared sources. Most of the WTTS exhibit no excess in these bands, but, in contrast with the case of the {\it J}$-${\it H}  vs. {\it H}$-${\it K}$_\mathrm{s}$ diagram, a significant subgroup of them ($37$ stars) exhibit quite large excesses in the $[3.4]-[4.6]$ color.

Based on the WISE $[3.4]-[4.6]$ color \citep{koenig2012} or, where they are available, {\it Spitzer IRAC} $[3.6]-[4.5]$ colors \citep{Gutermuth08} we identified a subsample of low-EW H$\alpha$ emission stars exhibiting infrared excess indicative of an optically thick accretion disk. We checked the {\it VizieR\/} catalogs and omitted the stars having known neighbors within the angular resolution of the WFGS$2$ instrument. The stars, selected in this manner and listed in Table~\ref{tab_atsorolt}, are apparently not accreting, in spite of their primordial accretion disks (Fig.~\ref{fig_varsed2}). Most of them are known variables (see Table~\ref{tabshort_wtts}). Comparison of their {\it EW\/}s, measured by various surveys and shown in columns 4--6 of Table~\ref{tab_atsorolt}, shows that 7 of the 37 stars exhibited much stronger H$\alpha$ emission in 2005, at the epochs of the observations of \citet{furesz} and \citet{dario2009}, that is, their accretion and wind activities are highly variable on a few years time scale.

\begin{deluxetable}{cccccc}
\tabletypesize{\scriptsize}

\tabletypesize{\scriptsize}
\tablecolumns{6}
\tablewidth{0pt}
\tablecaption{H$\alpha$ emission line stars classified as WTTS based on the H$\alpha$ equivalent width, but possessing 
optically thick disk based on mid-infrared color indices.\label{tab_atsorolt}}

\tablehead{
\colhead{2MASS}  &\colhead{[3.6]$-$[4.5]} &\colhead{W1$-$W2}&\multicolumn{3}{c}{EW (\AA)}\\
\cline{4-6}
\colhead{}  &\colhead{} &\colhead{}&\colhead{This survey}&\colhead{F\H{u}r\'esz et al.}&
\colhead{Da Rio et al.}
}
\startdata
J05334964$-$0536208	&	0.53		&	0.51	&	4.0 (0.3)&5.2&	\nodata\\
J05335210$-$0530284	&	\nodata		&	0.31	&	7.2 (1.0)&6.5&3.6	\\
J05340797$-$0536170	&	0.64		&	0.81	&	7.8 (0.2)&24.3&78.0	\\
J05340835$-$0514387	&	\nodata		&	0.40	&	$<$2.7&\nodata&\nodata	\\
J05342616$-$0526304	&	0.56		&	0.73	&	$>$5.7&42.2&230	\\
J05342650$-$0523239	&	\nodata		&	0.36	&	11.4 (1.1)&45&46	\\
J05342960$-$0547247	&	0.09		&	0.31	&	9.4 (0.7)&8.5&\nodata	\\
J05343203$-$0511248	&	\nodata		&	0.55	&	5.9 (0.8)&19.5&\nodata	\\
J05343417$-$0505170	&	0.16		&	0.53	&	3.6 (0.3)&4.7&	\nodata\\
J05344172$-$0536488	&	0.45		&	0.58	&	$>$6.5&12.8&30	\\
J05344239$-$0512381	&	\nodata		&	0.88	&	5.5 (0.3)&25.3&\nodata	\\
J05344244$-$0543256	&	0.32		&	0.50	&	8.0 (2.9)&3.9&\nodata	\\
J05344789$-$0530465	&	0.21		&	\nodata	&	9.2 (0.9)&22.2&10	\\
J05344815$-$0542289	&	0.24		&	0.54	&	6.7 (0.3)&3.1&23	\\
J05345825$-$0541498	&	0.39		&	0.68	&	5.3 (1.1)&\nodata&28	\\
J05345881$-$0547334	&	0.32		&	0.74	&	9.7 (3.5)&\nodata&\nodata	\\
J05350085$-$0509389	&	\nodata		&	0.34	&	9.2 (1.6)&\nodata&\nodata	\\
J05350532$-$0534285	&	0.37		&	0.46	&	7.7 (1.3)&7.2&	11.0\\
J05351205$-$0547296	&	0.36		&	0.49	&	$<$7.0&\nodata&\nodata	\\
J05351236$-$0543184	&	0.10		&	0.45	&	3.8 (0.4)&\nodata&\nodata\\
J05351464$-$0502251	&	0.26    	&	0.45	&	5.8 (0.7)&4.8&	\nodata\\
J05351715$-$0541538	&	0.28		&	0.48	&	5.7 (2.8)&\nodata&14	\\
J05353047$-$0549037	&	0.28		&	0.30	&	8.8 (0.7)&8.6&\nodata\\
J05353070$-$0518071	&	0.50		&	0.76	&	2.4 (0.2)&65.8&190	\\
J05353385$-$0538206	&	\nodata		&	0.33	&	4.9 (1.8)&\nodata&1.6	\\
J05353554$-$0506585	&	\nodata		&	0.78	&	2.8 (0.5)&5.1&	\nodata\\
J05354130$-$0538329	&	0.31		&	0.44	&	7.6 (1.6)&5.6&15	\\
J05355109$-$0507088	&	0.29		&	0.42	&	6.7 (0.4)&6.0&	\nodata\\
J05355232$-$0512569	&	0.23		&	0.63	&	4.8 (2.8)&\nodata&\nodata\\
J05355276$-$0512590	&	0.27		&	0.48	&	5.3 (0.8)&3.8&30	\\
J05360276$-$0515269	&	\nodata		&	0.43	&	11.8 (0.6)&13.3&11	\\
J05361010$-$0522050	&	\nodata		&	0.31	&	6.1 (0.7)&6.0&\nodata	\\
J05361975$-$0514386	&	0.30		&	0.37	&	10.0 (0.3)&3.0&\nodata	\\
J05362627$-$0518301	&	0.29		&	0.53	&	14.2 (2.6)&10.7&\nodata	\\
J05363167$-$0526356	&	\nodata		&	0.42	&	4.5 (0.8)&9.9&	\nodata\\
J05364005$-$0512231	&	0.25		&	0.37	&	4.5 (1.3)&3.1&	\nodata\\
J05364932$-$0533205	&	0.42		&	0.39	&	9.9 (2.8)&\nodata&\nodata	\\
\enddata

\end{deluxetable}

One star, exhibiting an extreme EW value around 500~\AA\  in Fig.~\ref{fig_ewhist}, deserves attention. It is V421~Ori, associated with the proplyd 280-931 \citep{Ricci}. \citet{dario2009} also measured very strong emission, EW(H$\alpha$) = 420~\AA\ from this object, in which the emission from the externally ionized disk contributes to the observed H$\alpha$.

\begin{figure*}
\centerline{
\includegraphics[width=4.0cm]{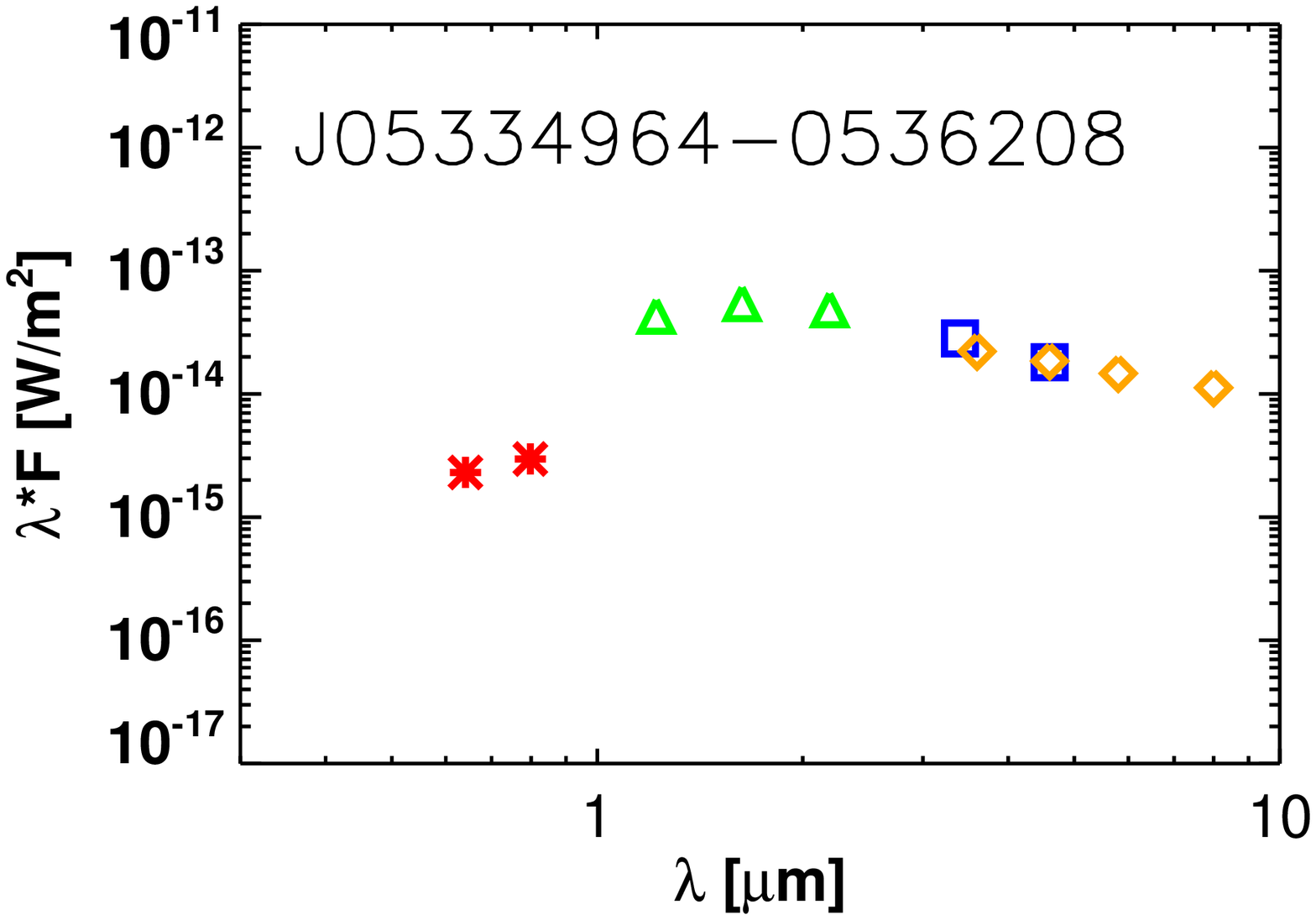}\hspace{0.3cm}\includegraphics[width=4.0cm]{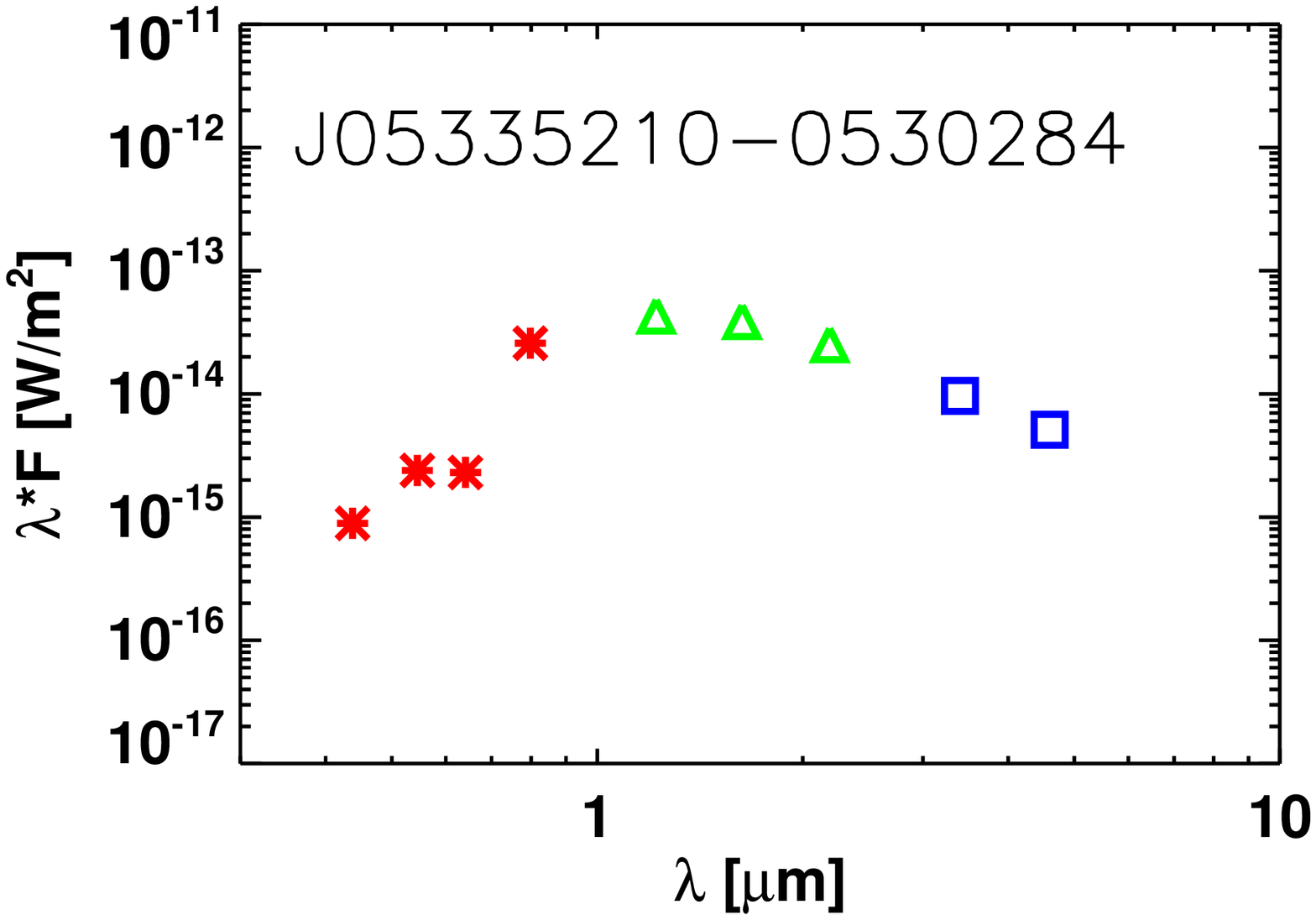}\hspace{0.3cm}\includegraphics[width=4.0cm]{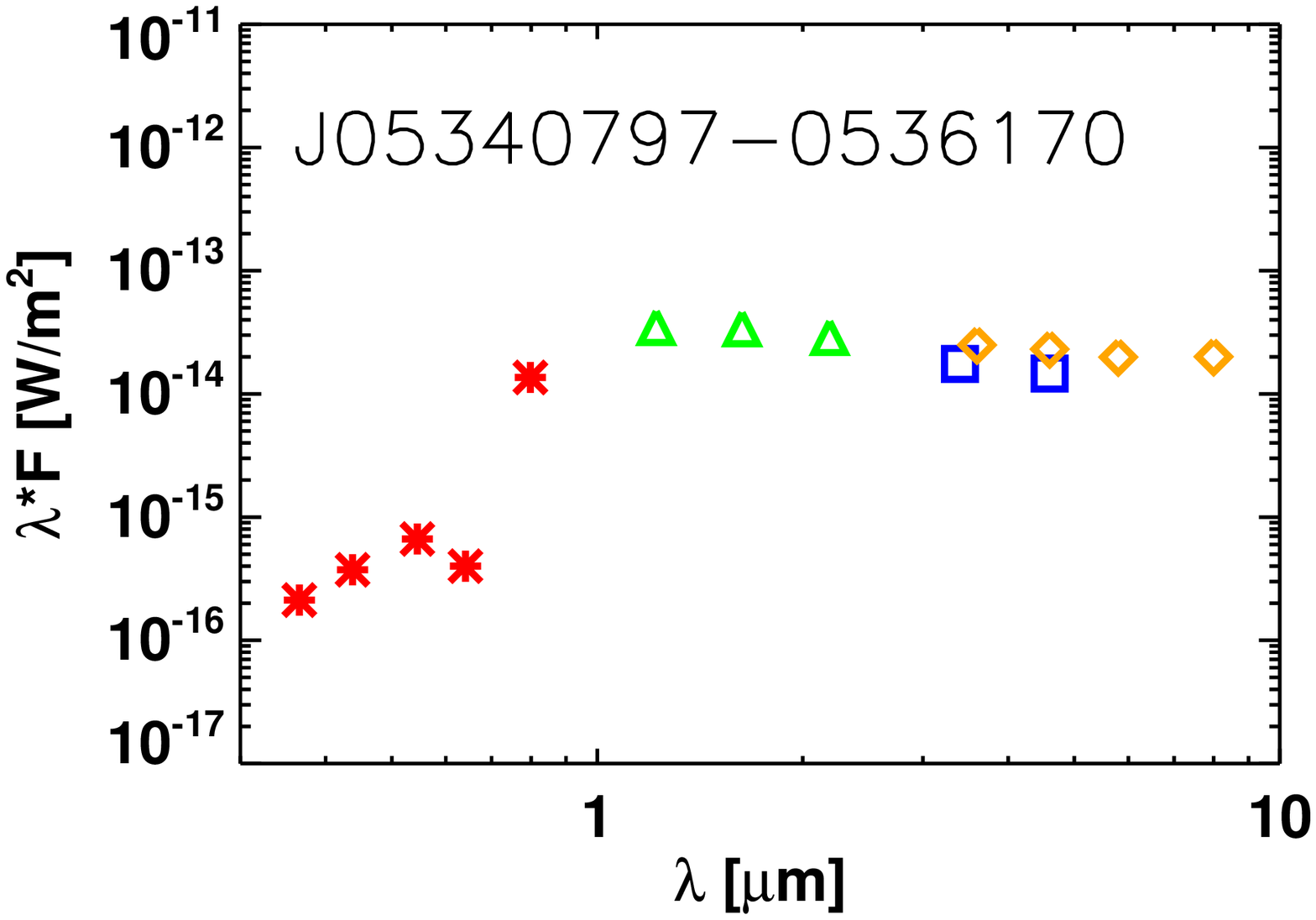}\hspace{0.3cm}\includegraphics[width=4.0cm]{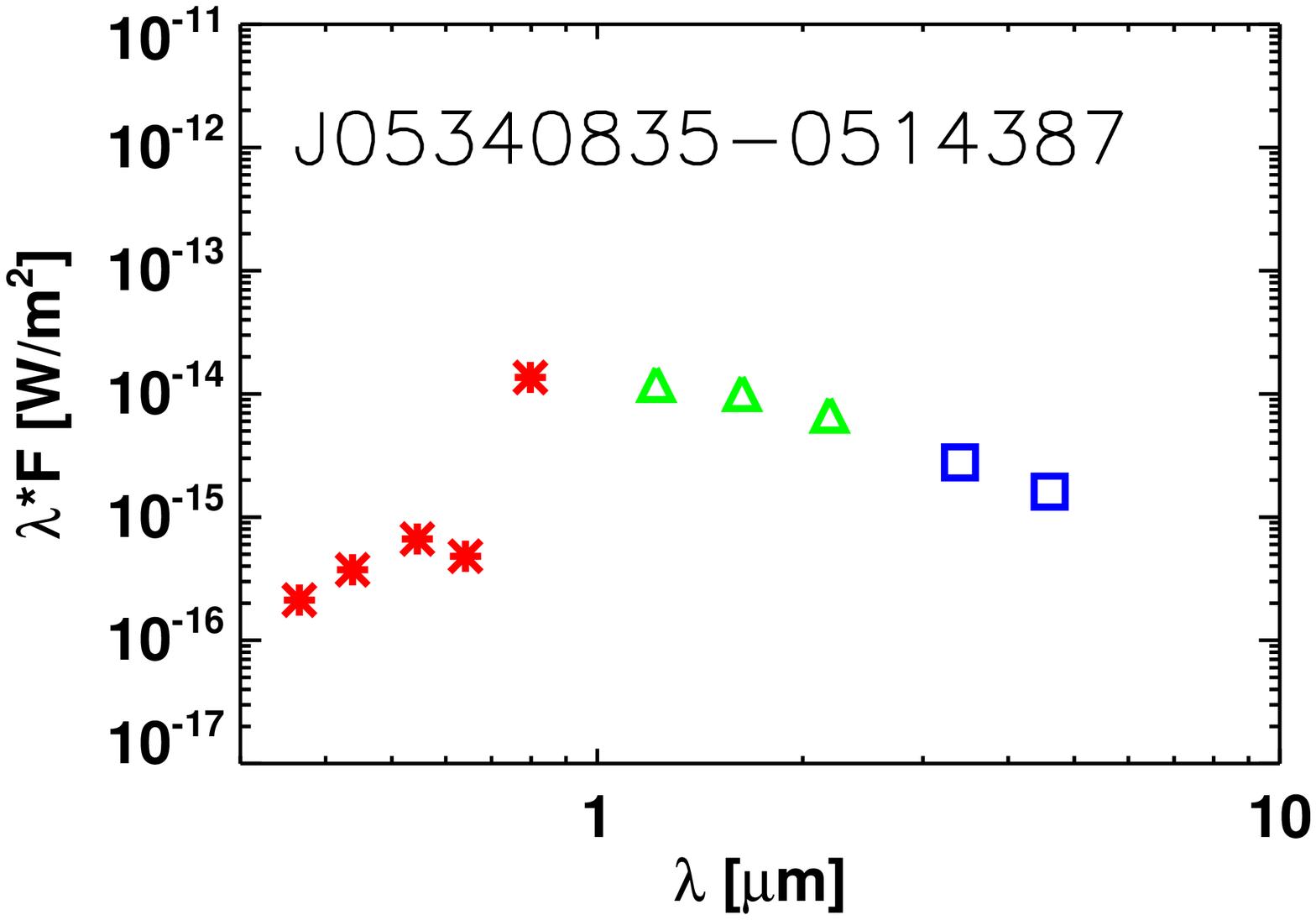}}
\centerline{
\includegraphics[width=4.0cm]{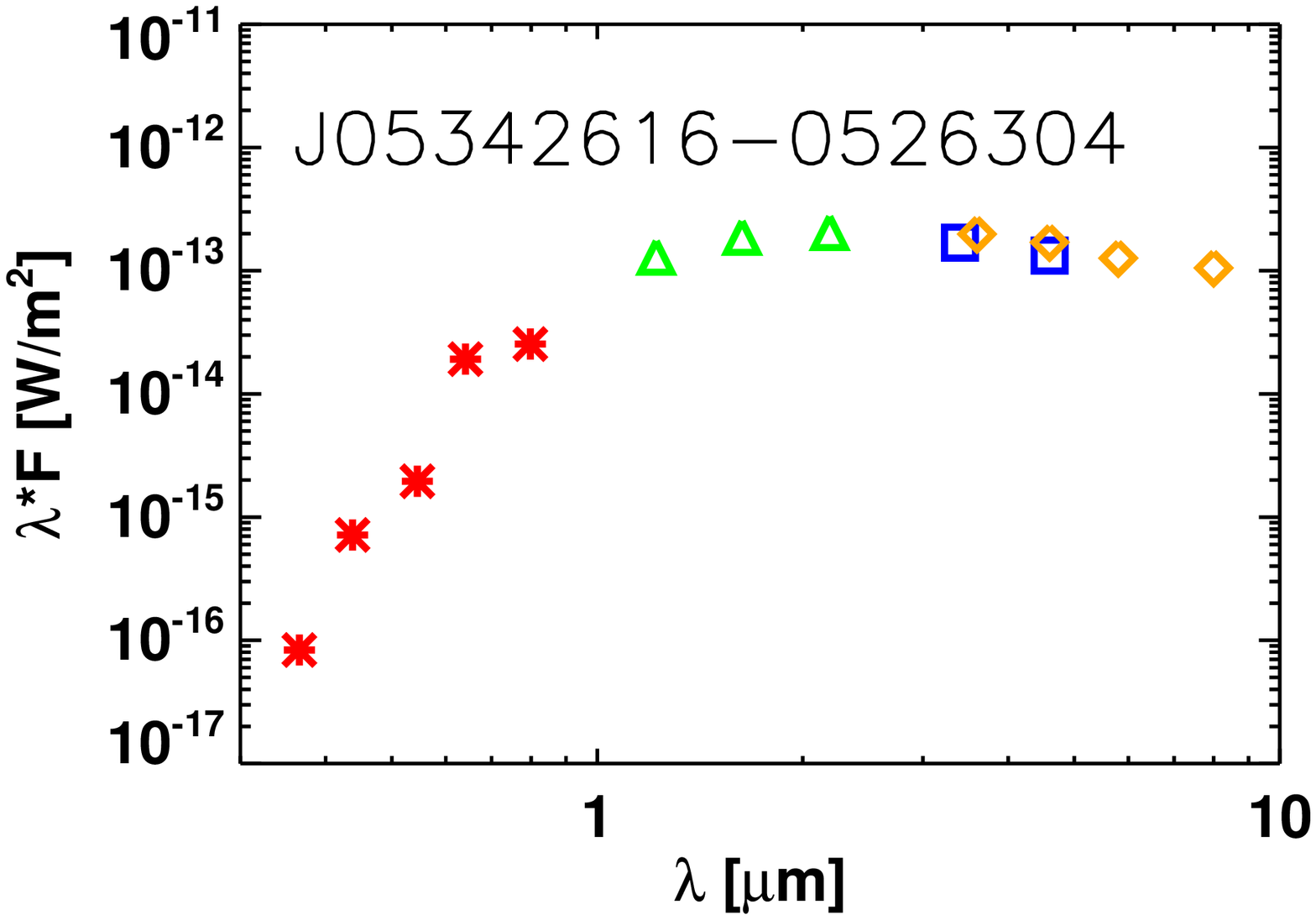}\hspace{0.3cm}\includegraphics[width=4.0cm]{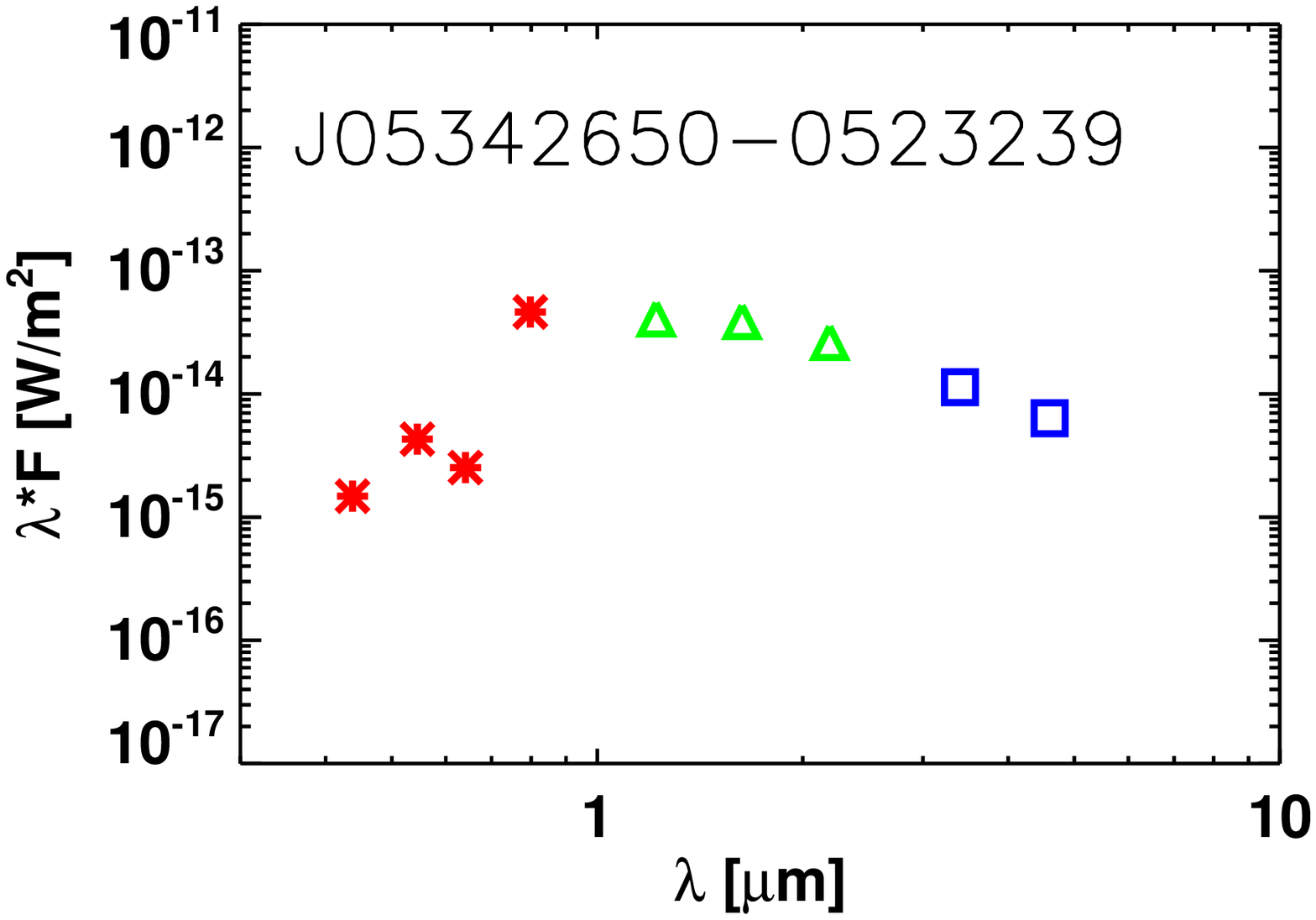}\hspace{0.3cm}\includegraphics[width=4.0cm]{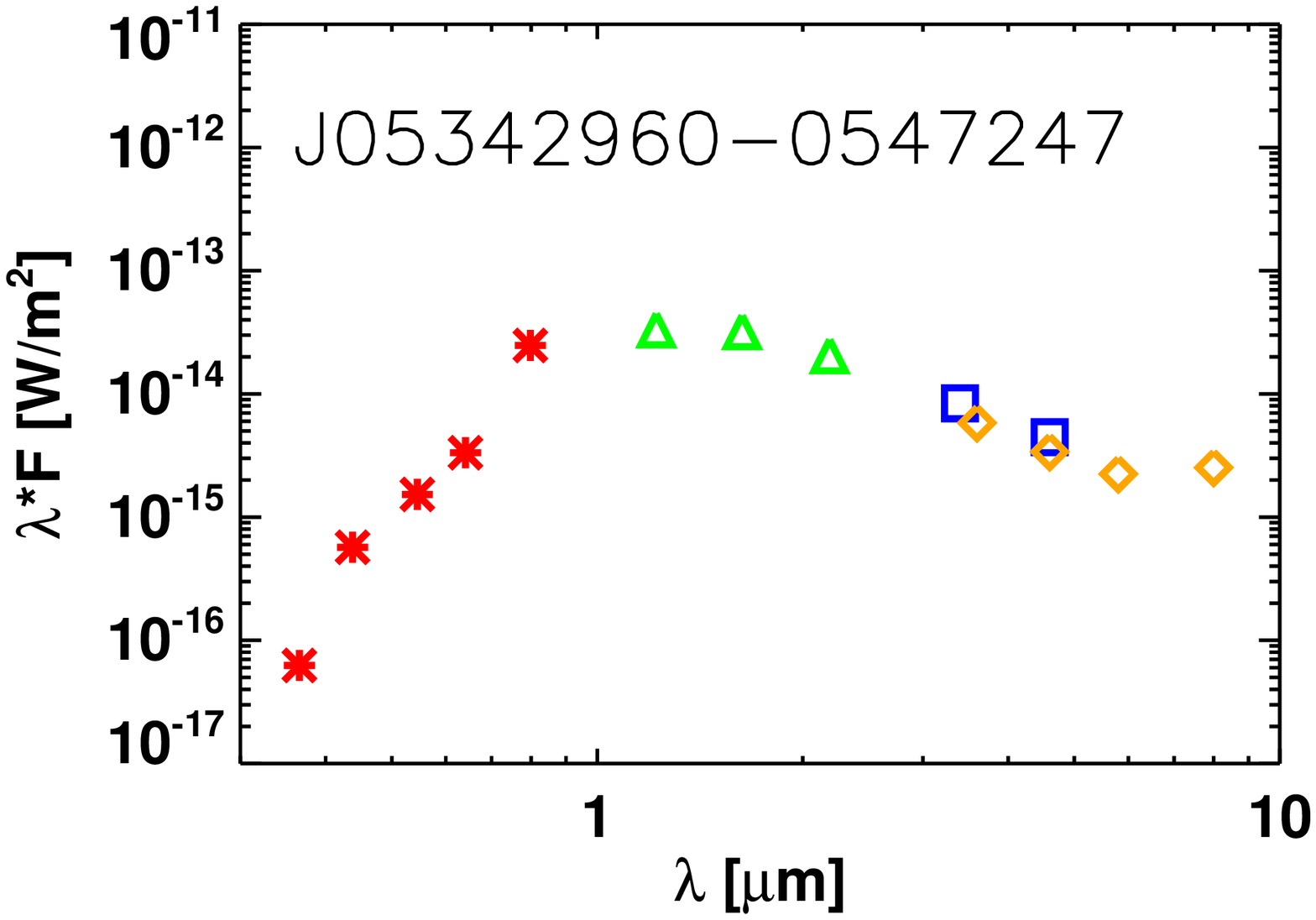}\hspace{0.3cm}\includegraphics[width=4.0cm]{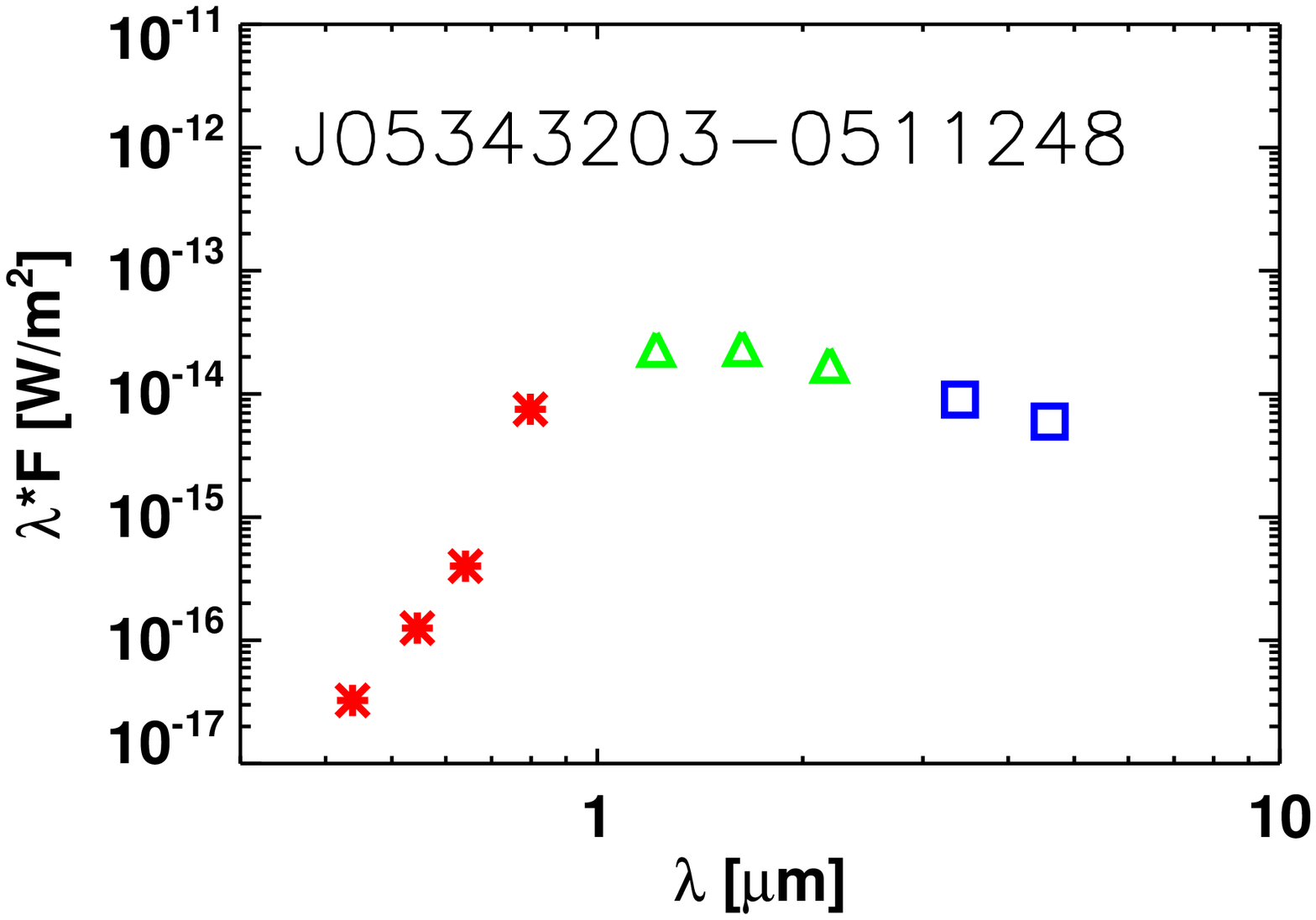}}
\centerline{
\includegraphics[width=4.0cm]{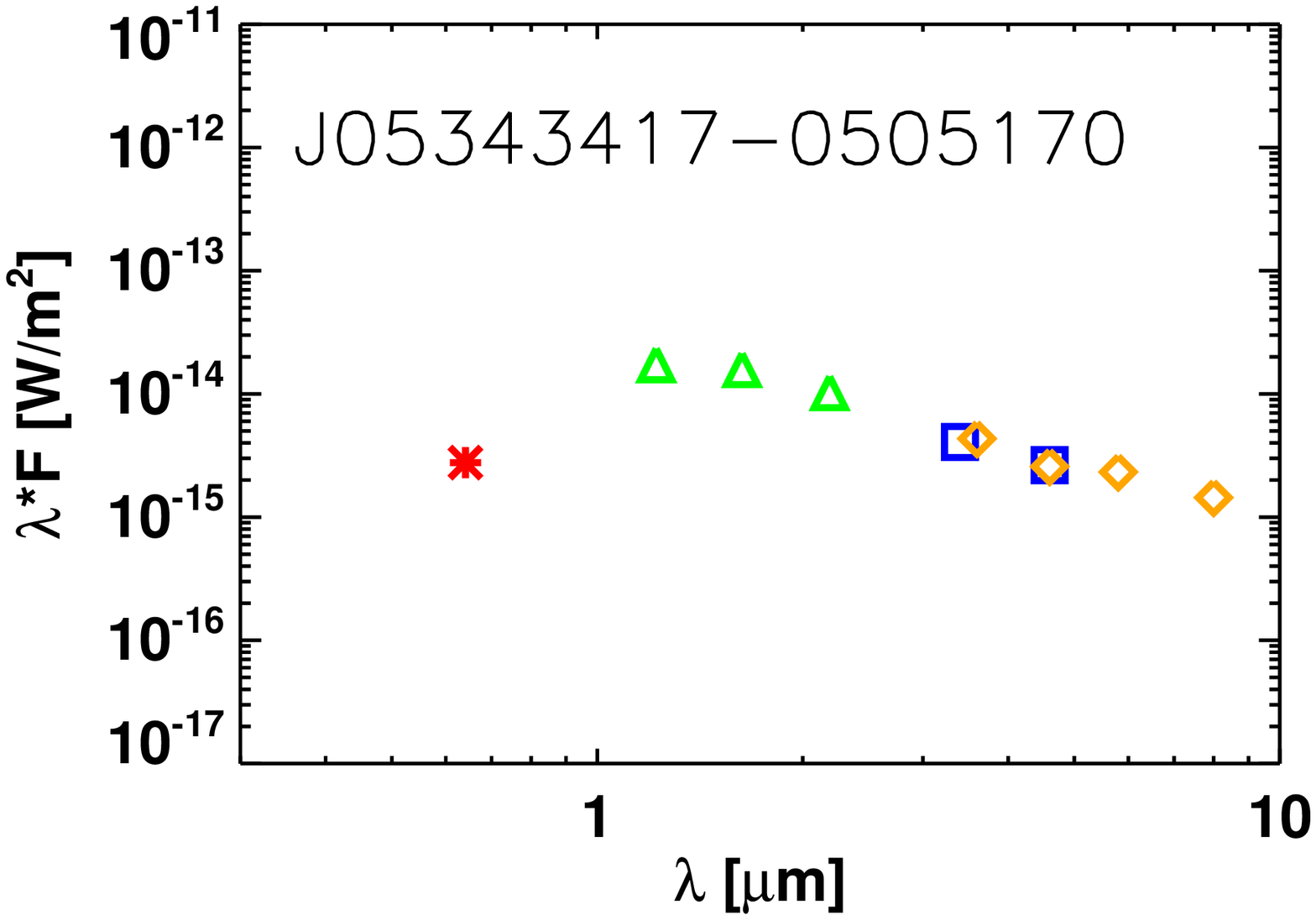}\hspace{0.3cm}\includegraphics[width=4.0cm]{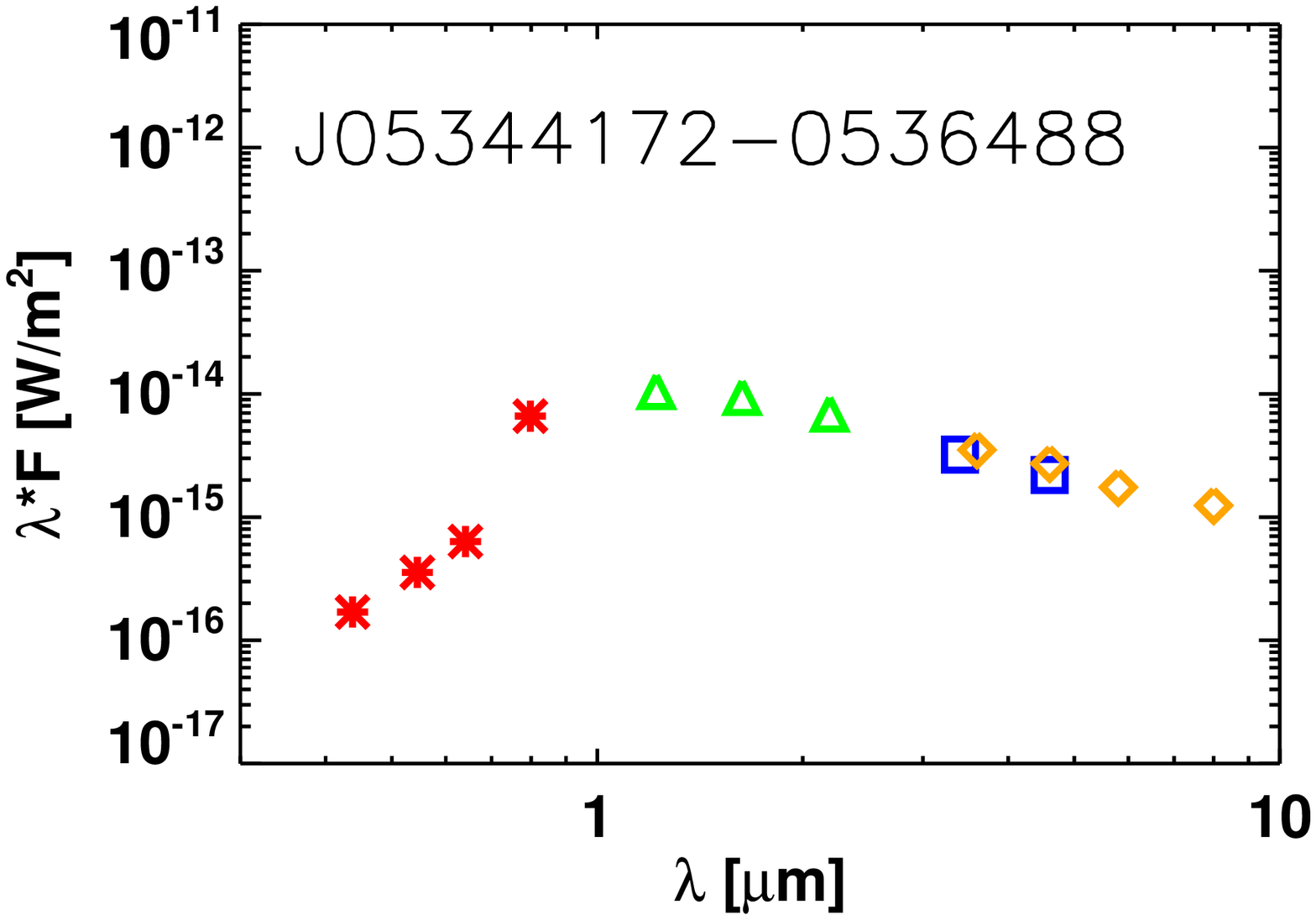}\hspace{0.3cm}\includegraphics[width=4.0cm]{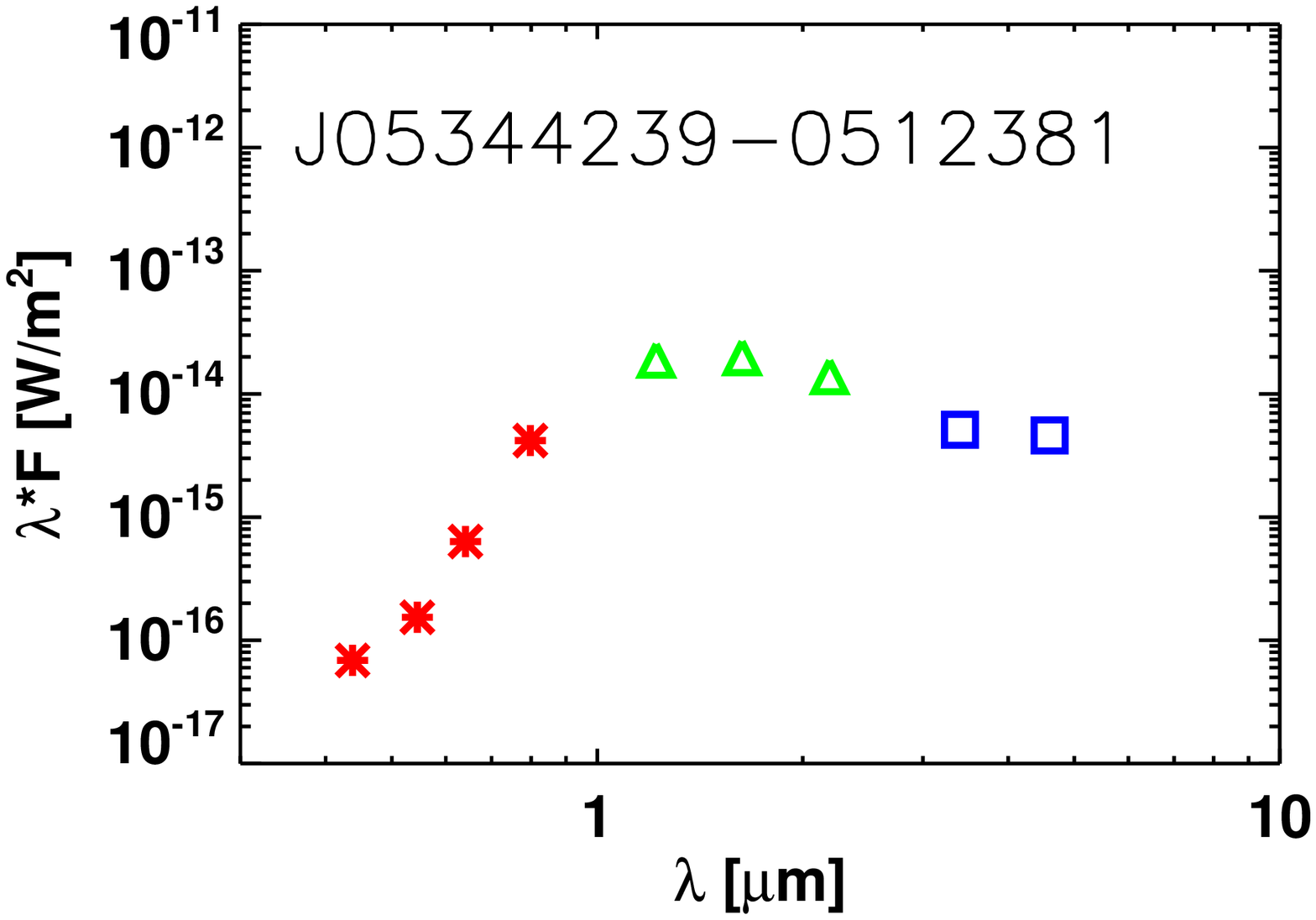}\hspace{0.3cm}\includegraphics[width=4.0cm]{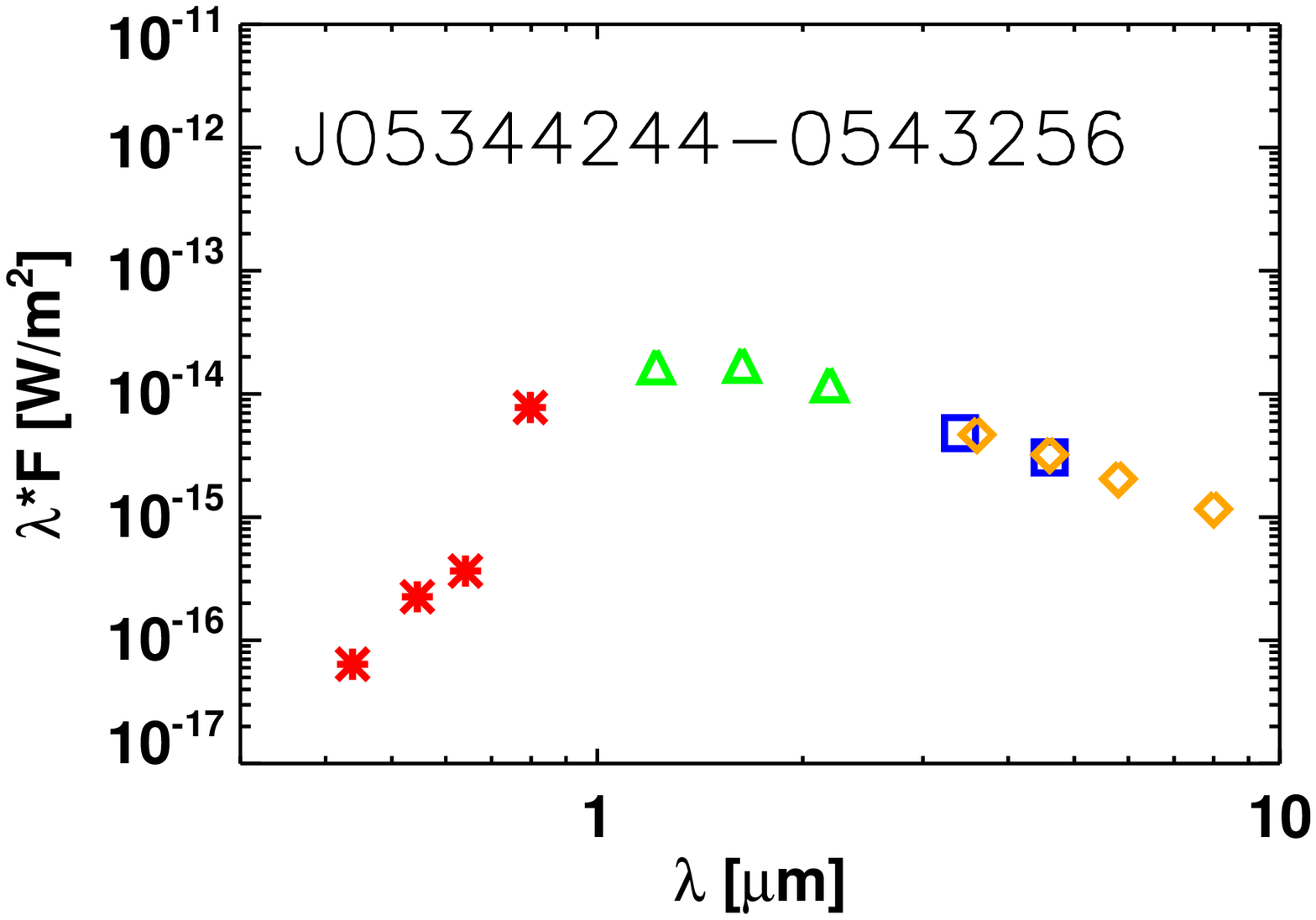}}
\centerline{
\includegraphics[width=4.0cm]{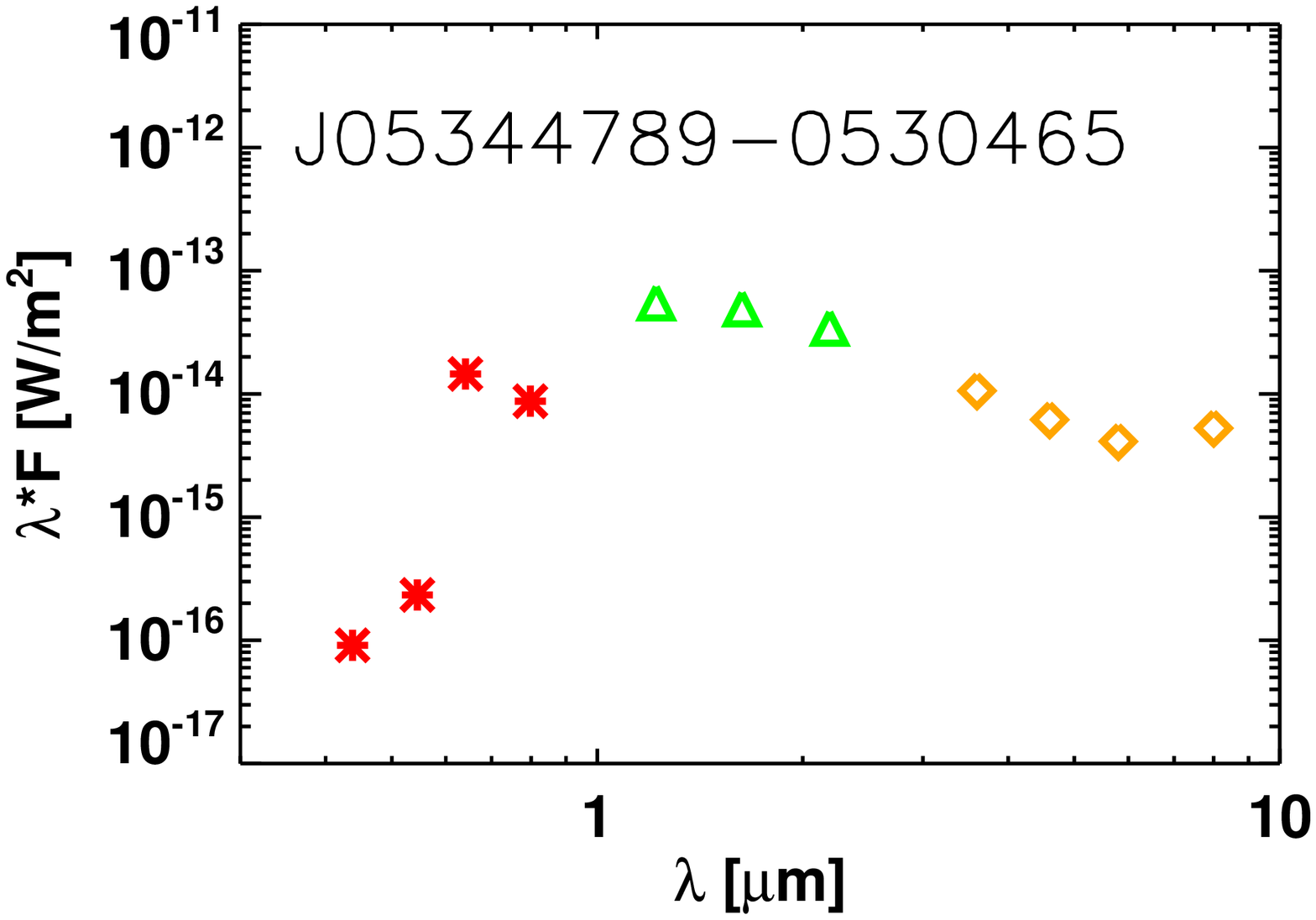}\hspace{0.3cm}\includegraphics[width=4.0cm]{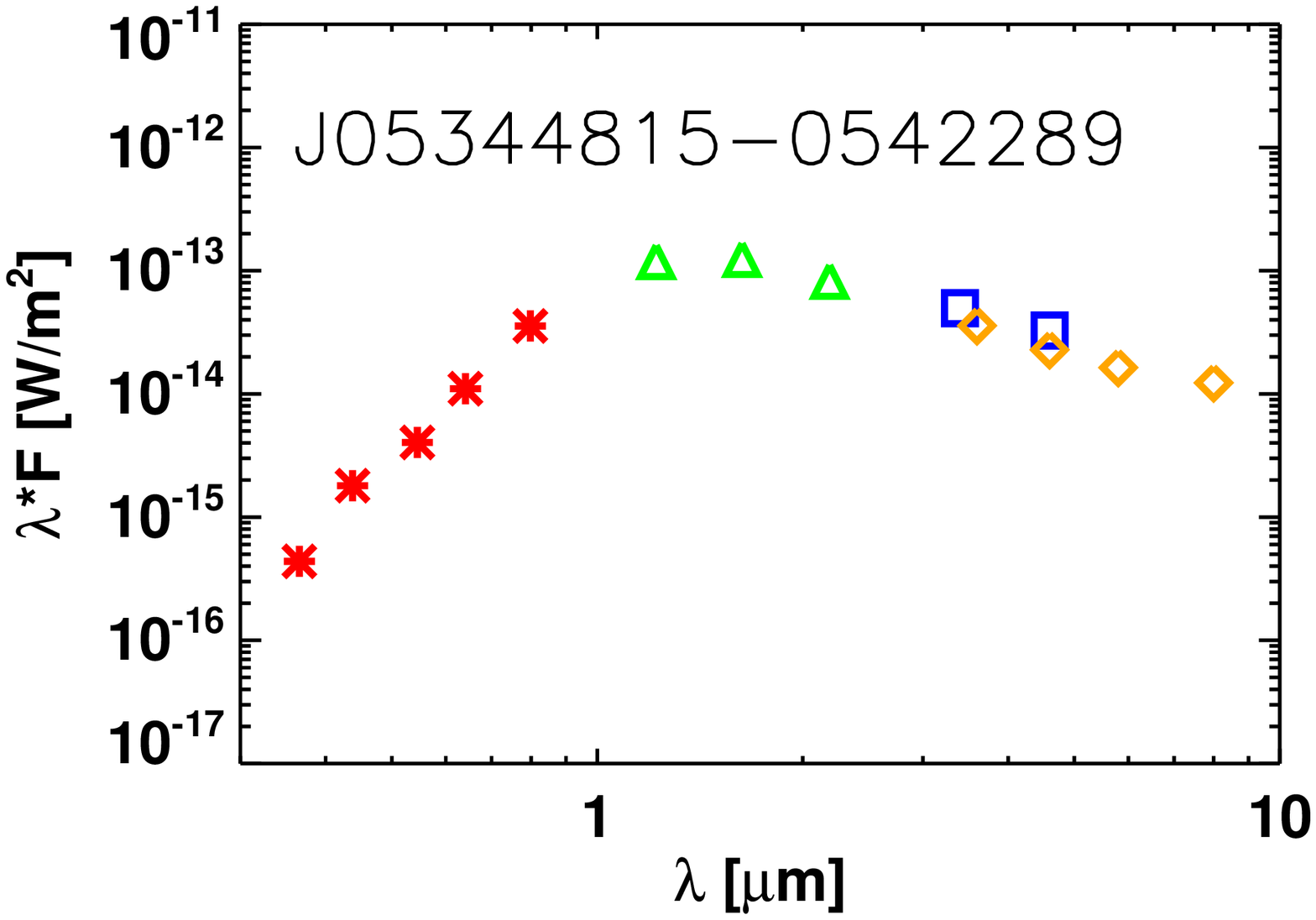}\hspace{0.3cm}\includegraphics[width=4.0cm]{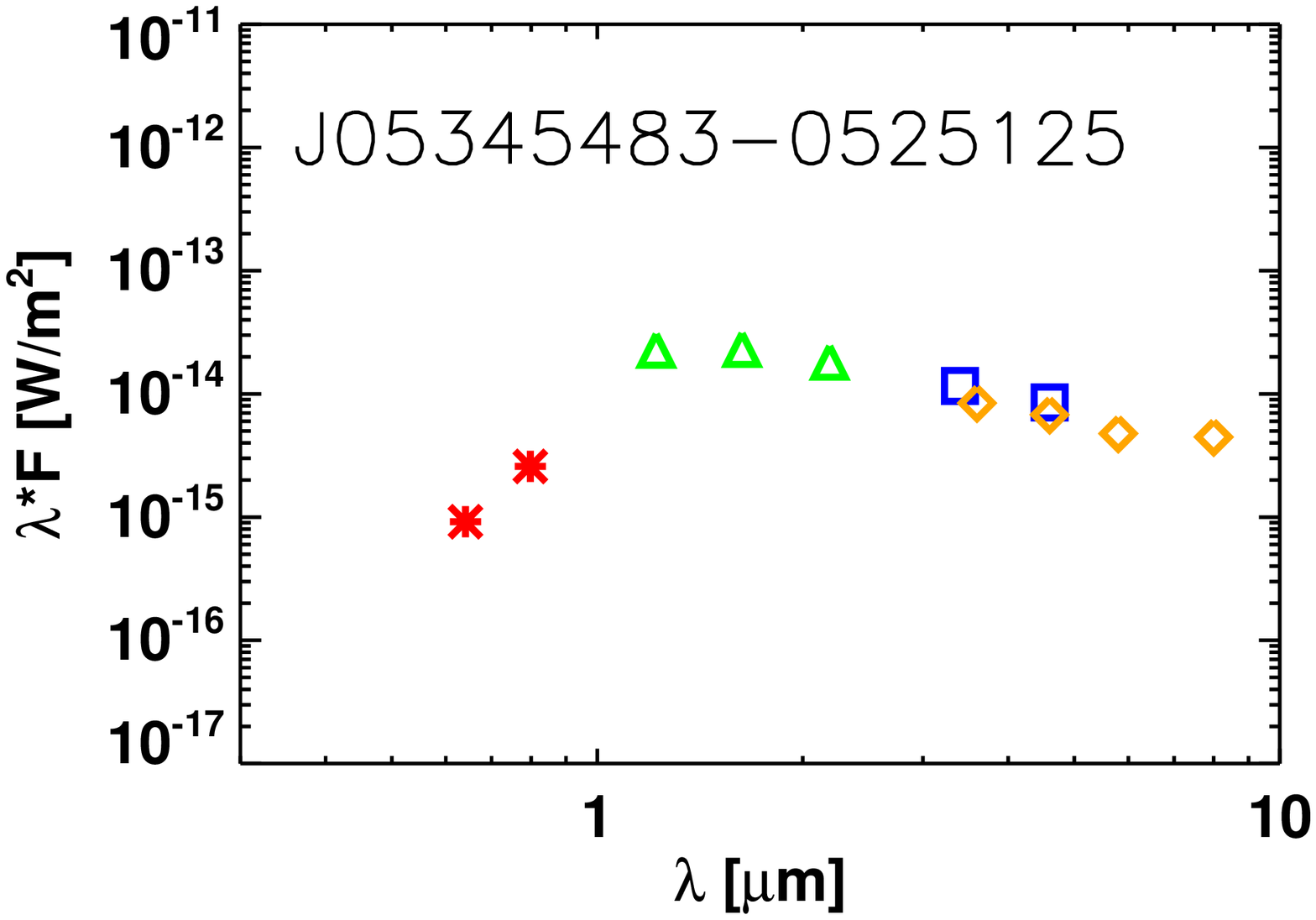}\hspace{0.3cm}\includegraphics[width=4.0cm]{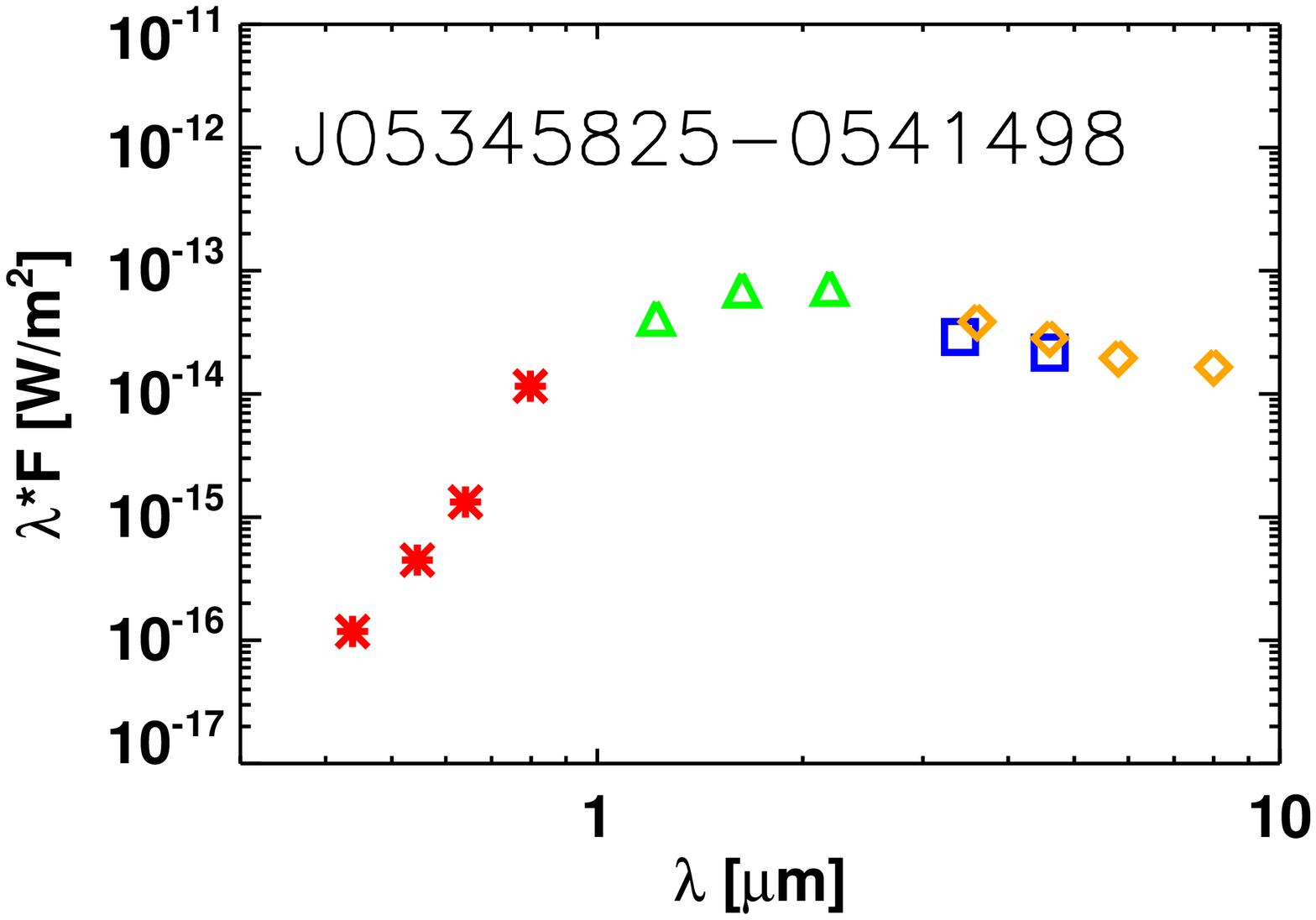}}
\centerline{
\includegraphics[width=4.0cm]{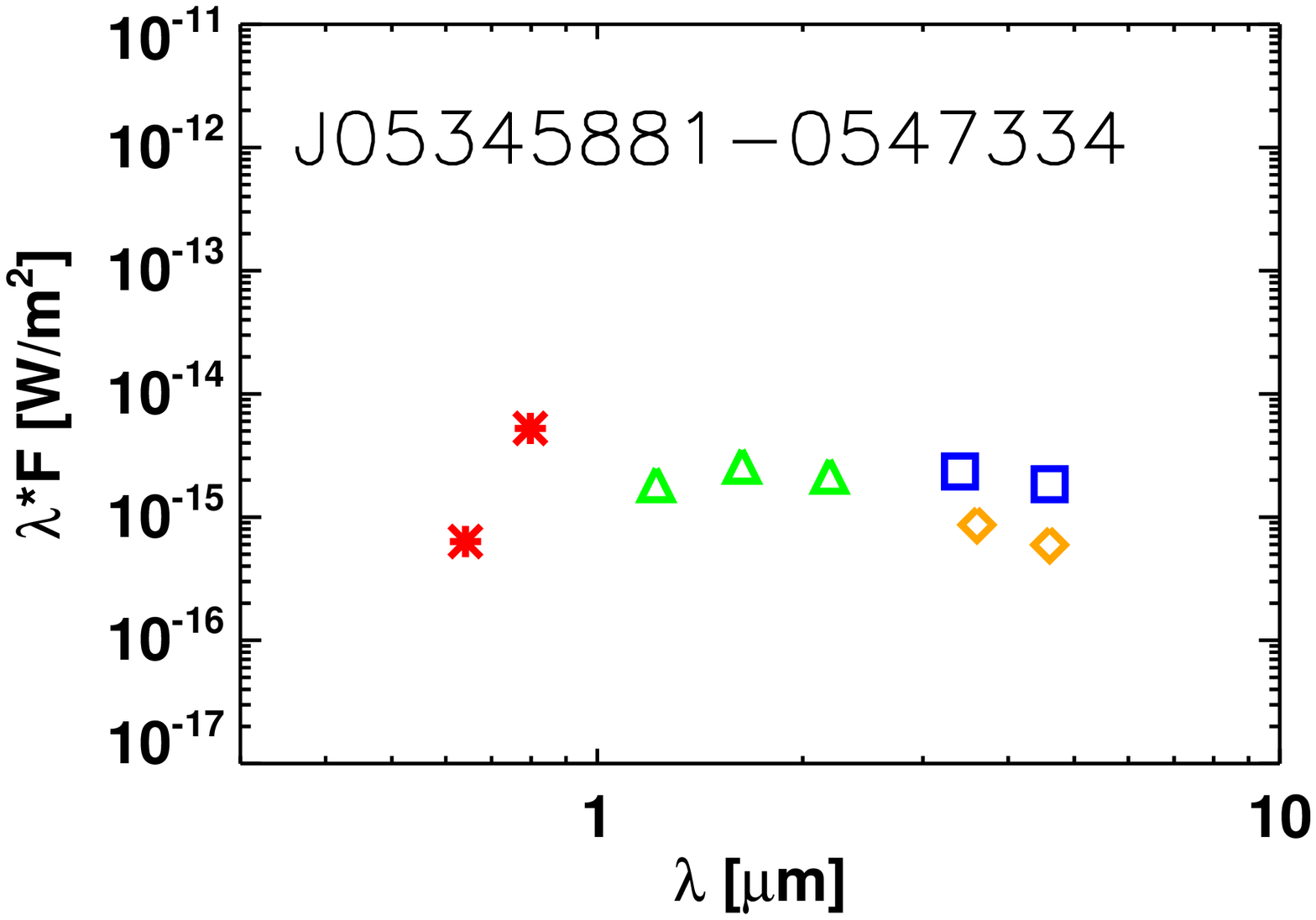}\hspace{0.3cm}\includegraphics[width=4.0cm]{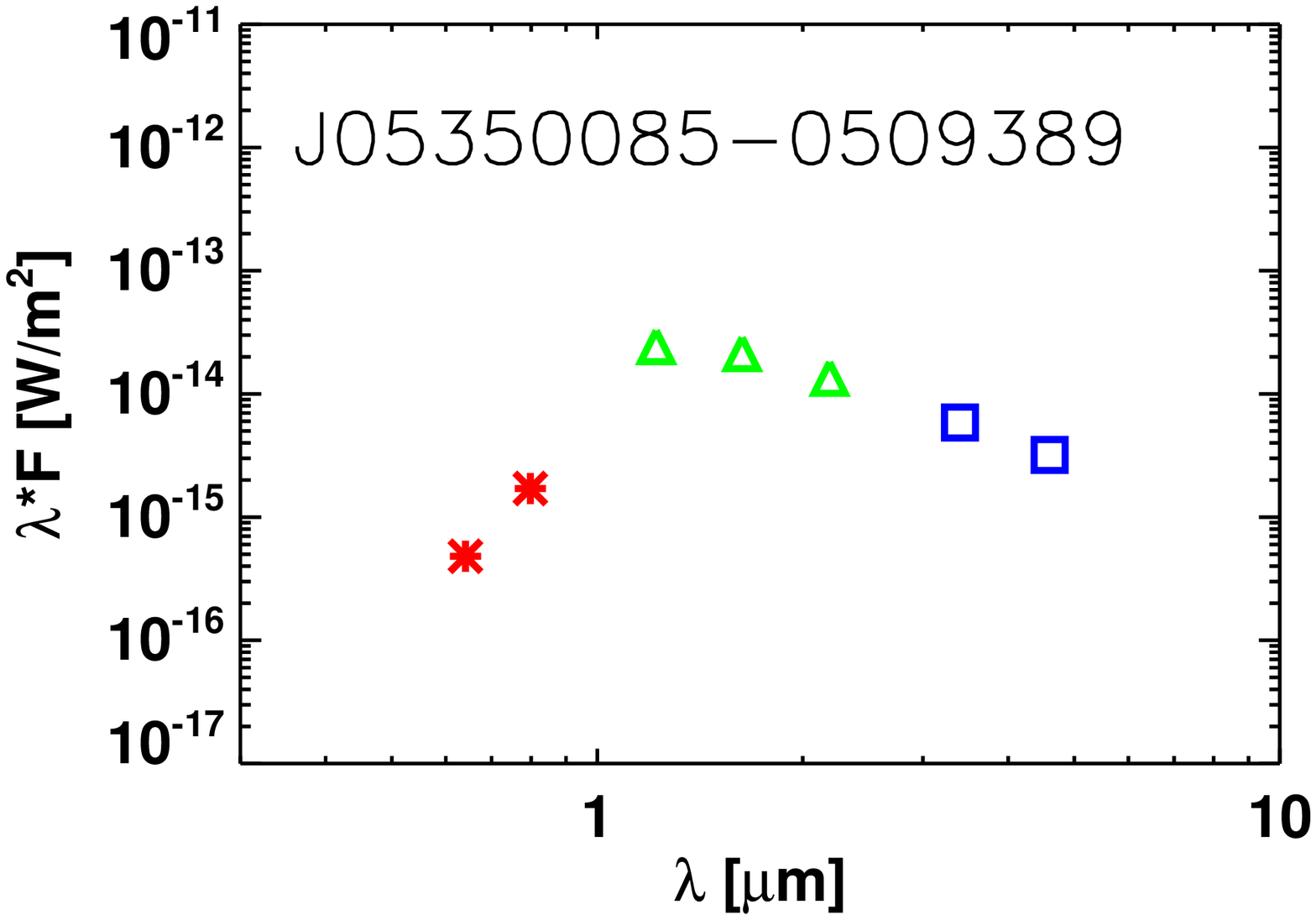}\hspace{0.3cm}\includegraphics[width=4.0cm]{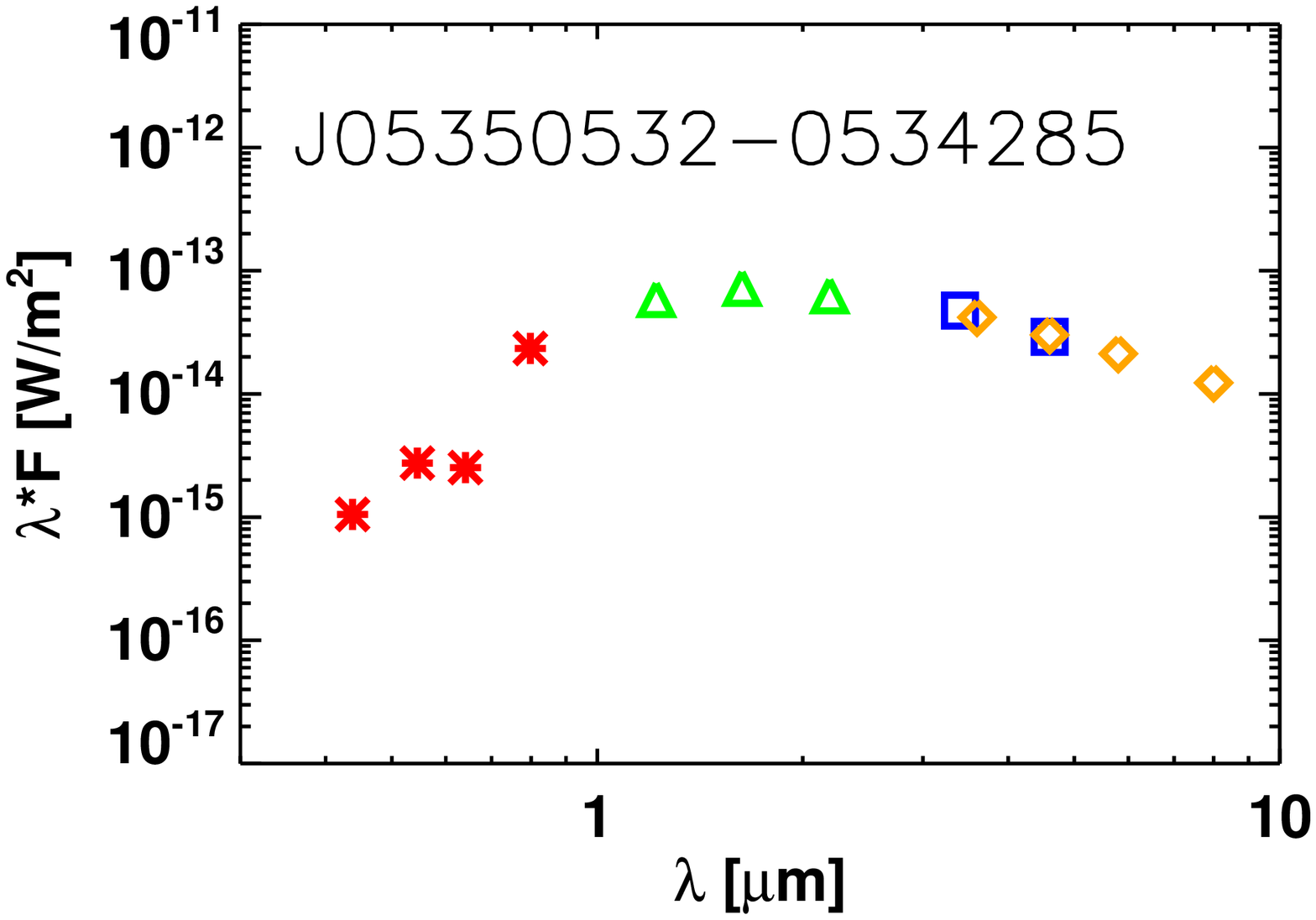}\hspace{0.3cm}\includegraphics[width=4.0cm]{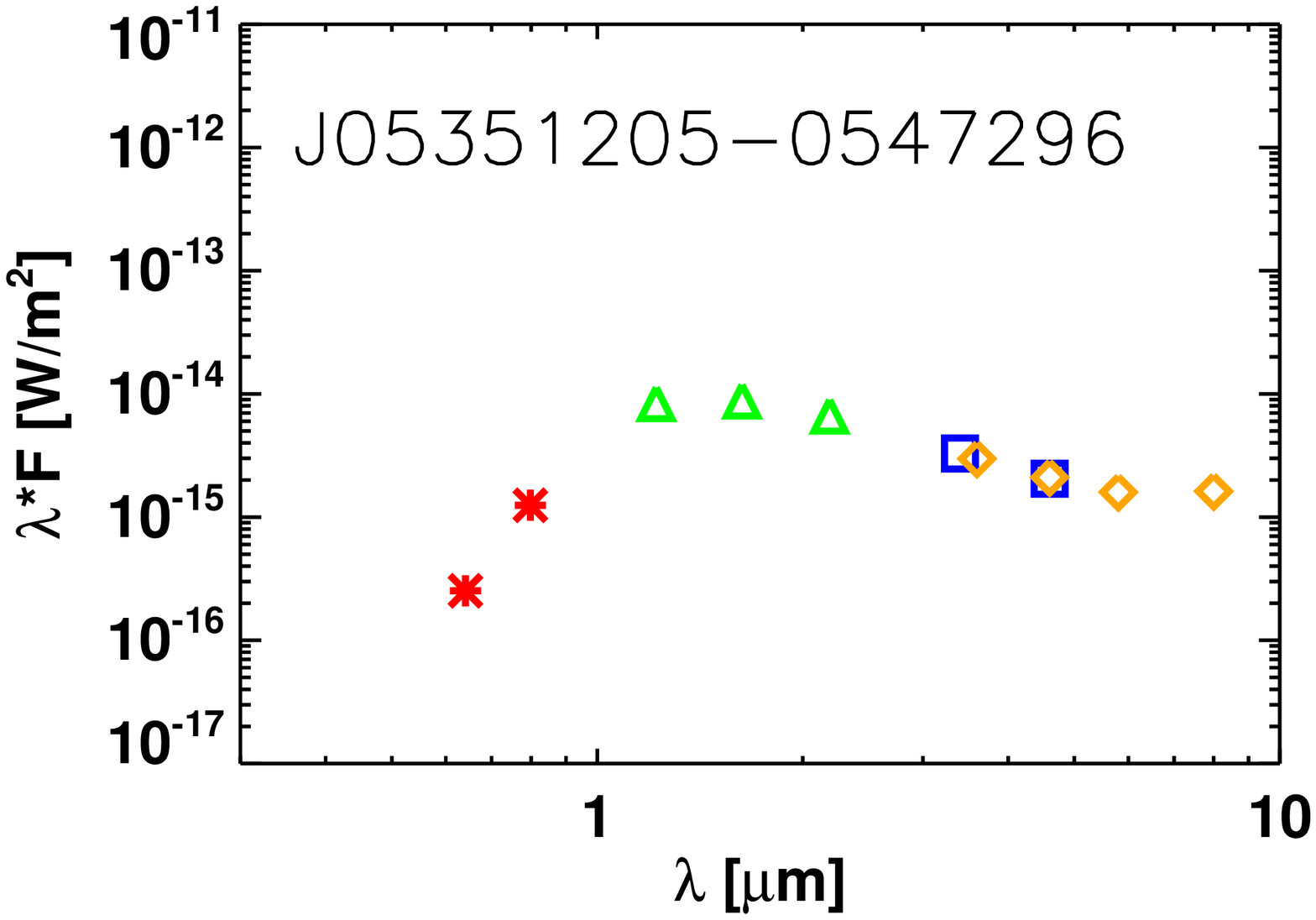}}
\centerline{
\includegraphics[width=4.0cm]{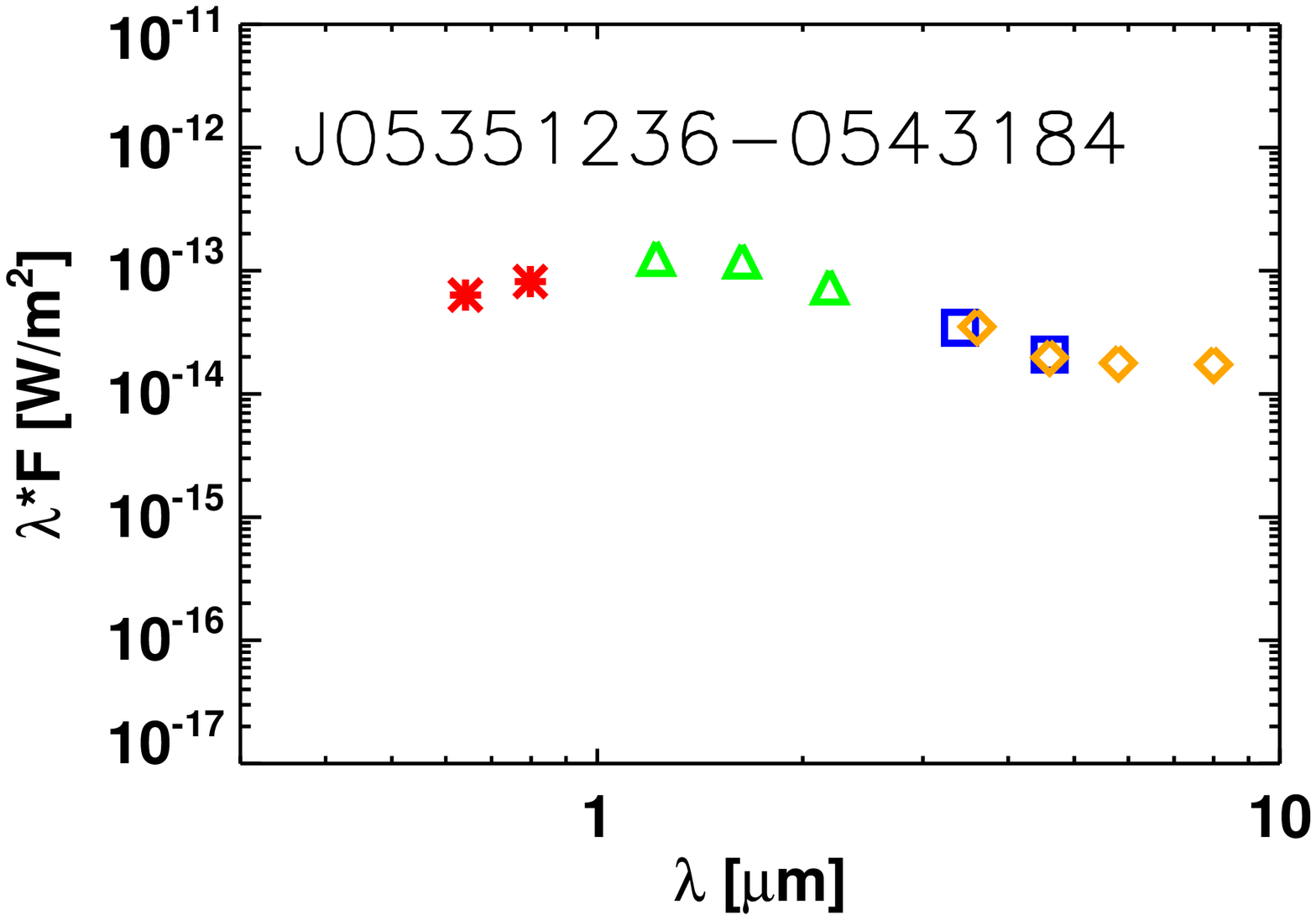}\hspace{0.3cm}\includegraphics[width=4.0cm]{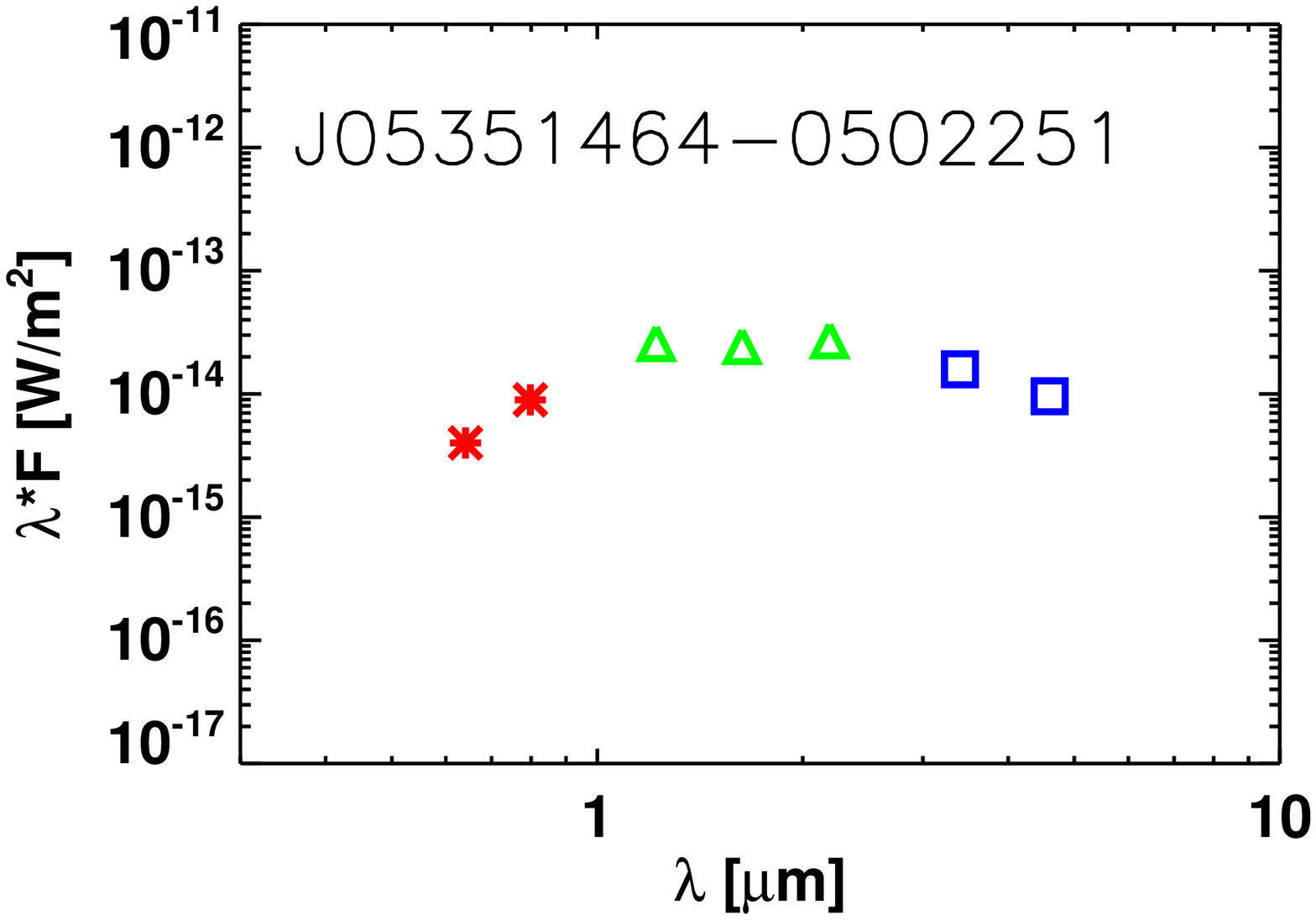}\hspace{0.3cm}\includegraphics[width=4.0cm]{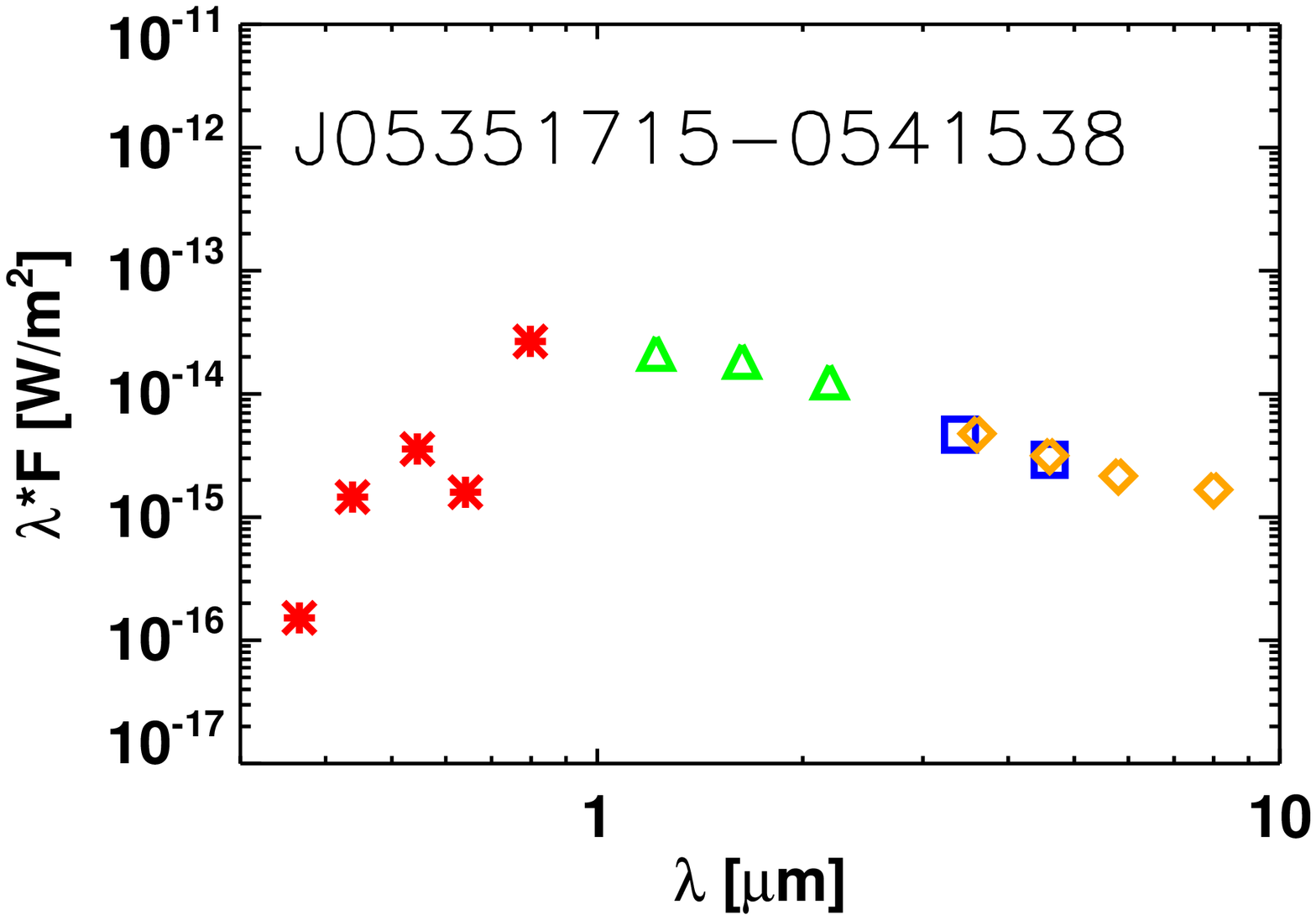}\hspace{0.3cm}\includegraphics[width=4.0cm]{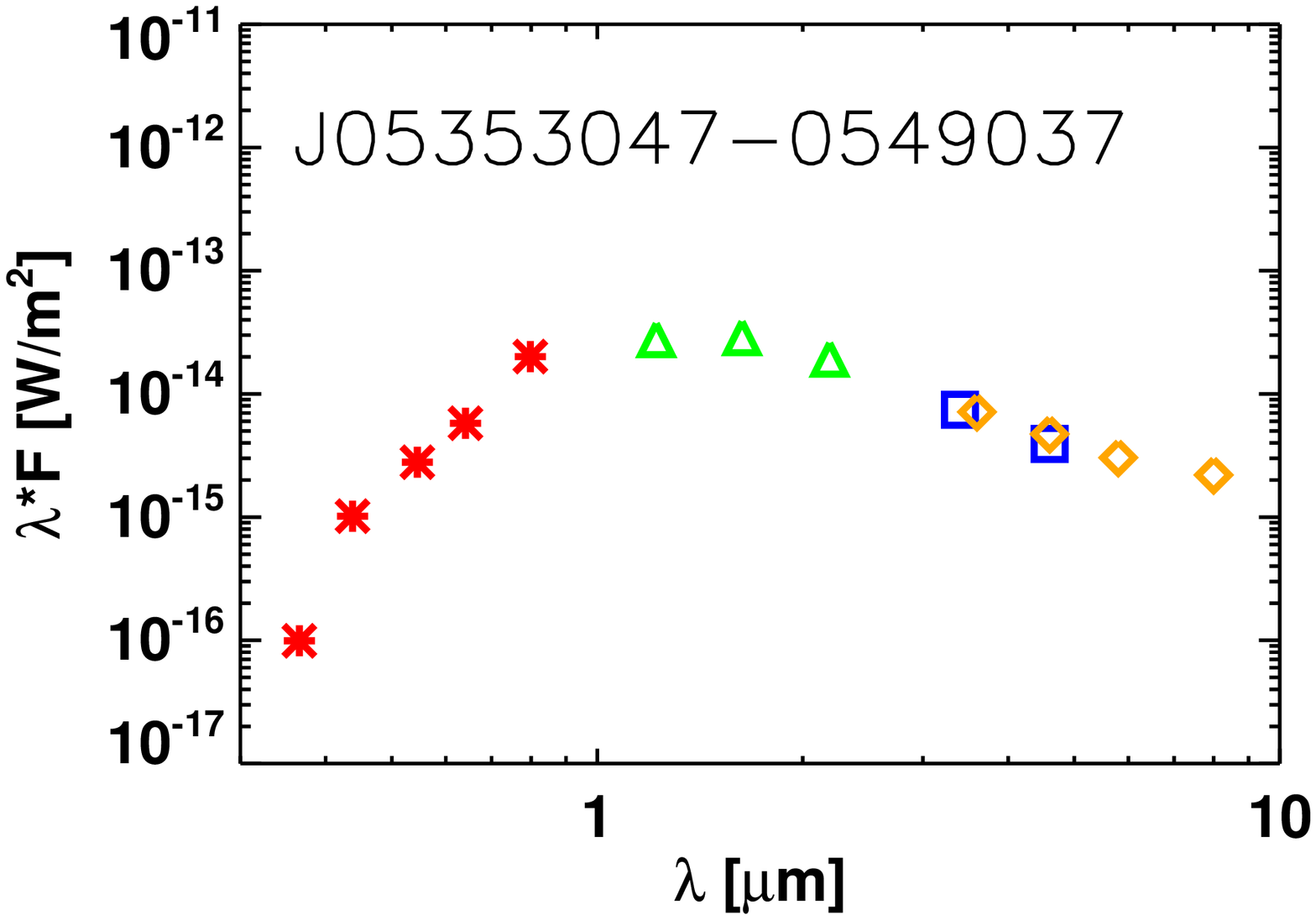}}
\caption{Spectral energy distribution for stars classified as WTTS, but exhibiting near- and mid-infrared excesses characteristic of CTTS. Symbols are same as in Figure~\ref{fig_varsed}.}
\label{fig_varsed2}
\end{figure*}
\addtocounter{figure}{-1}
\begin{figure*}
\centerline{
\includegraphics[width=4.0cm]{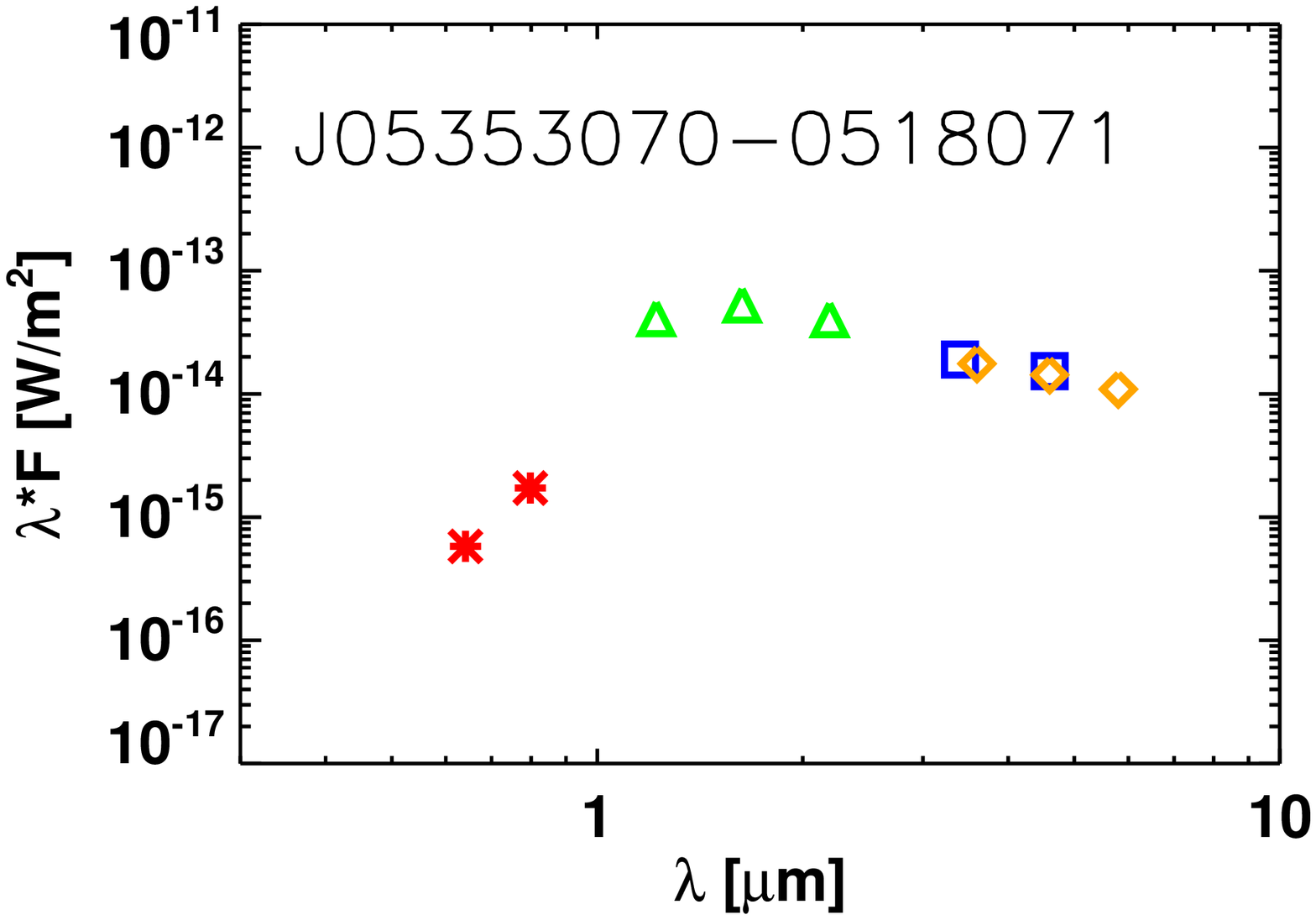}\hspace{0.3cm}\includegraphics[width=4.0cm]{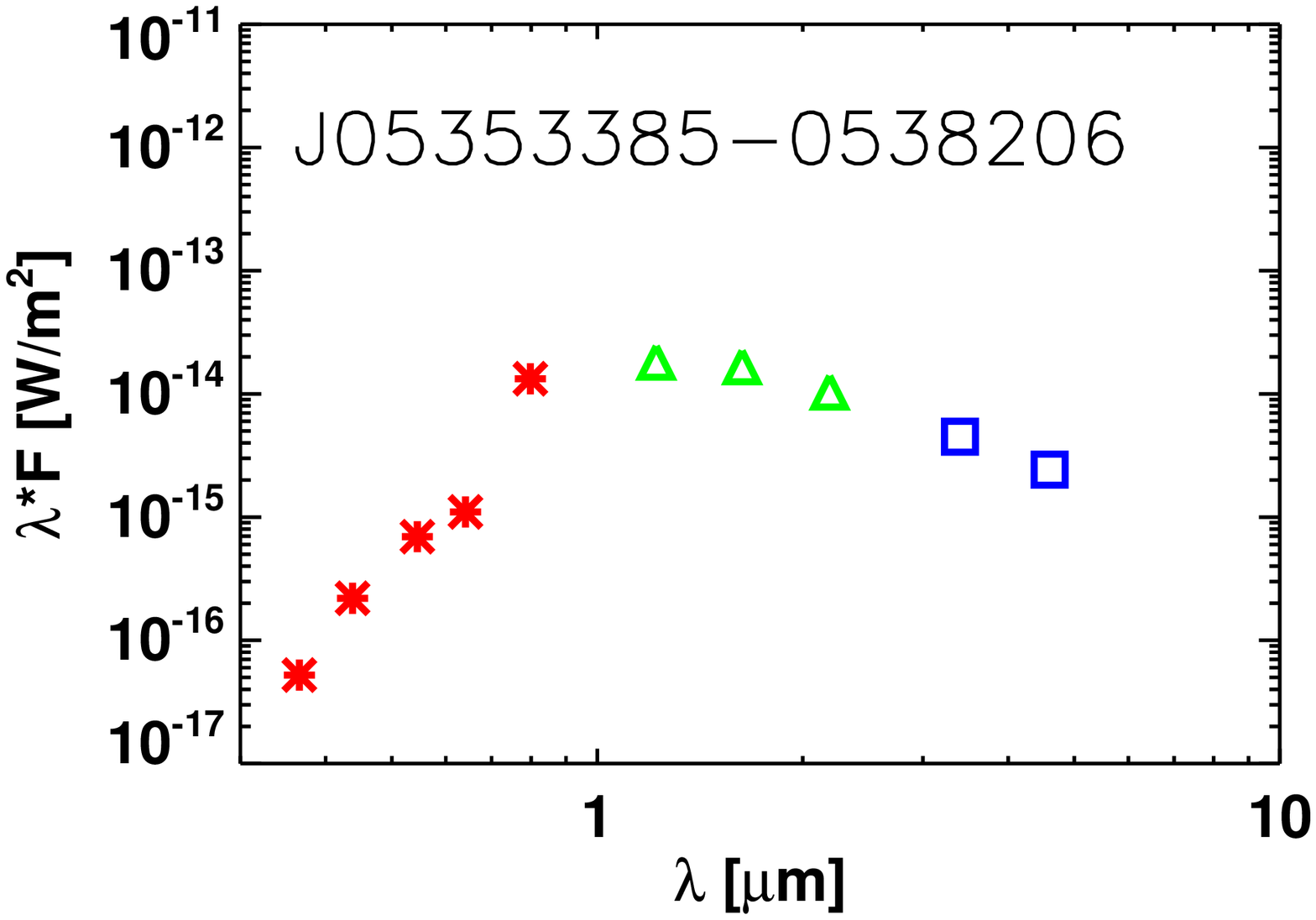}\hspace{0.3cm}\includegraphics[width=4.0cm]{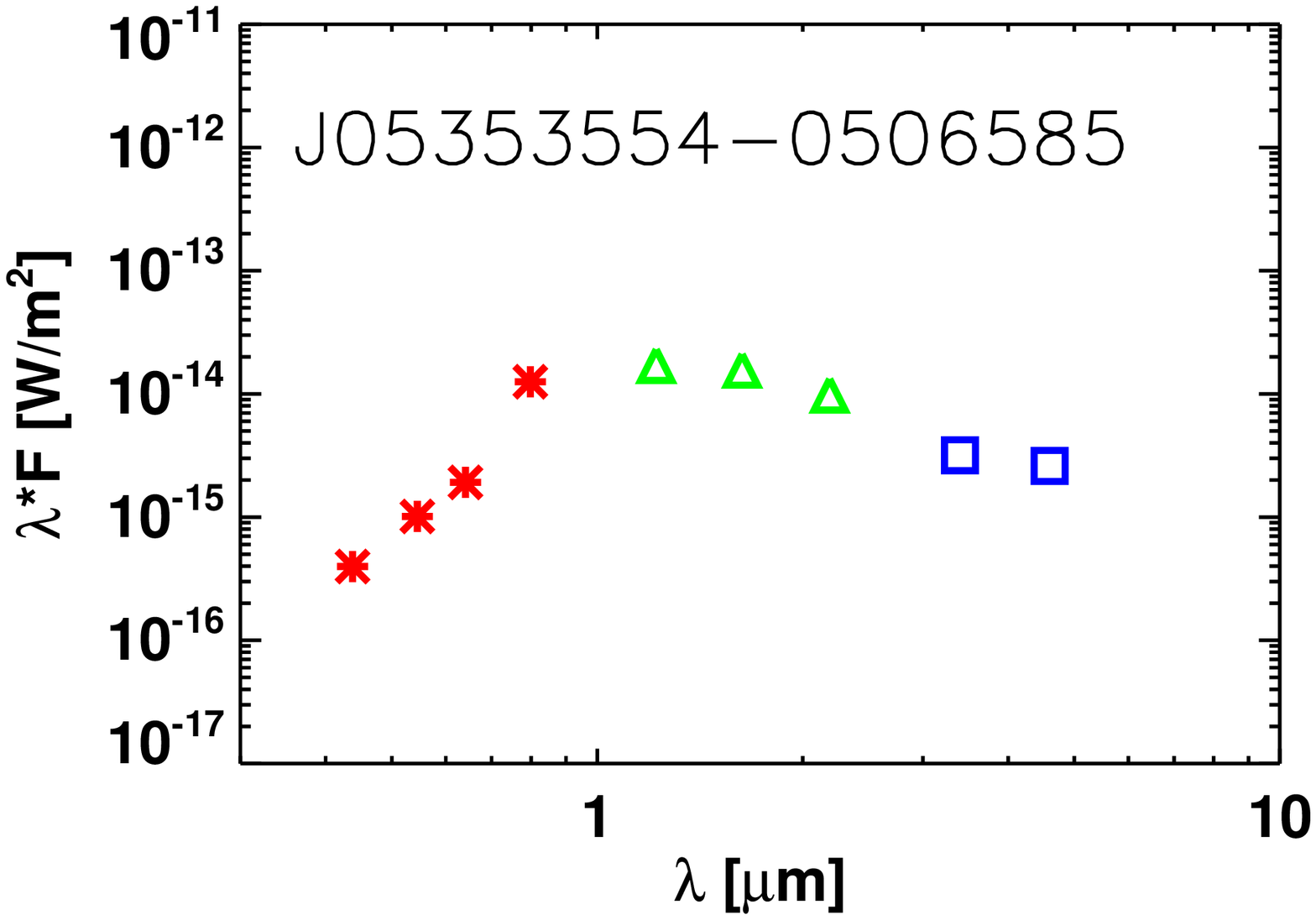}\hspace{0.3cm}\includegraphics[width=4.0cm]{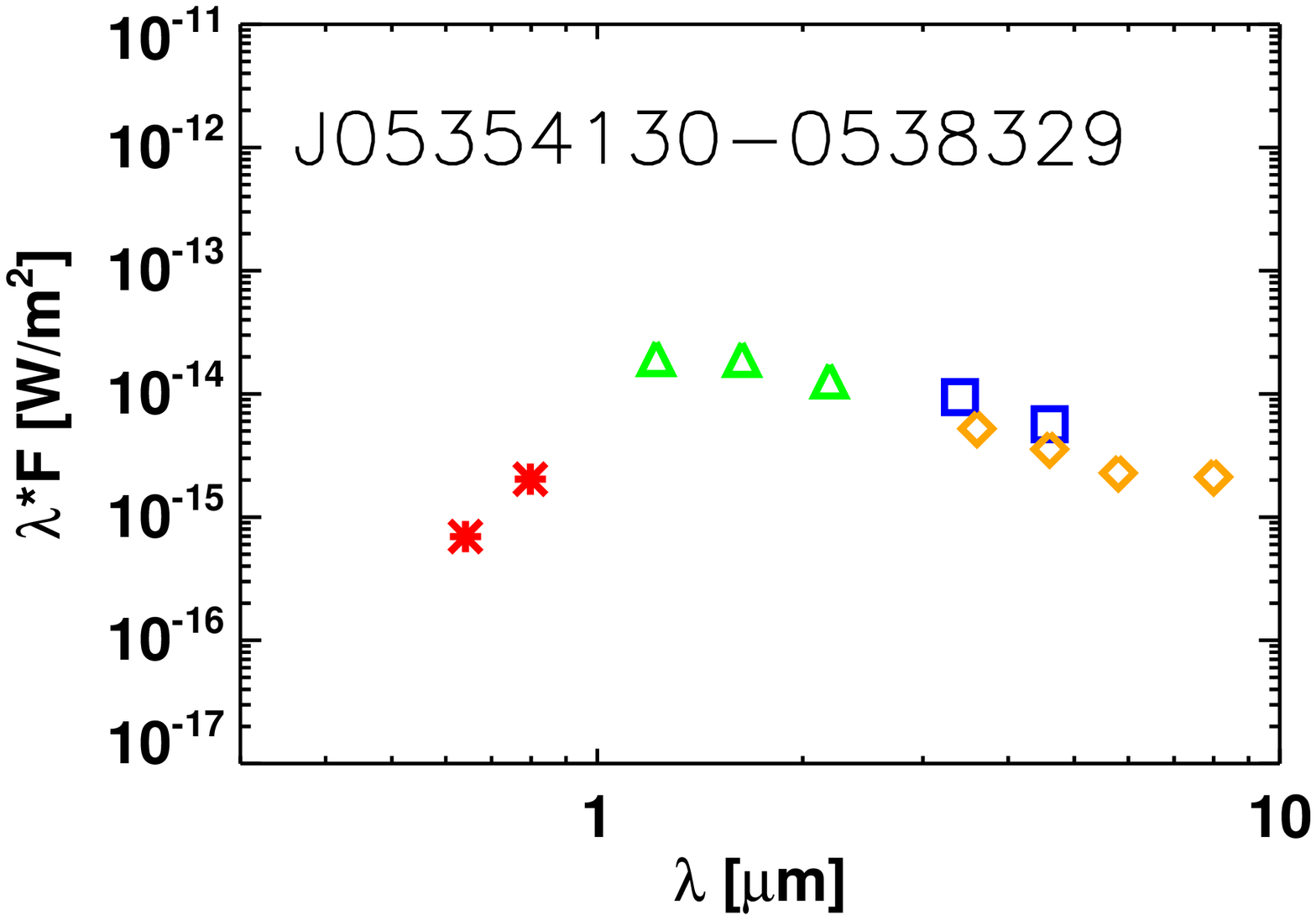}}
\centerline{
\includegraphics[width=4.0cm]{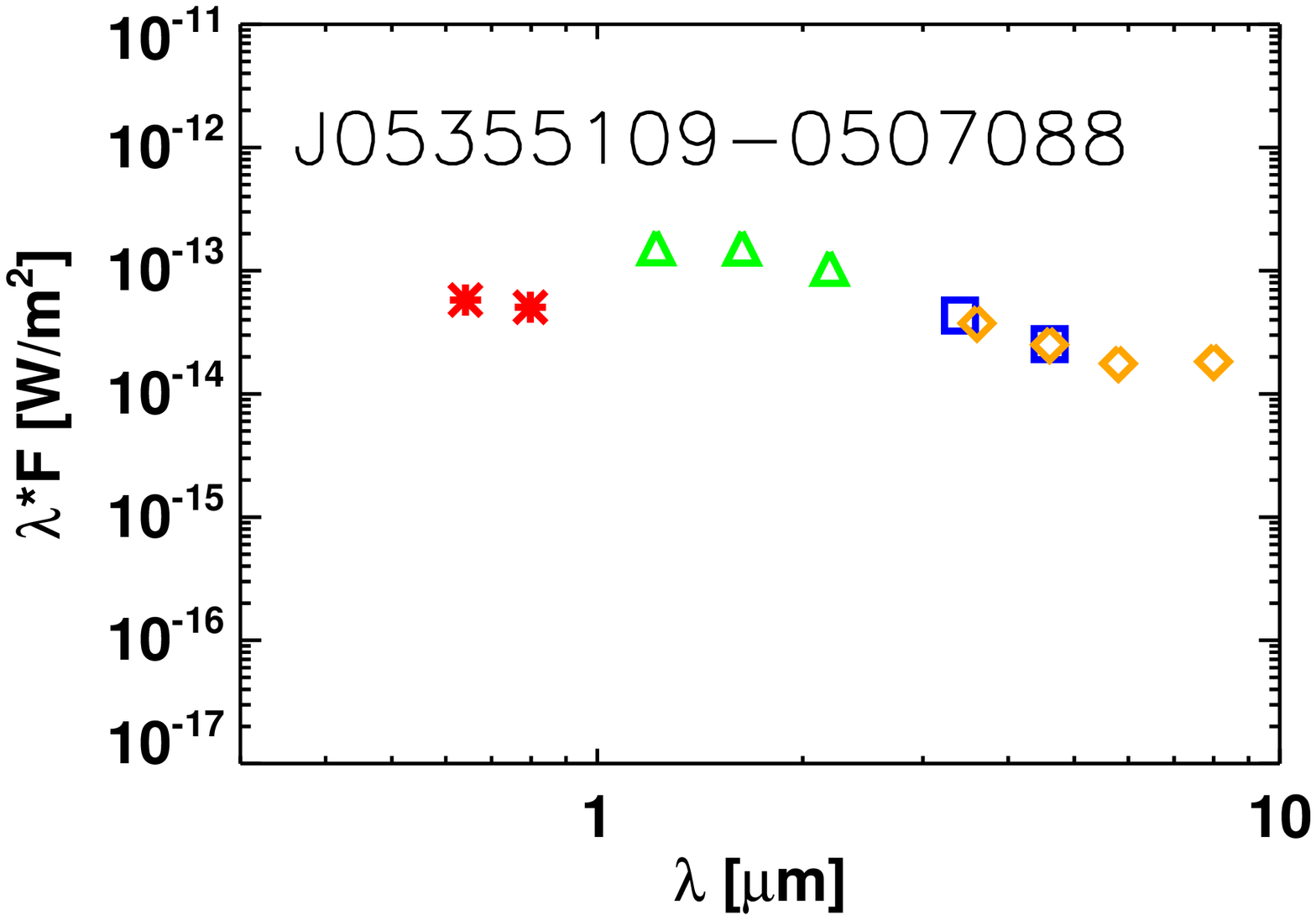}\hspace{0.3cm}\includegraphics[width=4.0cm]{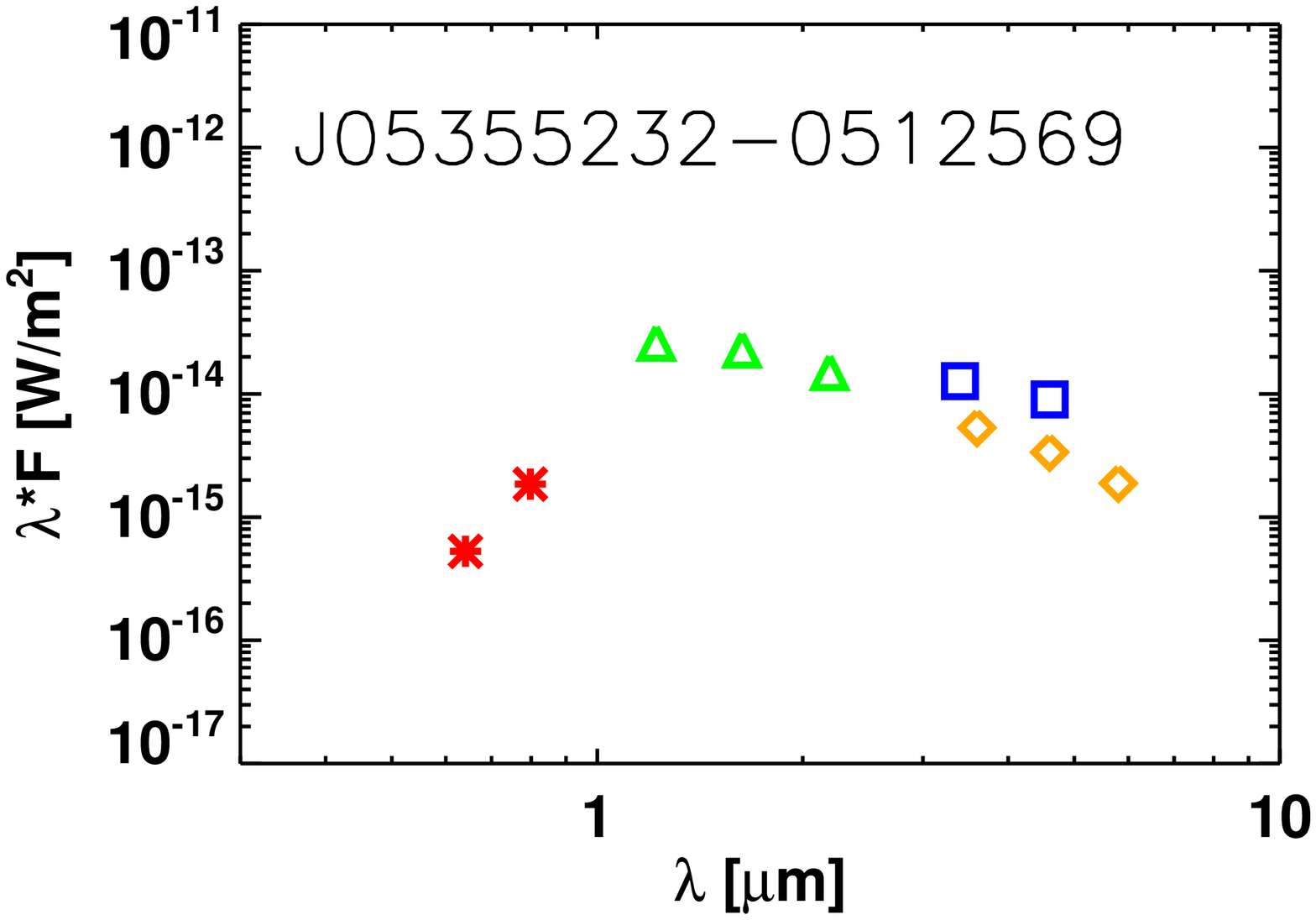}\hspace{0.3cm}\includegraphics[width=4.0cm]{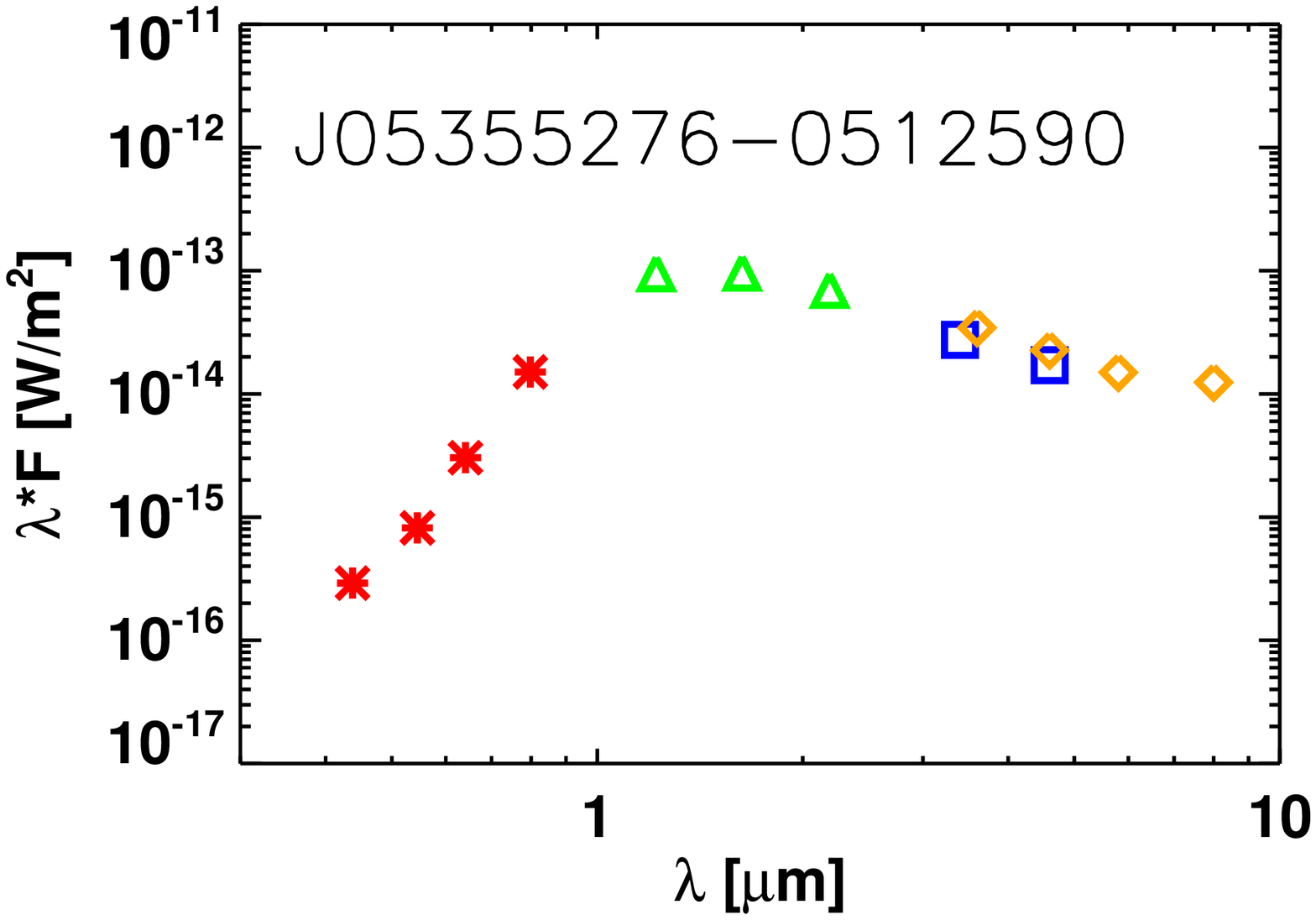}\hspace{0.3cm}\includegraphics[width=4.0cm]{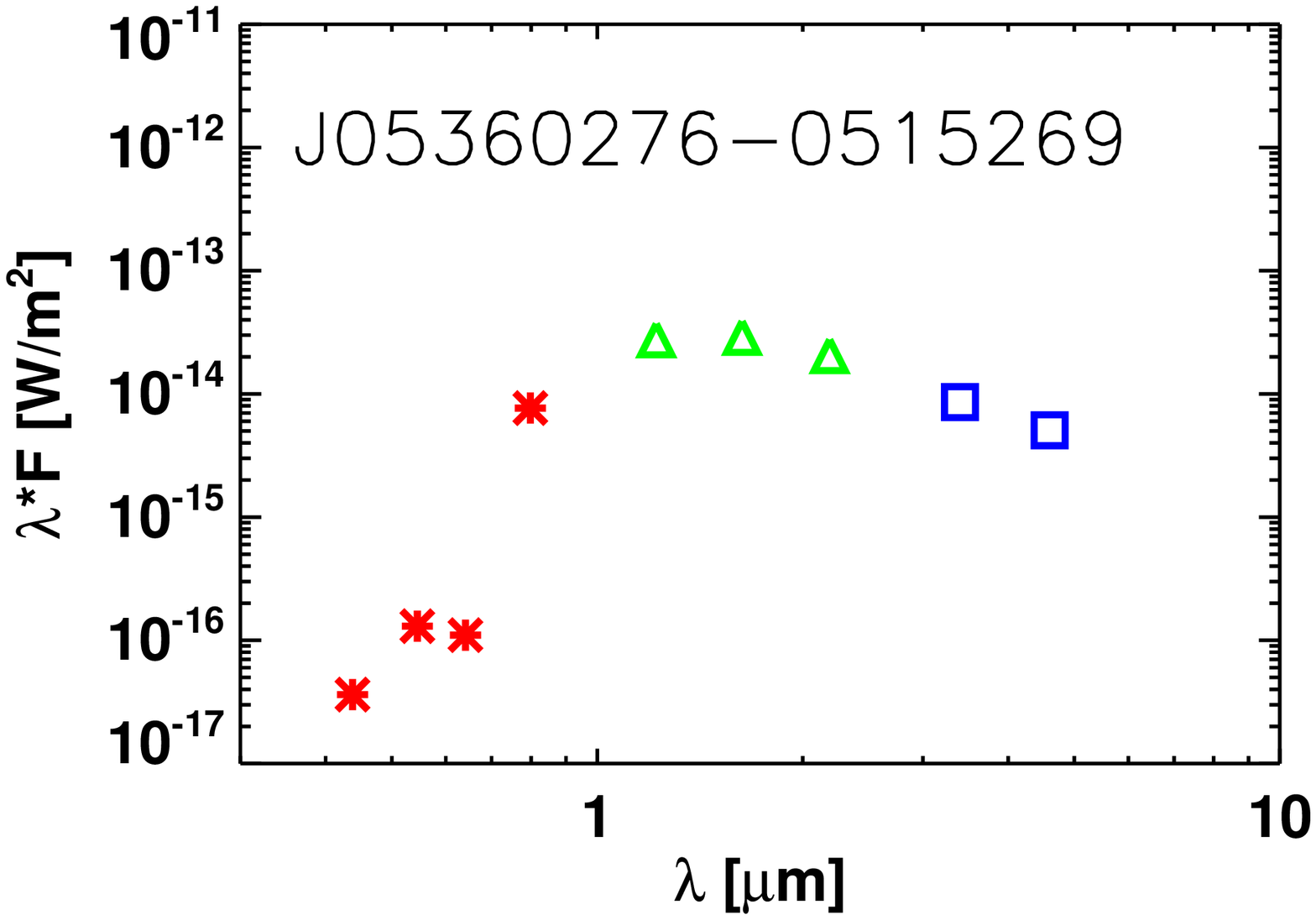}}
\centerline{
\includegraphics[width=4.0cm]{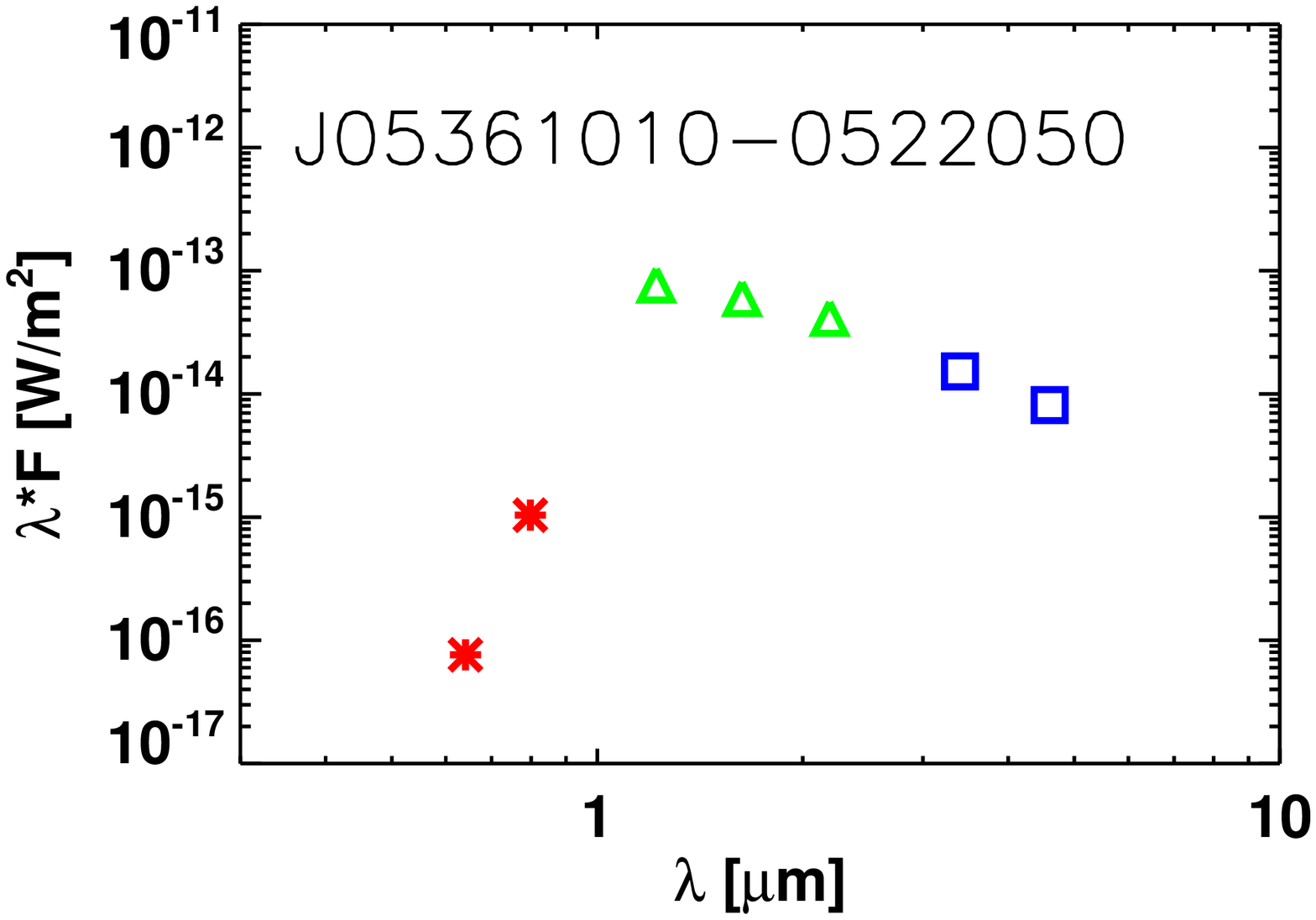}\hspace{0.3cm}\includegraphics[width=4.0cm]{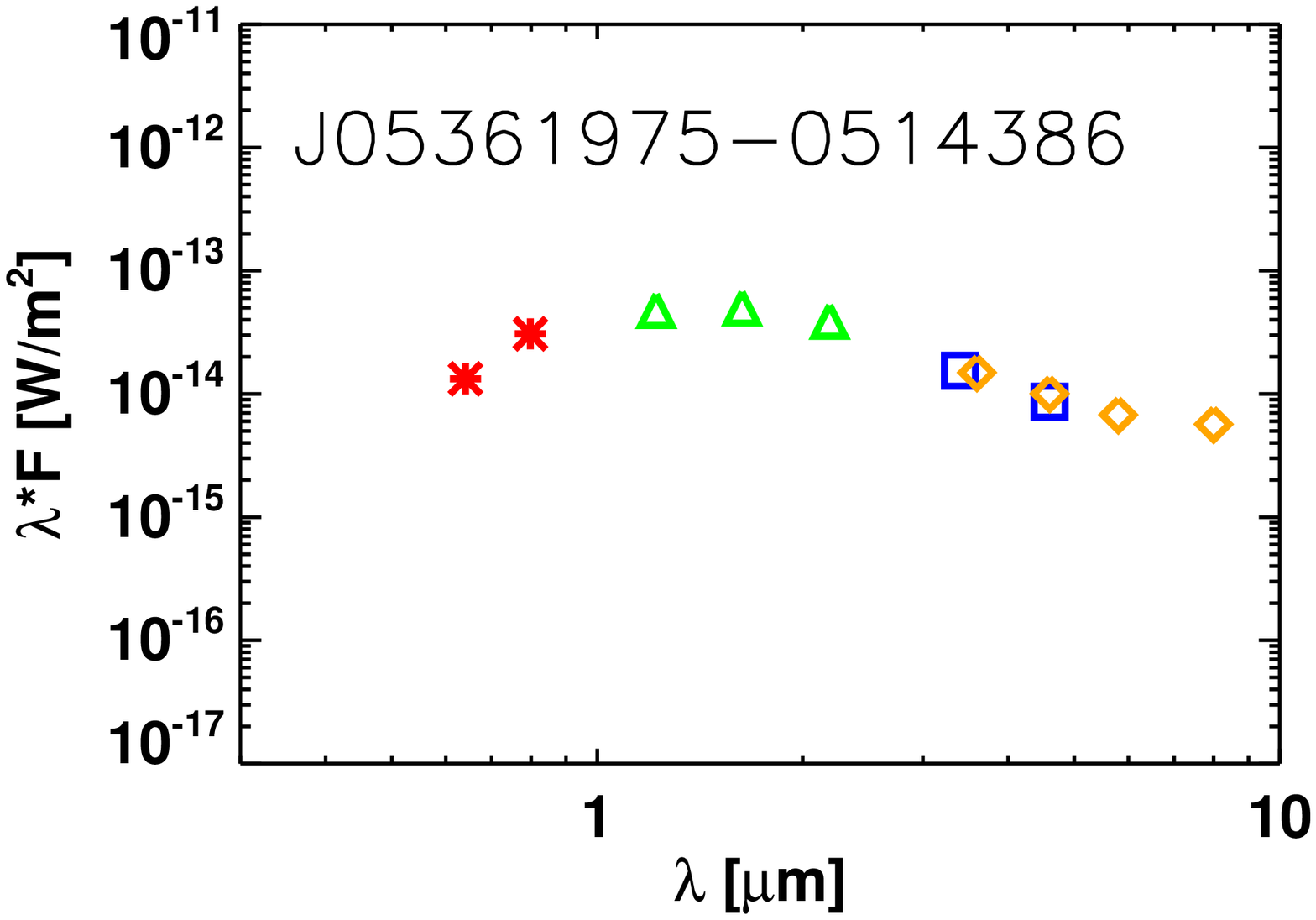}\hspace{0.3cm}\includegraphics[width=4.0cm]{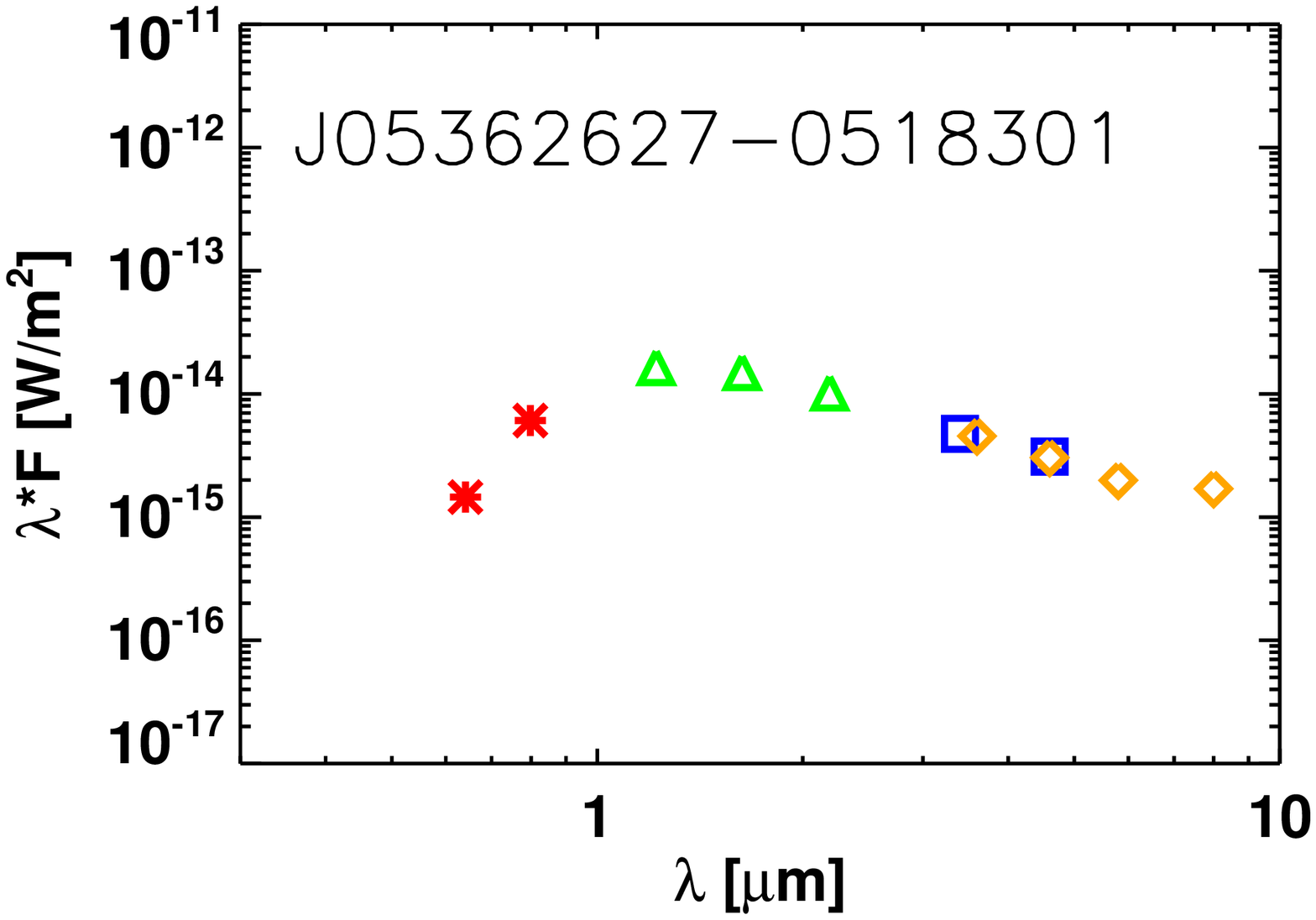}\hspace{0.3cm}\includegraphics[width=4.0cm]{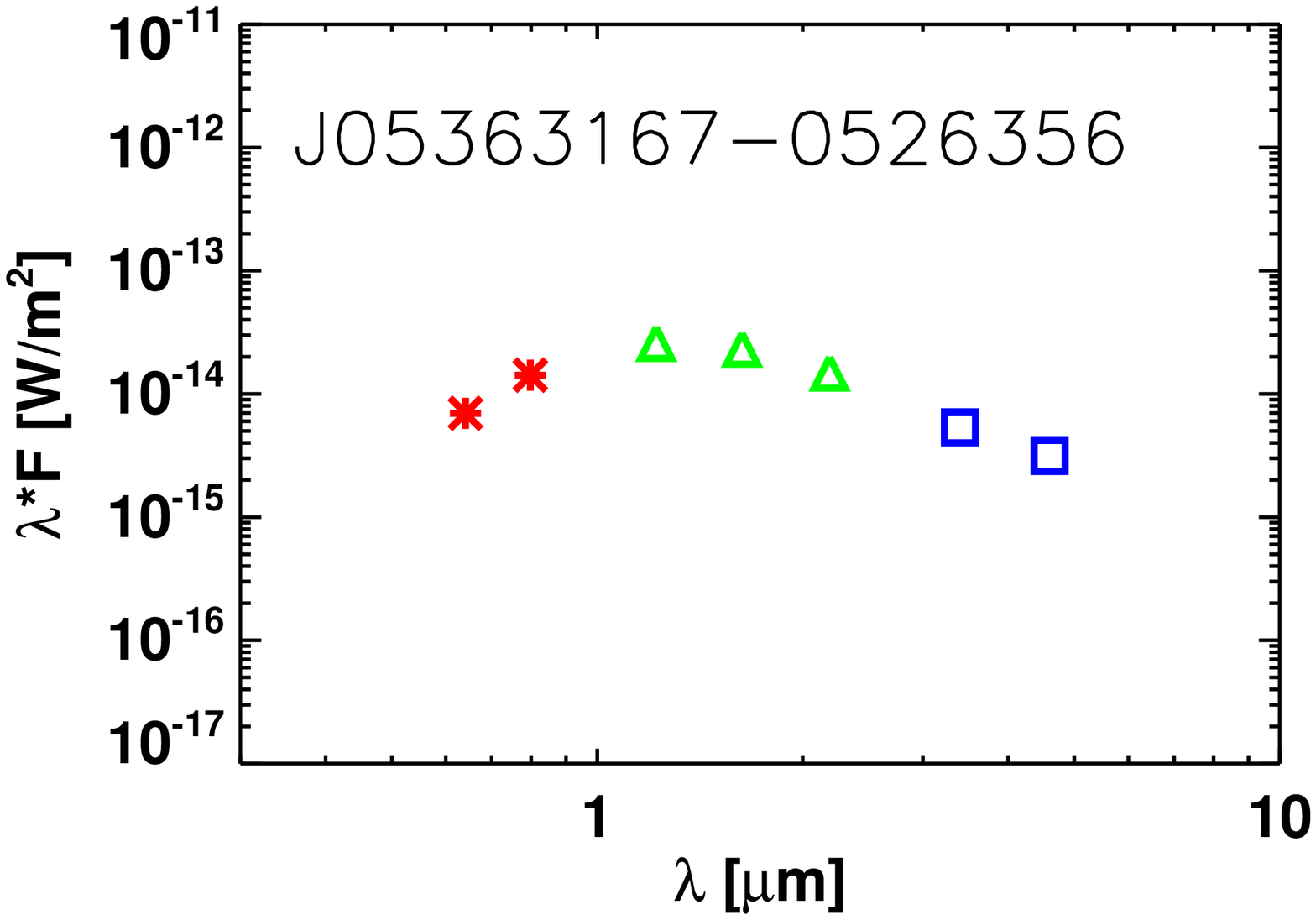}}
\centerline{
\includegraphics[width=4.0cm]{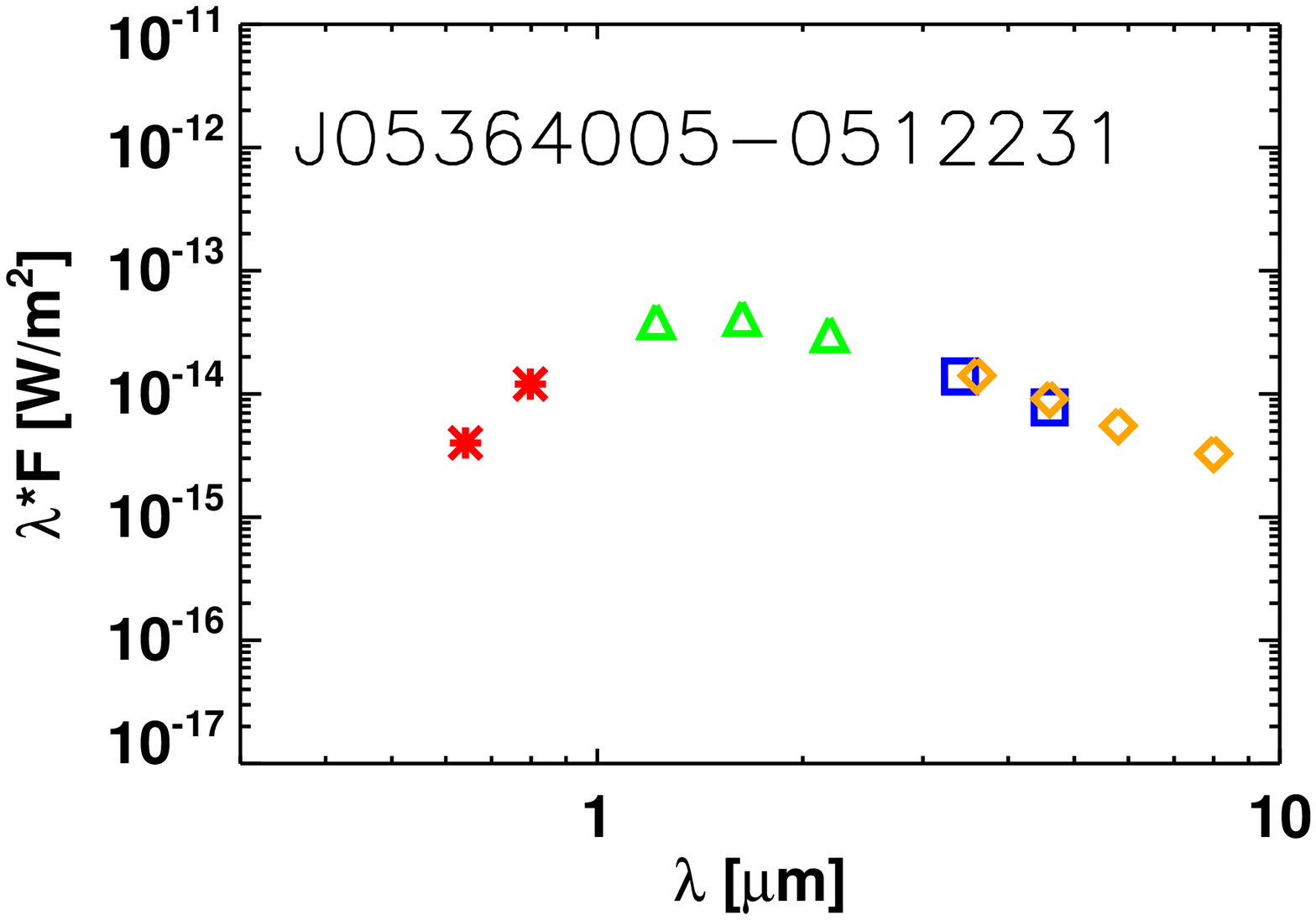}\hspace{0.3cm}\includegraphics[width=4.0cm]{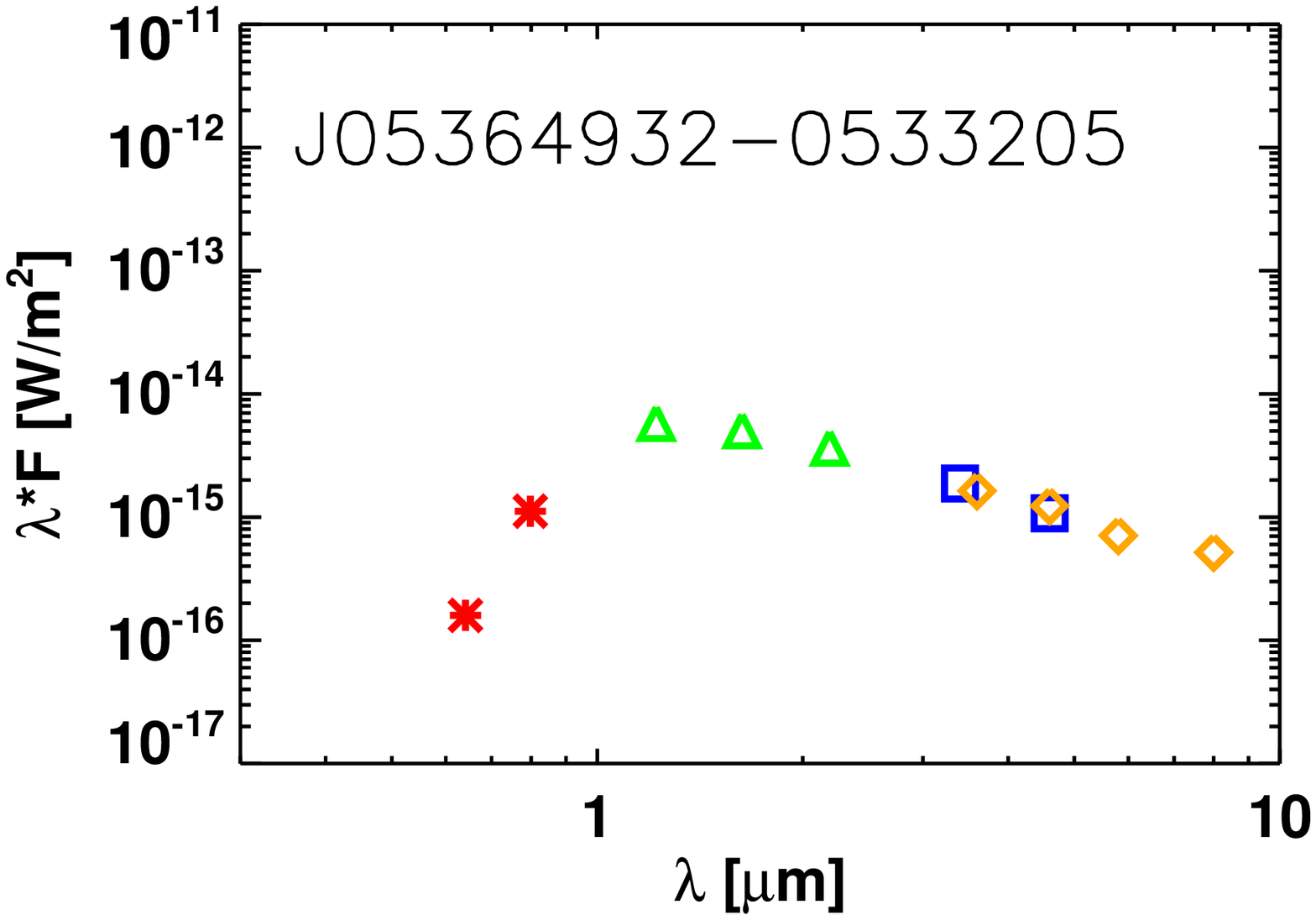}}
\caption{Continued.}
\label{fig_varsed2}
\end{figure*}

\subsection{Clustered and Distributed H$\alpha$ Emission Stars}
\label{sect_surf2}

\begin{figure}[!ht]
\begin{center}
\includegraphics[scale=0.8]{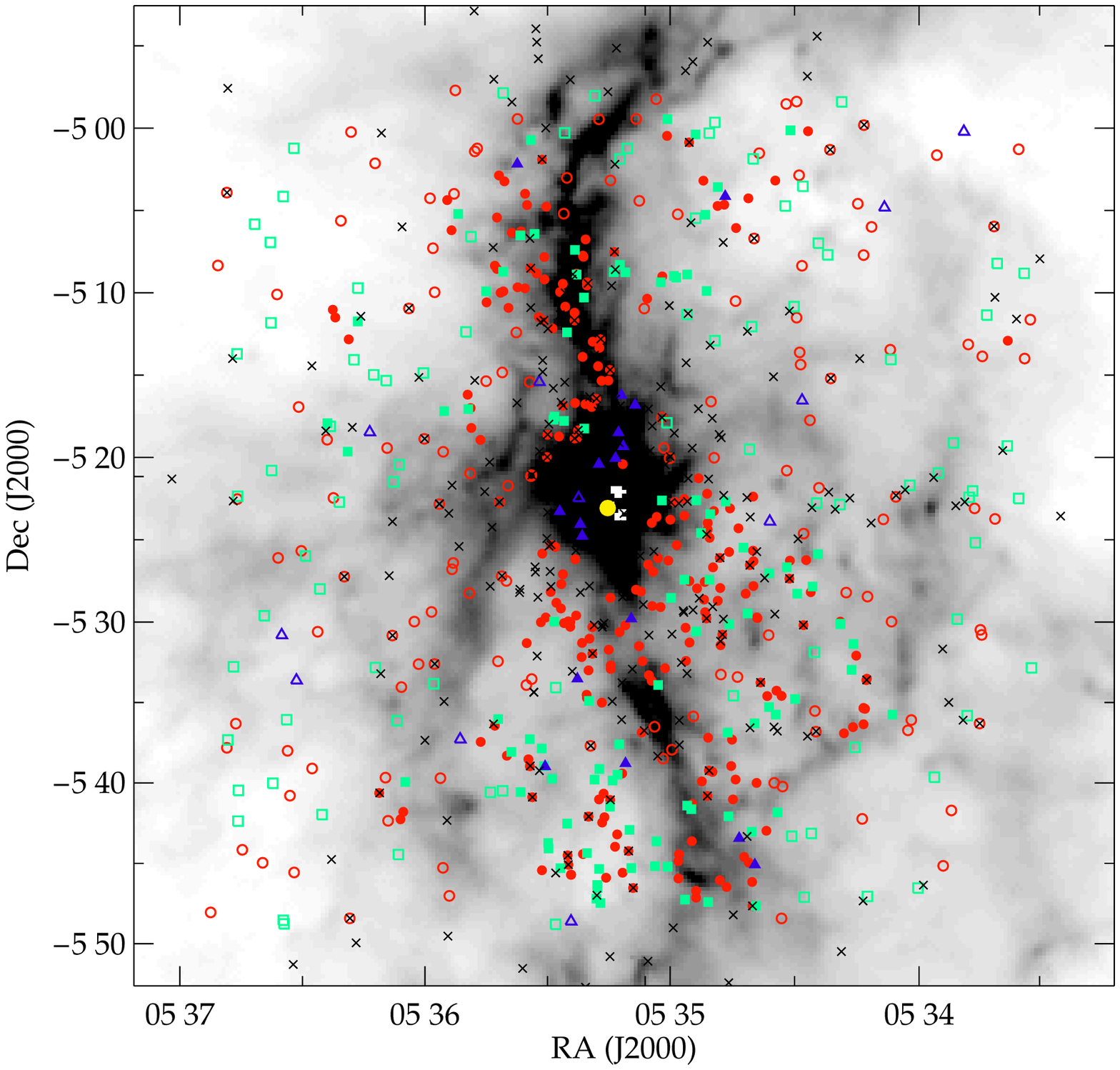}\vspace{0.5cm}
\figcaption{Distribution of the H$\alpha$ emission stars classified as clustered (filled symbols) and distributed (open symbols) based on the distance of their fourth nearest neighbors, overplotted on the  $500~\mu$m Herschel SPIRE image of the Orion region \citep{Andre2010}. Circles (red in the online version) indicate CTTS, squares (green in the online version) are for WTTS, and triangles (blue in the online version) represent sources of unknown equivalent width. The yellow dot symbol shows the position of $\theta^1$~Ori. It can clearly be seen that presence of the cold obscuring dust along the line of sight alone cannot be responsible for the observed inhomogeneities of the surface density of stars.
\label{fig_avmap}}
\end{center}
\end{figure}

To test whether the apparent clusters and voids in the surface distribution reflect patchy interstellar extinction, we plotted the cluster members and outliers, defined by their nearest neighbor distances, with different symbols onto the  $500~\mu$m {\it Herschel SPIRE\/} image of the Orion region, downloaded from the {\it Herschel Science Archive\/}. The data are part of the  Guaranteed Time Key Program ``Probing the origin of the stellar initial mass function: A wide-field Herschel photometric survey of nearby star-forming cloud complexes'' \citep[Herschel Gould Belt Survey, P. I.: P. Andr\'e,][]{Andre2010}. The structures in the $500~\mu$m radiation of the cold dust correlate well with those seen in the extinction maps of the region \citep{rowles2009,scandariato11}. We chose the far-infrared image to display the surface distribution of the absorbing dust since its angular resolution is higher than those of the available extinction maps. Figure~\ref{fig_avmap} suggests that variable extinction alone cannot account for the surface inhomogeneities of the stars. An apparent void can be seen next to the Trapezium where the bright background prevented us from detecting any emission line. The clustered stars (those having at least four neighbors within 2.16\arcmin) are found near the cold dust structures, suggesting that they are probably younger than the distributed population, and are closely associated with their natal clouds. Figure~\ref{fig_avmap} suggests that the lower average brightness of the distributed stars cannot result from a greater extinction along their line of sight. 

The difference in brightness between the clustered and distributed stars suggests that we observe two populations of pre-main sequence stars which differ from each other both in age and location in space. The lower average brightness of this scattered population suggests greater distance, higher age, or lower average mass of these stars. Since the presence of the massive molecular cloud, associated with the ONC, efficiently blocks the background stars, the greater distance is less probable than the older age and lower average mass. The increase in the brightness difference toward longer wavelengths points to a greater amount of circumstellar matter in cluster members with respect to the field stars, indicating a younger age of the clustered population. The presence of two overlapping populations toward the line of sight of ONC was suggested by the radial velocity survey of \citet{furesz}, and is demonstrated by the recent work of \citet{Alves12}. The distribution of our H$\alpha$ emission stars also reflects the presence of two, probably unrelated populations of pre-main sequence stars.

\subsection{Correlations of H$\alpha$ Equivalent Width  with Other Stellar Properties}
\label{sect_correl}

\subsubsection{H$\alpha$ Equivalent Width and Infrared Excess}
\label{sect_hair}

\begin{figure}[!ht]
\begin{center}
\includegraphics{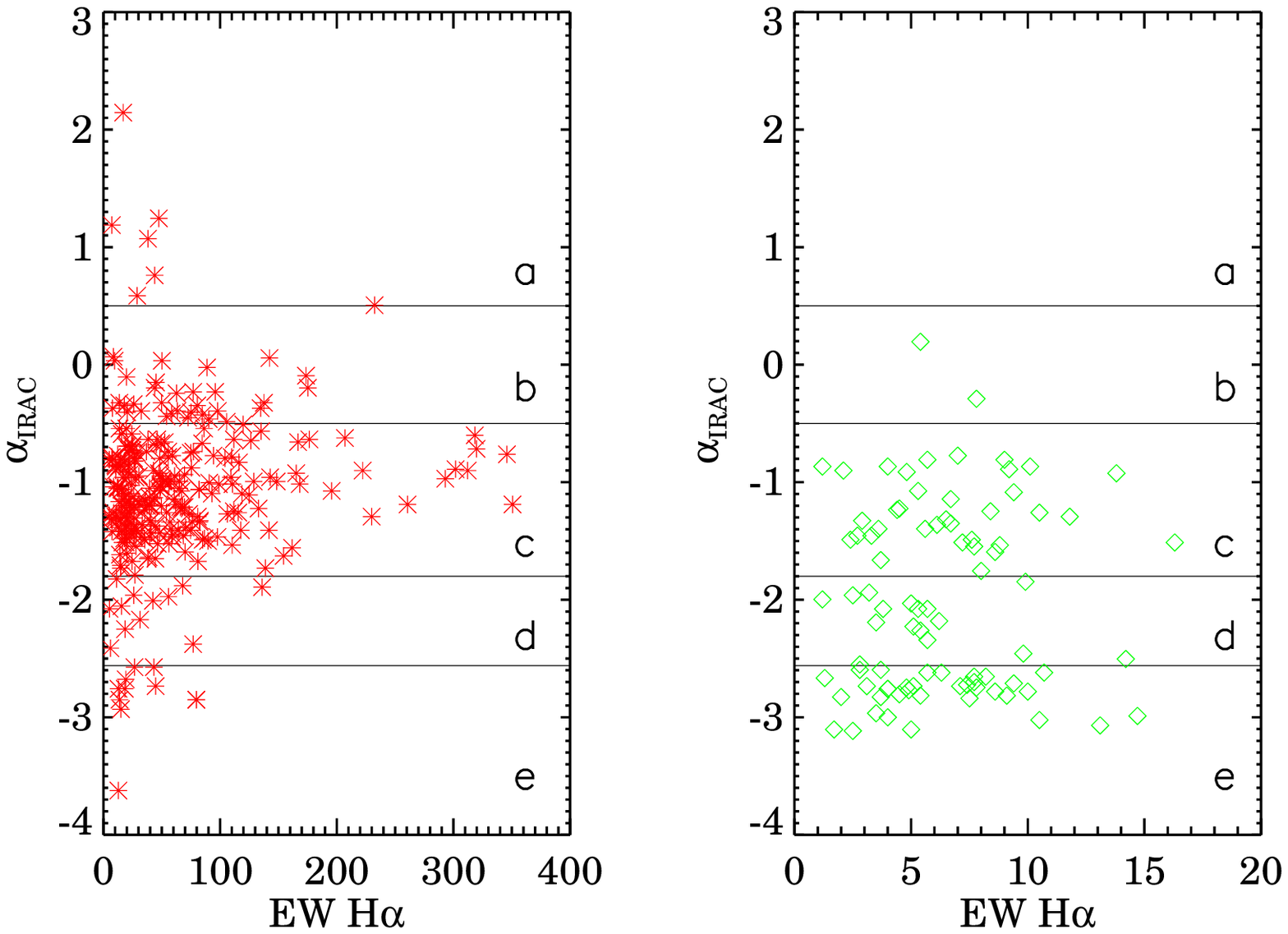}
\figcaption{Left: The H$\alpha$ equivalent width vs. $\alpha_{IRAC}$ for CTTSs. The horizontal lines indicate the separation between disk-less sources (e), sources with optically thin  disks (d), sources with optically thick disks (c), flat spectrum sources (b), and Class I sources (a), based on the SED classification criteria of \citet{lada2006} and \citet{index}. Right: The same for the WTTSs.
\label{fig_index}}
\end{center}
\end{figure}

The study of the slope of a young star's spectral energy distribution offers a great possibility to characterize the disk population of young stars. This method was applied e.g. by \citet{lada2006} and  \citet{index}. They measured the slope of the SED between 3.6\,$\mu$m and 8\,$\mu$m, and using this so-called $\alpha\sb{IRAC}$ index, sorted the disks into an evolutionary sequence. To get an idea of the evolutionary stages of the disk population in our sample, we applied this
method for the stars having {\it Spitzer\/}-data in the literature. Figure~\ref{fig_index} shows that most of our CTTS have $-1.8 < \alpha\sb{IRAC} < -0.5$, characteristic of optically thick accretion disks. The stars at  $\alpha\sb{IRAC} \lesssim -2.5$ exhibit low {\it EW\/}s, suggesting an overlap between our CTTS and WTTS classes. The WTTS sample clearly splits into a group of diskless stars at $\alpha\sb{IRAC} < -2.0$ and temporarily non-accreting disked stars at $-1.8 < \alpha\sb{IRAC} < -0.5$. 

\subsubsection{H$\alpha$ Equivalent Width and Rotation Period}
\label{sect_harot}

Theoretical predictions suggest that the rotation rate of young stars may be influenced by magnetic interactions with the accretion disk: the star is forced to corotate with the inner edge of the disk \citep[see e.g.][]{konigl1991}. If this disk-locking is real, classical T~Tauri stars should have longer rotational periods than weak-line T Tauri stars. 

\begin{figure}[!ht]
\begin{center}
\includegraphics{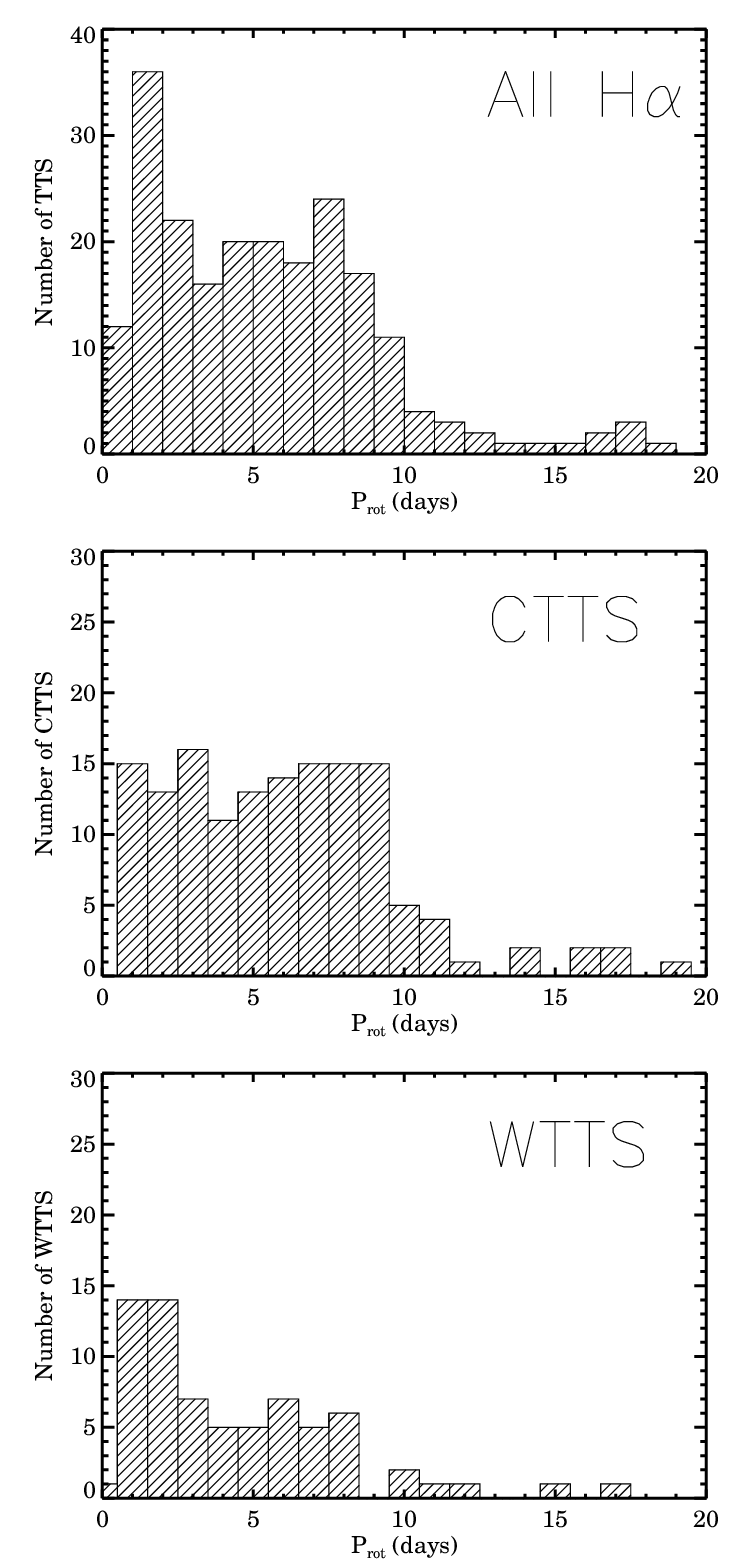}
\figcaption{Rotational period histograms for all H$\alpha$ emission stars (top), for CTTS (middle) and WTTS (bottom). Rotation period data are adopted from  \citet{wherbst}, \citet{cieza2007}, and \citet{flaccomio2005}.
\label{fig_rothist}}
\end{center}
\end{figure}

Of the $587$ TTS listed in Tables \ref{tabshort_ctts}, \ref{tabshort_wtts}, and \ref{tab_unc},  $214$ were found to be periodic variables by \citet{wherbst}, \citet{cieza2007}, and \citet{flaccomio2005}. Of the periodic variables, $144$ are CTTS and $70$ WTTS.  The period histograms for the whole H$\alpha$ emission star sample, both CTTSs and WTTSs, are shown in the upper, middle, and lower panel of Fig.~\ref{fig_rothist}, respectively, using the rotation data from  \citet{wherbst}, \citet{cieza2007}, and \citet{flaccomio2005}, binned into one day intervals. It can be seen clearly, that there are two peaks at $2$ and $8$ days in the top panel. \citet{wherbst} showed a similar bimodal distribution for stars having masses greater than $0.25$ M$_{\sun}$ in their rotational study. There is only one weakly defined peak at $8$ days in the middle panel, while in the bottom panel there is another peak at $2$ days. Applying the statistical Welch's test we found a significant difference between the rotation periods of CTTS and WTTS. The mean rotation period of the CTTS sample is statistically longer than that of the  WTTS group, as expected.

\subsubsection{H$\alpha$ Equivalent Width and X-Ray Emission}
\label{sect_hax}

X-ray luminosity of main-sequence stars shows a tight correlation with rotational period. This correlation confirms the existence of a solar-like dynamo as the origin of the X-ray activity \citep{pizzolato2003}.
Pre-main sequence stars show much greater X-ray emission than main sequence stars. The origin of this emission is still debated, but presumably originates from the coronal magnetic activity \citep{preibischa}. There are still some unanswered questions about X-ray emission of pre-main sequence stars.

Studies based on observations from the {\it Chandra} X-ray Observatory indicated that classical T Tauri stars in the Orion Nebula Cluster may have lower levels of X-ray activity than weak-line T Tauri stars  \citep{flaccomio2005,preibischb}. Similar results were published for the Taurus--Auriga association \citep{neuhauser1995,stelzer2001}. \citet{telleschi} found a significant difference in the X-ray luminosity functions of CTTS and WTTS based on the measure of XMM-Newton Extended Survey of Taurus molecular cloud (XEST) in the Taurus--Auriga molecular cloud. The CTTS are weaker in X-rays by about a factor of two. In the open cluster IC 348, \citet{stelzer12} found, that for the lowest examined masses (0.1-0.25 M$\sb{\odot}$) there is an important difference between accreting and non-accreting stars. Explanations for this phenomenon are either disk-locking \citep{neuhauser97} or distortion of the magnetic field structure \citep{preibischb}. 
In contrast with these results, \citet{feigelson2002} found no evidence for the influence of an accretion disk on the activity levels of ONC stars.

To test this relationship in our sample which contains both classical and weak-line T Tauri stars, we used X-ray data from the {\it XMM-Newton Serendipitous Source Catalogue}. 
A simple way to compare the X-ray luminosity for CTTS and WTTS is the statistical Welch's test. The {\it t} variable is used for testing the significance of the difference between the corresponding groups. Applying this test for comparison between the X-ray luminosity for CTTS and WTTS, shows that there is no difference between the mean of X-ray luminosity of our CTTS and WTTS. However, since our H$\alpha$ emission stars represent a strongly biased and incomplete sampling of the cluster population, this result does not conflict with those established for the whole ONC.


We examined the ratio of detected and non-detected stars within the {\it XMM-Newton\/} field of view, and found that whereas $151$ of the $342$ CTTS in the {\it XMM\/}-field were detected ($44.2$ \%), $106$ of the $154$ WTTS ($68.8$\%) were above the detection threshold of the X-ray observatory. A Chi-square test demonstrated that this difference is significant at a reasonably high level. According to this test the hypothesis that the difference in the detection rate (detected/non-detected) is independent of the type can be rejected with a very low error probability. In other words the fraction of X-ray-detected WTT type objects is significantly greater than that of the CTT group.

We examined the differences between X-ray detected and non-detected objects in other measured quantities, such as EW(H$\alpha$), rotation period, {\it WISE\/}, {\it Spitzer\/}, and {\it 2MASS\/} magnitudes. First we studied the whole sample and then the case of the CTT and WTT groups, separately. The results, summarized in Table~\ref{stat}, show that the EW values of the stars not detected in X-rays are greater than those of the detected ones. The rotational periods are somewhat longer for the X-ray-detected T~Tau stars, the differences, however, exceed only slightly the standard error of the mean. As far as the {\it 2MASS\/}, {\it Spitzer\/}, and {\it WISE\/} fluxes are concerned, the X-ray detected objects are systematically brighter in all bands. The brightness difference between the X-ray detected and non-detected objects decreases with the increasing wavelength. We applied Welch's test to study the significance of these values. We carried out this test for each parameter separately. The results demonstrate that except for the rotational period all the other variables have significant differences between the X-ray detected and non-detected stars. The most striking difference appears in  {\it EW\/}(H$\alpha$). The greater infrared fluxes of the stars above the X-ray detection limit suggest that they are more massive than those below the detection limit.

\begin{deluxetable}{cccccccccccccc}

\tabletypesize{\scriptsize}
\tablecolumns{14}
\tablewidth{0pt}
\tablecaption{Comparison of H$\alpha$ emission stars detected and not detected by {\it XMM--Newton} (each located within the field of view of {XMM}). \label{stat}}
\tablehead{
\colhead{Quantity} & \colhead{Class}  &  \multicolumn{3}{c}{Mean}  & \multicolumn{3}{c}{Std. Dev.} & \multicolumn{3}{c}{Std. Error Mean} & \multicolumn{3}{c}{Number of stars}\\
\cline{1-14}
& & \colhead{All} & \colhead{CTT} & \colhead{WTT} & \colhead{All} & \colhead{CTT} & \colhead{WTT} & \colhead{All} & \colhead{CTT} & \colhead{WTT} &  \colhead{All} & \colhead{CTT} & \colhead{WTT} \\}
\startdata
{\it 2MASS2\/} J  & non detected & 13.15 & 13.25 & 12.77 & 1.21 & 1.26 & 0.87 & 0.08 & 0.08 & 0.11 & 238 & 190 & 48\\
         & detected     & 12.14 & 12.03 & 12.28 & 1.00 & 1.08 & 0.85 & 0.06 & 0.08 & 0.10 & 256 & 150 & 106 \\
\cline{1-14}
{\it 2MASS2\/} H  & non detected & 12.28 & 12.35 & 12.14 &  1.23 & 1.32 & 0.85 &  0.08 & 0.08 & 0.12 & 238 & 190 & 48 \\
         & detected     & 11.31 & 11.15 & 11.60 &  1.01 & 1.15 & 0.88 & 0.06 & 0.08 & 0.10 & 256 & 150 & 106 \\
\cline{1-14}
{\it 2MASS2\/} K$_\mathrm{s}$  & non detected &  11.79 & 11.81 & 11.87 &  1.28 & 1.35 & 0.88 & 0.09 & 0.08 & 0.12 & 238 & 190 & 48  \\
                               & detected     &  10.91 & 10.67 & 11.33 &  1.16 & 1.25 & 0.89 & 0.10 & 0.09 & 0.10 & 256 & 150 & 106  \\
\cline{1-14}
{\it Spitzer\/} [3.6] &  non detected & 10.78 & 10.77 & 10.93 & 1.79 & 1.83 & 1.30 & 0.13 & 0.14 & 0.33 & 179 & 163 & 16  \\
                      &  detected     & 9.64  & 9.65  & 9.63  & 1.44 & 1.21 & 2.34 & 0.12 & 0.11 & 0.49 & 143 & 120 & 23  \\ 
\cline{1-14}
{\it Spitzer\/} [4.5] &  non detected & 10.37 & 10.35 & 10.54 & 1.79 & 1.82 & 1.40 & 0.13 & 0.14 & 0.35 & 179 & 163 & 16  \\ 
                      &  detected     & 9.27  & 9.23  & 9.40  & 1.50 & 1.29 & 2.33 & 0.13 & 0.12 & 0.49 & 143 & 120 & 23  \\
\cline{1-14}
{\it Spitzer\/} [5.8] &  non detected & 9.90  & 9.91  & 9.83 & 1.82 & 1.84 & 1.66 & 0.14 & 0.15 & 0.43 & 169 & 154 & 15  \\
                      &  detected     & 8.93  & 8.90  & 9.07 & 1.53 & 1.35 & 2.28 & 0.13 & 0.14 & 0.48 & 141 & 118 & 23  \\
\cline{1-14}
{\it Spitzer\/} [8.0] &  non detected & 8.99  & 9.02  & 8.64 & 1.91 & 1.83 & 2.73 & 0.16 & 0.16 & 0.79 & 144 & 132 & 12  \\
                      &  detected     & 8.12  & 8.06  & 8.41 & 1.60 & 1.46 & 2.16 & 0.14 & 0.14 & 0.45 & 135 & 112 & 23  \\ 
\cline{1-14}
 {\it Spitzer\/} [24] &  non detected & 5.89  & 5.90  & 5.84 & 1.84 & 1.84 & 2.04 & 0.18 & 0.19 & 0.68 & 100 & 91 & 9  \\
                      &  detected     & 4.97  & 4.90  & 5.31 & 1.56 & 1.52 & 1.76 & 0.15 & 0.16 & 0.40 & 109 & 90 & 19  \\
\cline{1-14}
{\it WISE W1\/}  & non detected & 11.01 & 10.93 & 11.27 & 1.34 & 1.43 & 0.92 & 0.10 & 0.12 & 0.14 & 183 & 143 & 40 \\
         & detected     & 10.37 & 9.99 & 10.92  & 1.40 & 1.47 & 0.95 &  0.09 & 0.12 & 0.12 &  233 & 137  & 96 \\
\cline{1-14}
{\it WISE W2\/}  & non detected & 10.47 & 10.34 & 10.90 & 1.42 & 1.50 & 0.95 & 0.10 & 0.13 & 0.14 & 183 & 143 & 40 \\
         & detected     & 9.96  & 9.47  & 10.66 & 1.54 & 1.59 & 0.98 & 0.10 & 0.14 & 0.12 & 233 & 137 & 96 \\
\cline{1-14}
Rot. period  & non detected & 5.33 & 5.72 & 4.21 & 3.78 & 3.72 & 3.91 & 0.43 & 0.49 & 0.98 &  78 & 58 & 20 \\
             & detected     & 5.56 & 6.14 & 4.84 & 3.72 & 3.63 & 3.64 & 0.33 & 0.43 & 0.59 &  131 & 72 & 59 \\
\cline{1-14}
EW(H$\alpha$)  & non detected & 60.29 & 74.17 & 5.63 & 70.72 & 72.93 & 2.59 &  5.31 & 4.92 & 0.35 & 237 & 189 & 48 \\
               & detected     & 31.52 & 49.76 & 5.70 & 52.62 & 62.65 & 2.73 & 5.11 & 6.42 & 0.32 & 252 & 150 & 106 \\
    
\enddata
\end{deluxetable}
																	
\section{SUMMARY}\label{sect_sum}

We identified $587$ stars with H$\alpha$ emission in a one square degree area centered on the ONC, $99$ of which have not appeared in previous H$\alpha$ surveys. We determined the equivalent width  of the line for $559$ stars, and based on it classified $372$ stars as classical T Tauri and $187$ as weak line T Tauri star. The Halpha survey is strongly biased towards the minority population of CTTs and misses the majority population
of WTTs reported in the Chandra X-ray sample. Compared with published data we found 2--3-fold variations in the equivalent width of the H$\alpha$ line for the greater part of our sample. In a few cases the variations are 10--20-fold.

We examined the surface distribution of the H$\alpha$ emission stars and, based on the angular distance of their $4th$ nearest neighbors, defined a clustered and distributed population. We compared the properties of the clustered and distributed sources and found the mean brightness of the clustered stars to be statistically greater in all photometric bands. The clustered population is associated with cold dust structures, suggesting that these stars are younger than the dispersed population. The dispersed stars may be lower mass foreground objects, possibly associated with another, older subsystem of the Orion star forming complex.

We studied the correlation between the equivalent width of the H$\alpha$ line and other properties of the stars and found that
\begin{itemize}
\item[a)] According to the slope of their spectral energy distributions over the 3.6--8\,$\mu$m wavelength region, our CTTSs show Class~II SEDs, characteristic of optically thick accretion disks. The non-accreting WTTS sample splits into a diskless (Class~III) and a disked (Class~II) subgroup, suggesting that phases of very low accretion rate occur during the evolution of primordial protoplanetary disks. 
\item[b)] The CTTS of our sample have longer rotational period than WTTS, in accordance with theoretical results.
\item[c)] We examined the X-ray counterparts of the H$\alpha$ emission stars in the {\it XMM-Newton\/} data base. We found that, although the detection threshold in the X-ray regime is the same for the CTT and WTT classes, a much higher fraction of WTTS have been detected in X-rays than of CTTS. We found that except for the rotational period all measured variables have significant differences between the X-ray detected and non-detected stars. The most striking difference appears in the {\it EW\/}(H$\alpha$): the X-ray detected CTTS show much lower mean equivalent width than the non-detected ones. 
\end{itemize}

\section{ACKNOWLEDGMENTS}
This research is based on observations with the $2.2$-m telescope of the University of Hawaii and we thank Colin Aspin for his interest and support. This publication makes use of data products from the Wide-field Infrared Survey Explorer, which is a joint project of the University of California, Los Angeles, and the Jet Propulsion Laboratory/California Institute of Technology, founded by the National Aeronautics and Space Administration, and data products from the Two Micron All Sky Survey, 
which is a joint project of the University of Massachusetts and the Infrared Processing and Analysis Center/California Institute of Technology, founded by 
the National Aeronautics and Space Administration and the National Science Foundation. This work also makes use of observations made with the \textit{Spitzer Space Telescope}, which is operated by the Jet Propulsion Laboratory, California Institute of Technology under a contract with NASA.  This research was supported by the Hungarian OTKA grant K81966, the National Aeronautics and Space Administration through the NASA Astrobiology Institute under Cooperative Agreement No. NNA09DA77A issued through the Office of Space Science, the MB08C 81013 Mobility-grant of the Mag Zrt, “Lend\"ulet” grant LP2012-31/2012.
  This research has made use of the VizieR catalogue access tool, CDS, Strasbourg, France. This research has made use of data from the Herschel Gould Belt survey (HGBS) project (http://gouldbelt-herschel.cea.fr). The HGBS is a Herschel Key Programme jointly carried out by SPIRE Specialist Astronomy Group 3 (SAG 3), scientists of several institutes in the PACS Consortium (CEA Saclay, INAF-IFSI Rome and INAF-Arcetri, KU Leuven, MPIA Heidelberg), and scientists
of the Herschel Science Center (HSC). We thank an anonymous referee for helpful suggestions.


\begin{thebibliography}{99}
\bibitem[Alves \& Bouy(2012)]{Alves12} Alves, J., \& Bouy, H. 2012, \aap, 547, A97
\bibitem[Andr\'e et al.(2010)]{Andre2010}Andr\'e, P., et al. 2010, \aap, 518, L102
\bibitem[Barrado \& Mart\'{\i}n(2003)]{barrado2003} Barrado y Navascu\'es, D., \&  Mart\'{\i}n, E. L. 2003, \aj, 126, 2997
\bibitem[Cardelli et al.(1989)]{cardelli1989}	Cardelli, Ja. A., Clayton, G. C., \&  Mathis, J. S. 1989, \apj, 345, 245
\bibitem[Carpenter et al.(2001)Carpenter, Hillenbrand \& Skrutskie]{Carpenter} Carpenter, J. M., Hillenbrand, L. A., \& Skrutskie, M. F. 2001, \aj, 121, 3160
\bibitem[Cieza \& Baliber (2007)]{cieza2007} Cieza, L., \& Baliber, N. 2007, \apj, 671, 605
\bibitem[Cutri et al.(2003)]{2mass} Cutri, R. M., et al. 2003, VizieR On-line Data Catalog: II/246
\bibitem[Cutri et al.(2012)]{wise2012} Cutri, R. M., et al. 2012, VizieR On-line Data Catalog: II/311
\bibitem[Dahm et al.(2005)]{dahm2005} Dahm, S. E., \&  Simon, T. 2005, \aj, 129, 829
\bibitem[Da Rio et al.(2009)]{dario2009} Da Rio, N., et al. 2009,  \apj, 183, 261
\bibitem[Da Rio et al.(2010)]{dario2010} Da Rio, N., et al.  2010, \apj, 722, 1092
\bibitem[Epchtein et al.(1997)]{denis} Epchtein, N., et al. 1997, Msngr, 87, 27 
\bibitem[Feigelson et al.(2002)]{feigelson2002}Feigelson, E. D., et al. 2002, \apj, 574, 258
\bibitem[Feigelson et al.(2005)]{feigelson2005}Feigelson, E. D., et al. 2005, \apjs, 160, 379 
\bibitem[Flaccomio et al.(2005)]{flaccomio2005} Flaccomio, E., et al. 2005, \apjs, 160, 450
\bibitem[F\H{u}r\'esz et al.(2008)]{furesz}F\H{u}r\'esz, G., Hartmann, L. W., Megeath, S. T., Szentgy\"orgyi, A. H., \& Hamden, E. T. 2008, \apj, 676, 1109
\bibitem[Getman et al.(2005)]{getman}	Getman, K. V., et al. 2005, \apjs, 160, 319
\bibitem[Gras-Vel\'azquez and Ray (2005)]{excess}Gras-Vel\'azquez, A., \& Ray, T. P. 2005, \aap, 443, 541 
\bibitem[Gutermuth et al.(2008)]{Gutermuth08} Gutermuth, R. A., et al., 2008, \apjs, 674, 336
\bibitem[Haro(1953)]{Haro53} Haro, G. 1953, \apj, 117, 73
\bibitem[Herbig(1960)]{herbig62} Herbig, G. H. 1960, \apj, 131, 632
\bibitem[Herbst et al.(2002)]{wherbst} Herbst, W., Bailer-Jones, C. A. L., Mundt, R., Meisenheimer, K., \& Wackermann, R. 2002, \aap, 396, 513
\bibitem[Herbst et al.(1994)]{herbst1994} Herbst, W., Herbst, D. K., \& Grossman, E. J. 1994, \aj, 108, 1906
\bibitem[Hillenbrand(1997)]{hillenbrand1997} Hillenbrand, L. A. 1997, \aj, 113, 1733
\bibitem[Hillenbrand et al.(1998)]{hillenbrand1998} Hillenbrand, L. A., et al. 1998, \aj, 116, 1816
\bibitem[Koenig et al.(2012)]{koenig2012}Koenig, X. P., et al. 2012, \apj, 744, 130
\bibitem[K\"onigl(1991)]{konigl1991} K\"onigl, A. 1991, \apj, 370, L39
\bibitem[Lada \& Adams(1992)]{lada1992} Lada, C. J., \&  Adams, F. C. 1992, \apj, 393, 278
\bibitem[Lada et al.(2006)]{lada2006} Lada, C. J., et al.  2006,  \aj, 151, 1574
\bibitem[Lawrence et al.(2007)]{Lawrence} Lawrence, A., et al. 2007, \mnras, 379, 1599
\bibitem[Mart\'{\i}n(1998)]{martin1998}Mart\'{\i}n, E. L. 1998, \aj, 115, 351
\bibitem[Megeath et al.(2012)]{megeath12} Megeath, S. T., et al. 2012, \aj, 144, 192
\bibitem[Menten et al.(2007)]{distance} Menten, K. M., Reid, M. J., Forbrich, J., \&  Brunthaler, A. 2007, \aap, 474, 515
\bibitem[Meyer, Calvet \& Hillenbrand(1997)Meyer et al.]{Meyer97} Meyer, M. R., Calvet, N., \& Hillenbrand, L. A. 1997, \aj, 114, 288
\bibitem[Muench et al. (2008)]{muench2008}Muench, A., Getman, K., Hillenbrand, L., Preibisch, T. 2008, Handbook of Star Forming Regions, Volume I: The Northern Sky ASP Monograph Publications, Vol. 4. Edited by Bo Reipurth, p.483
\bibitem[Nakano et al.(2012)]{nakano2012} Nakano, M., Sugitani, K., Watanabe, M., Fukuda, D., Ishihara, D., \&  Ueno, M. 2012, \aj, 143, 61
\bibitem[Neuh\"auser et al.(1995)]{neuhauser1995}Neuh\"auser, R., Sterzik, M. F., Schmitt, J. H. M. M., Wichmann, R., \& Krautter, J. 1995, \aap, 297, 391
\bibitem[Neuh\"auser(1997)]{neuhauser97} Neuh\"auser, R. 1997,  Sci., 276, 1363
\bibitem[O'Dell et al. (2008)]{odell2008}O'Dell, C. R., Muench, A., Smith, N., Zapata, L. 2008, Handbook of Star Forming Regions, Volume I: The Northern Sky ASP Monograph Publications, Vol. 4. Edited by Bo Reipurth, p.544
\bibitem[P\'al(2012)]{pal2012} P\'al, A. 2012, \mnras, 421, 1825
\bibitem[Parsamian \& Chavira(1982)]{Parsam} Parsamian, E. S., \& Chavira, E. 1982, Bol. Inst. Tonantzintla, 3, 69 (VizieR On-line Catalog II/309)
\bibitem[Pizzolato et al. (2003)]{pizzolato2003} Pizzolato, N., Maggio, A., Micela, G., Sciortino, S., \& Ventura, P. 2003, \aap, 397, 147
\bibitem[Preibisch et al.(2005a)]{preibischa} Preibisch, Th., et al. 2005a, \apjs, 160, 401
\bibitem[Preibisch et al.(2005b)]{preibischb} Preibisch, Th., et al. 2005b, \apjs, 160, 582
\bibitem[Rebull(2001)]{rebull2001} Rebull, L. M. 2001, \aj, 121, 1676
\bibitem[Rebull et al.(2000)]{rebull2000} Rebull, L. M., et al. 2000, \aj, 119, 3026
\bibitem[Rebull et al.(2006)]{rebull2006} Rebull, L. M., Stauffer, J. R., Megeath, S. T., Hora, J. L., \&  Hartmann, L. 2006, \apj, 646, 297
\bibitem[Ricci et al.(2008)]{Ricci} Ricci, L., Robberto, M., \& Soderblom, D. R. 2008, \aj, 138, 2138
\bibitem[Rowles \& Froebrich(2009)]{rowles2009} Rowles, J., \& Froebrich, D. 2009, \mnras, 395, 1640
\bibitem[Samus et al.(2007)]{GCVS} Samus, N. et al. 2007--2012, {\it General Catalogue of Variable Stars\/}, {\it VizieR Online Catalog\/} B/gcvs
\bibitem[Scandariato et al.(2011)]{scandariato11} Scandariato, G., Robberto, M., Pagano, I., \& Hillenbrand, L. A. 2011, \aap, 533, A38
\bibitem[Siess, Dufour, \& Forestini(2000)Siess et al.]{siess} Siess, L., Dufour, E., \& Forestini, M. 2000, \aap, 358, 593  
\bibitem[Stelzer et al.(2001)]{stelzer2001} Stelzer, B., \& Neuh\"auser, R. 2001, \aap, 377, 53
\bibitem[Stelzer et al. (2012)]{stelzer12}Stelzer, B., Preibisch, T., Alexander, F., Mucciarelli, P.,  Flaccomio, E., Micela, G., \& Sciortino, S. 2012, \aap, 537, A135
\bibitem[Teixeira et al.(2012)]{index} Teixeira, P. S., Lada, C. J., Marengo, M., \&  Lada, E. A. 2012, \aap, 540, A83
\bibitem[Telleschi et al. (2007)]{telleschi} Telleschi, A., G\"udel, M., Briggs, K. R., Audard, M. 2007, \aap, 468, 425 
\bibitem[Watson et al.(2009)]{xmm} Watson, M. G., et al. 2009, \aap, 493, 339 (VizieR Online Catalogue IX/41)
\bibitem[White \& Basri(2003)]{WB2003} White, R. J., \& Basri, G. 2003, \apj, 582, 1109
\bibitem[Wiramihardja et al.(1991)]{kiso}Wiramihardja, S. D., Kogure, T., Yoshide, S., Nakano, M., Ogura, K., \&  Iwata, T. 1991, \pasj, 43, 27
\bibitem[Wright et al.(2010)]{WISE} Wright, E. L., et al. 2010, \aj, 140, 1868 
\end{thebibliography}
\end{document}